\magnification=\magstep1
\vsize=47.2pc
\overfullrule=0pt
%%%%%%%%%%%%%%%%%%%%%%%%%%%%
\def\jstat{1234}
\def\journal{1}
\ifnum\journal=\jstat \baselineskip=18pt \fi
%%%%%%%%%%%%%%%%%%%%%%%%%%%%
\def\lb{\lbrack}
\def\rb{\rbrack}
\def\min{{\rm min}}
\def\q#1{\lb#1\rb}
\def\bibitem#1{\parindent=8mm\item{\hbox to 6 mm{$\q{#1}$\hfill}}}
\def\mn{\medskip\smallskip\noindent}
\def\sn{\smallskip\noindent}
\def\bn{\bigskip\noindent}
%%%%%%%%%%%%%%%%%%%%%%%%%%%%%%%%%%%%%%%%%%%%%%%%%%%%%%%%%%%%%%%%%%%%
% some fonts
\font\extra=cmss10 scaled \magstep0 \font\extras=cmss10 scaled 750

%%%%%%%%%%%%%%%%%%%%%%%%%%%%%%%%%%%%%%%%%%%%%%%%%%%%%%%%%%%%%%%%%%%%
% check if AMS fonts are present
\def\yesans{y }
\message{Do you have the AMS fonts 2 (y/n) ? }
\read-1 to \amsfontspresent
\ifx\amsfontspresent\yesans
%%%%%%%%%%%%%%%%%%%%%%%%%%%%%%%%%%%%%%%%%%%%%%%%%%%%%%%%%%%%%%%%%%%%
% Some Math symbols (AMS version)

\def\Real{\hbox{\amsmath R}}
\def\Complex{\hbox{\amsmath C}}
\def\Zed{\hbox{\amsmath Z}}

%%%%%%%%%%%%%%%%%%%%%%%%%%%%%%%%%%%%%%%%%%%%%%%%%%%%%%%%%%%%%%%%%%%%
%%% Modification
% Some extra Math symbols

\def\Parity{\hbox{\fraktur P}}
\else
\def\Parity{\wp}
%%%%%%%%%%%%%%%%%%%%%%%%%%%%%%%%%%%%%%%%%%%%%%%%%%%%%%%%%%%%%%%%%%%%
% Special fonts (handmade version)
\setbox111 = \hbox{{{\extra R}}}
\setbox112 = \hbox{{{\extra F}}}
\setbox3 = \hbox{{{\extra C}}}
\def\R{\hskip2.0 true pt{{\extra R}}\hskip-\wd111\hskip-2.0 true pt{{\extra F}
}\hskip-\wd112\hskip2.0 true pt\hskip\wd111}
\def\Real{\hbox{{\extra\R}}}
\def\C{{{\extra C}}\hskip-\wd3\hskip2.5 true pt{{\extra I}}\hskip-\wd2
\hskip-2.5 true pt\hskip\wd3}
\def\Complex{\hbox{{\extra\C}}}
\setbox4=\hbox{{{\extra Z}}}
\setbox5=\hbox{{{\extras Z}}}
\setbox6=\hbox{{{\extras z}}}
\def\Z{{{\extra Z}}\hskip-\wd4\hskip 2.5 true pt{{\extra Z}}}

\def\Zed{\hbox{{\extra\Z}}}

\fi
\setbox22 = \hbox{{{\extra 1}}}
\def\One{{{\extra 1}}\hskip-\wd22\hskip 2.5 true pt{{\extra 1}}}
\def\id{\hbox{{\extra\One}}}
%%%%%%%%%%%%%%%%%%%%%%%%%%%%%%%%%%%%%%%%%%%%%%%%%%%%%%%%%%%%%%%%%%%%
\def\mod{\phantom{l} {\rm mod} \phantom{l}}
\def\vac{\mid \!v \rangle\, }
\def\avac{\langle v \! \mid\, }
\def\state#1{\mid \! #1 \rangle\, }
\def\rrangle{\rangle \hskip -1pt \rangle}
\def\llangle{\langle \hskip -1pt \langle}
\def\astate#1{\langle #1 \! \mid\, }
\def\pstate#1{\Vert #1 \rrangle\, }
\def\apstate#1{\llangle #1 \Vert\, }
\def\norm#1{\Vert #1 \Vert}
\def\abs#1{\vert #1 \vert}
% Section headings

\def\chapsubtitle#1{\leftline{\bf #1}
\vskip-10pt
\line{\hrulefill}}
% Useful abbreviations
\def\vl{\hskip 1pt \vrule}
\def\normvac{\langle v \! \mid \!v \rangle\, }
\def\ab{\bar{\alpha}}
\def\a{\alpha}
\def\si{\sigma}
\def\Ga{\Gamma}
\def\om{\omega}
\def\la{\lambda}
\def\vphi{\varphi}

\def\D{{\cal D}}
\def\E{{\cal E}}
\def\EE{\hbox{{\extra E}}}
\def\H{{\cal H}}
\def\HN{\H_N}

\def\O{{\cal O}}

\def\bsin#1{\sin{\left(#1\right)}}
\def\bcos#1{\cos{\left(#1\right)}}
\def\cPh{\bcos{\textstyle{P \over 2}}}
\def\phit{{\phi \over 3}}
\def\Ph{{P \over 2}}
\def\re{{\rm Re}}
\def\lotimes{\ {\buildrel \leftarrow \over \otimes}\ }
\def\Ddots{\mathinner{\mkern1mu\raise1pt\hbox{.}\mkern2mu
       \raise4pt\hbox{.}\mkern2mu\raise7pt\vbox{\kern7pt\hbox{.}}\mkern1mu}}
% Macros for drawing picture
\def\mskp{\mkern-6mu}
\def\hr{\hbox{\raise2.5pt\hbox to 1.0cm{\hrulefill}}}
\def\Hr#1{\mathop{\hr}\limits_{#1}}
\def\hrr{\hbox{\raise1pt\hbox to 1.0cm{\hrulefill
 }\hskip-1.0cm\raise4pt\hbox to 1.0cm{\hrulefill}}}
\def\Hrr#1{\mathop{\hrr}\limits_{#1}}
\def\hrf{\hbox{\raise1pt\hbox to 1.0cm{\hrulefill
 }\hskip-1.0cm\raise2pt\hbox to 1.0cm{\hrulefill
 }\hskip-1.0cm\raise3pt\hbox to 1.0cm{\hrulefill
 }\hskip-1.0cm\raise4pt\hbox to 1.0cm{\hrulefill}}}
\def\Hrf#1{\mathop{\hrf}\limits_{#1}}
\def\bull{\bullet}
\def\bu#1{\mathop\bull\limits^{#1}}
\def\cab{{\cal C}}
\def\ccab{{\bar{\cal C}_{3}}}
\def\cnab#1{{\bar{\cal C}_{#1}}}
\def\chab{{\widehat{\cal C}}}
\def\Rab{{\cal R}}
%
%%%%%%%%%%%%%%%%%%%%%%%%%%%%%%%%%%%%%%%%%%%%%%%%%%%%%%%
% Definition of sections
\def\secA{1}
\def\secB{2}
\def\secC{3}
\def\secD{4}
\def\secE{5}
\def\secG{6}
\def\secH{7}
\def\secI{8}
\def\secJ{9}
\def\appA{A}
\def\appB{B}
\def\appC{C}
%%%%%%%%%%%%%%%%%%%%%%%%%%%%%%%%%%%%%%%%%%%%%%%%%%%%%%%
% Definitions of references
\def\ostlund{1}
\def\rietalA{2}
\def\rietalB{3}
\def\hkn{4}
\def\gehri{5}
\def\dogra{6}
\def\perk{7}
\def\onsager{8}
\def\yang{9}
\def\baxterA{10}
\def\baxterB{11}
\def\albertiniA{12}
\def\albertiniB{13}
\def\mccoyadv{14}
\def\perkadv{15}
\def\scm{16}
\def\roanA{17}
\def\daviesA{18}
\def\daviesB{19}
\def\roanB{20}
\def\ahn{21}
\def\fatzamA{22}
\def\alcaraz{23}
\def\fateev{24}
\def\lykyanov{25}
\def\cardy{26}
\def\zamPA{27}
\def\zamPB{28}
\def\zamPC{29}
\def\fatzamB{30}
\def\mussardo{31}
\def\weA{32}
\def\gehlenph{33}
\def\lett{34}
\def\dkcoy{35}
\def\kedem{36}
\def\han{37}
\def\chrihen{38}
\def\zam{39}
\def\baym{40}
\def\tang{41}
\def\hela{42}
\def\camp{43}
\def\thesis{44}
\def\jones{45}
\def\yildirim{46}
\def\cardyB{47}
\def\schuetz{48}
\def\automos{49}
\def\kogut{50}
\def\krallm{51}
\def\albcoy{52}
\def\lehmann{53}
\def\gehkal{54}
\def\albCONF{55}
\def\reedsimon{56}
\def\katobook{57}
\def\rellich{58}
\def\kato{59}
%%%%%%%%%%%%%%%%%%%%%%%%%%%%%%%%%%%%%%%%%%%%%%%%%%%%%%%%%%%%%%
%
% Title Page
%
%%%%%%%%%%%%%%%%%%%%%%%%%%%%%%%%%%%%%%%%%%%%%%%%%%%%%%%%%%%%%%
%%% fonts for title page
\font\large=cmbx10 scaled \magstep3
\font\bigf=cmr10 scaled \magstep2
\pageno=0
\def\folio{
\ifnum\pageno<1 \footline{\hfil} \else\number\pageno \fi}
\phantom{not-so-FUNNY}
\rightline{ BONN--TH--94--21\break}
\rightline{ hep-th/9409122\break}
\rightline{ September 1994\break}
\rightline{ revised April 1995\break}
\ifnum\journal=\jstat
 \vskip 1.0truecm
\else
 \vskip 2.0truecm
\fi
\centerline{\large A Perturbative Approach}
\vskip 6pt
\centerline{\large \phantom{g} to the \phantom{g}}
\vskip 6pt
\centerline{\large Chiral Potts Model}
\vskip 1.0truecm
\centerline{\bigf A.\ Honecker}
\bigskip\medskip
\centerline{\it Physikalisches Institut der Universit\"at Bonn}
\centerline{\it Nu{\ss}allee 12, 53115 Bonn, Germany}
\vskip 1.1truecm
\centerline{\bf Abstract}
\vskip 0.2truecm
\noindent
The massive high-temperature phase of the chiral Potts quantum chain is
studied using perturbative methods. For the $\Zed_3$-chain we present
high-temperature expansions for the groundstate energy and the dispersion
relations of the two single-particle states as well as two-particle states
at general values of the parameters. We also present a perturbative argument
showing that a large class of massive $\Zed_n$-spin quantum
chains have quasiparticle spectra with $n-1$ fundamental particles.
It is known from earlier investigations that --at special values of the
parameters-- some of the fundamental particles exist only for limited ranges
of the momentum. In these regimes our argument is not rigorous as one
can conclude from a discussion of the radius of convergence of the
perturbation series.
\sn
We also derive correlation functions from a perturbative evaluation of the
groundstate for the $\Zed_3$-chain. In addition to an exponential decay we
observe an oscillating contribution. The oscillation length seems to be
related to the asymmetry of the dispersion relations. We show that this
relation is exact at special values of the parameters for general $\Zed_n$
using a form factor expansion.
\vfill
\eject
%%%%%%%%%%%%%%%%%%%%%%%%%%%%%%%%%%%%%%%%%%%%%%%%%%%%%%%
\chapsubtitle{\secA.\ Introduction}
\mn
In this paper we discuss the chiral Potts model in its spin quantum
chain formulation. The first chiral Potts model that was
introduced in 1981 by Ostlund in order to describe incommensurate
phases of physisorbed systems $\q{\ostlund}$ was a classical 2D spin model.
The associated quantum chain Hamiltonians were obtained in 1981-82 by
Rittenberg et al.\  $\q{\rietalA}\q{\rietalB}$.
Because this chain was not self-dual the location of the
critical manifold was difficult. In 1983, Howes, Kadanoff
and denNijs introduced a self-dual $\Zed_3$-symmetric chiral quantum
chain $\q{\hkn}$, which however, does not correspond to a two-dimensional
model with positive Boltzmann weights. Soon afterwards, von Gehlen and
Rittenberg noticed that the remarkable property of the first gap of this model
being linear in the inverse temperature also applies to the second
gap and can be generalized to arbitrary $\Zed_n$ $\q{\gehri}$. Furthermore,
the authors of $\q{\gehri}$ showed that the Ising-like form of the eigenvalues
is related to this $\Zed_n$-Hamiltonian satisfying the Dolan-Grady
%%% Modification
% additional reference
integrability condition $\q{\dogra}$ -- or equivalently $\q{\perk}$ Onsager's
algebra $\q{\onsager}$. It was then shown by
%%% Modification
% further names included
Au-Yang, Baxter, McCoy, Perk et al.\
that this integrability
property -- nowadays called `superintegrability' -- can be implemented in a 2D
classical model with Boltzmann weights defined on higher genus Riemann
surfaces that satisfy a generalized Yang-Baxter relation. In the sequel
the chiral Potts model attracted much attention because of these mathematical
aspects, i.e.\ on the one hand the generalized Yang-Baxter relations
$\q{\yang - \roanA}$ and on the other hand because of Onsager's algebra
$\q{\perk}\q{\daviesA - \ahn}$.
In this paper we present new results showing that the model is also
`physically' very interesting although it is not directly related to a
realistic 2D physisorbed system.
\sn
Our observations will apply to general $\Zed_n$-spin quantum chains:
The superintegrable $\Zed_n$-chiral Potts quantum chains can be
generalized (not necessarily demanding integrability) to include
further known integrable models, in particular the conformally
invariant models of Fateev and Zamolodchikov with ${\cal WA}_{n-1}$-symmetry
$\q{\fatzamA - \lykyanov}$.
Recently, Cardy introduced an integrable chiral perturbation of these
models $\q{\cardy}$. The $\Zed_n$-spin quantum chains describe
both this perturbation as well as the integrable thermal perturbations
of the conformal field theories (see e.g.\ $\q{\zamPA-\mussardo}$).
\bn
In previous papers we provided numerical evidence that the low-lying
excitations in the zero momentum sectors can be explained in terms of
$n-1$ fundamental particles for $n=3$, $4$ at general values
of the parameters $\q{\weA}\q{\gehlenph}$ and checked for $n=3$ that
this quasiparticle picture extends to general momenta $\q{\lett}$.
For the superintegrable $\Zed_3$-chiral Potts model
McCoy et al.\ have derived a quasiparticle picture of the
complete spectrum using Bethe ansatz techniques $\q{\dkcoy}$.
Recently, they argued that this quasiparticle picture should in
general be valid for the integrable $\Zed_3$-chiral Potts
quantum chain $\q{\kedem}$.
In this paper we will show that both results can be combined
into the general statement that the massive high-temperature
phases of general chiral Potts quantum chains have quasiparticle
spectra. In fact, this quasiparticle picture will in certain cases
give small corrections to the additivity of energy in the momentum zero
sectors observed in $\q{\weA}$.
\sn
The massive low-temperature phases of the $\Zed_n$-spin quantum
chains exhibit spectra that are dual to those in the
high-temperature phases, the main difference being that the r\^ole
of charge and boundary conditions is interchanged $\q{\han}$.
Therefore, our results about the massive high-temperature phase
can be transferred to the massive low-temperature phase
using duality.
\medskip
The outline of this paper is as follows. In section {\secB} we recall
some well-known facts about the chiral Potts quantum chain and
introduce basic notions. Section {\secC} presents a short summary
of perturbation theory which is applied in section {\secD} to the
dispersion relations of the lowest excitations of the $\Zed_3$-chain.
In section {\secE} we derive the main statement of our paper:
The quasiparticle structure of the massive high-temperature phase.
Details of the proof are shifted to an appendix.
%%% Modification
% finite-size corrections mentioned
This argument can also be used in order to obtain some control on the
finite-size effects.
In section {\secG} we
apply perturbation expansions and form factor decompositions to
the correlation functions, our main interest being the oscillatory
contribution. Then we specialize to vanishing chiral
angles and discuss some of the results obtained previously in more
detail. The final section {\secI} where we discuss the radius of
convergence of the perturbation series completes our investigation.
\bn
\chapsubtitle{\secB.\ The chiral Potts quantum chain}
\mn
This section summarizes well-known basic facts about $\Zed_n$-spin
quantum chains. We also introduce some notions that will be useful
later on. For more details see e.g.\ the review $\q{\chrihen}$.
\mn
A general $\Zed_n$-spin quantum chain with $N$ sites is defined by
the Hamiltonian:
$$H^{(n)}_N = - \sum_{j=1}^N \sum_{k=1}^{n-1} \ab_k \si_j^k
                           + \la \a_k \Ga_j^k \Ga_{j+1}^{n-k}
                           + \hat{\la} \hat{\a}_k \Xi_j^k \Xi_{j+1}^{n-k} .
             \eqno{(\rm \secB.1)}$$
For reasons to be explained below we will in all subsequent sections
set $\hat{\la} = 0$, i.e.\ we will neglect the extra term in (\secB.1)
introduced in ref.\ $\q{\ahn}$ and will consider
$$H^{(n)}_N = - \sum_{j=1}^N \sum_{k=1}^{n-1} \ab_k \si_j^k
                                     + \la \a_k \Ga_j^k \Ga_{j+1}^{n-k}
             \eqno{(\rm \secB.2)}$$
instead.
$\si_j$, $\Ga_j$ and $\Xi_j$ freely generate a finite dimensional
associative algebra with involution by the following relations
($1 \le j,l \le N$):
$$\eqalign{
\si_j \si_l &= \si_l \si_j \ , \qquad
\si_j \Ga_l = \Ga_l \si_j \om^{\delta_{j,l}}_{} \ , \cr
\Ga_j \Ga_l &= \Ga_l \Ga_j \ , \qquad
\si_j^n = \Ga_j^n = \Xi_j^n = \left(\Xi_j \Ga_j\right)^n = \id \ , \cr
\Xi_j \Xi_l &= \Xi_l \Xi_j \ , \qquad
\si_j \Xi_l = \Xi_l \si_j \om^{\delta_{j,l}}_{} \ , \qquad
\underbrace{\Xi_j \Ga_j \Xi_j \Ga_j \ldots}_{n \ {\rm operators}}
 \ne \id \ , \cr
\si_j^{+} &= \si_{n}^{n-1} \ , \qquad
\Ga_j^{+} = \Ga_{j}^{n-1} \ , \qquad
\Xi_j^{+} = \Xi_{j}^{n-1} \cr
} \eqno{(\rm \secB.3)}$$
where $\om$ is the primitive $n$th root of unity
$\om = e^{2 \pi i \over n}_{}$. In the following we will consider only
periodic boundary conditions for $H^{(n)}_N$, i.e.\ $\Ga_{N+1} = \Ga_1$.
\sn
The Hamiltonian (\secB.1) contains $3n-1$ parameters: The temperature-like
parameters $\la$ and $\hat{\la}$ which we choose to be real and the complex
constants $\ab_k$, $\a_k$ and $\hat{\a}_k$.
$H^{(n)}_N$ is hermitean iff $\ab_k = \ab_{n-k}^{*}$, $\a_k = \a_{n-k}^{*}$
and $\hat{\a}_k = \hat{\a}_{n-k}^{*}$.
\medskip
The algebra (\secB.3) is conveniently represented in
$$\HN := \underbrace{
\Complex^n \otimes \Complex^n \otimes \ldots \otimes \Complex^n}_{
N \phantom{l} {\rm times}}              \eqno{(\rm \secB.4)}$$
labeling the standard basis of $\Complex^n$ by $\{e_0, \ldots, e_{n-1}\}$.
Then a basis for (\secB.4) is given by:
$$\state{i_1 \ldots i_N} := e_{i_1} \otimes \ldots \otimes e_{i_N}
\ , \qquad 0 \le i_j \le n-1.   \eqno{(\rm \secB.5)}$$
Now the following operation in the space (\secB.4) is a faithful
irreducible representation $r$ of the algebra (\secB.3):
$$\eqalign{
r(\si_j) \state{i_1 \ldots i_j \ldots i_N}
            &= \om^{i_j} \state{i_1 \ldots i_j \ldots i_N} \ , \cr
r(\Ga_j) \state{i_1 \ldots i_j \ldots i_N}
            &= \state{i_1 \ldots (i_j+1 \mod n) \ldots i_N} \ , \cr
r(\Xi_j) \state{i_1 \ldots i_j \ldots i_N}
            &= \cases{
     -\state{i_1 \ldots (i_j+1) \ldots i_N}, &if $i_j < n-1$; \cr
     \state{i_1 \ldots 1 \ldots i_N}, &if $i_j = n-1$ . \cr } \cr
}         \eqno{(\rm \secB.6)}$$
The involution is the adjoint operation with
respect to the standard scalar product in the tensor product of $\Complex^n$.
\medskip
The Hamiltonian (\secB.1) commutes with the $\Zed_n$ charge operator
$\hat{Q} := \prod_{j=1}^N \si_j$ acting on the vectors (\secB.5) as
$$r(\hat{Q}) \state{i_1 \dots i_N} = \om^{\left(\sum_{j=1}^N i_j\right)}_{}
     \state{i_1 \dots i_N}     \eqno{(\rm \secB.7)}$$
which shows that the eigenvalues of $\hat{Q}$
have the form $\om^Q$ with $Q$ integer. Thus, $H^{(n)}_N$ has $n$ charge
sectors which we shall refer to by $Q=0,$ $\ldots,$ $n-1$.
\medskip
$H^{(n)}_N$ also commutes with the translation operator $T_N$ that
acts on the basis vectors (\secB.5) in the following way:
$$r(T_N) \state{i_1 i_2 \dots i_{N}} = \state{i_2  \ldots i_N i_1}.
     \eqno{(\rm \secB.8)}$$
The eigenvalues of $T_N$ are $N$th roots of unity. We label them by
$e^{i P}$ and call $P$ the `momentum'. We choose $0 \le P < 2 \pi$
corresponding to the first Brillouin zone
and have $P \in \{0, {2 \pi \over N}, \ldots, {2 \pi (N-1) \over N} \}$.
Note that the states
$$\eqalign{
\pstate{i_1 i_2 \ldots i_{N-1} i_{N}}_P^{} :=
  {1 \over \sqrt{{\cal N}}} {\bigg (} &\state{i_1 i_2 \ldots i_{N-1} i_{N}}
       + e^{i P} \state{i_N i_1 i_2 \ldots i_{N-1}}
       + \ldots \cr
      &+ e^{i P (N-1)} \state{i_2 \ldots i_{N-1} i_N i_1} {\bigg )} \cr
}\eqno{(\rm \secB.9)}$$
are eigenstates of $T_N$ with eigenvalue $e^{i P}$.
${\cal N}$ is a suitable normalization constant.
If the state $\state{i_1 \ldots i_{N} }$ has no symmetry
(i.e.\ $T_N^k \state{i_1 \ldots i_{N} }
\ne \state{i_1 \ldots i_{N} }$ for all $0<k<N$), one has ${\cal N} = N$.
This will apply to most cases below where we need (\secB.9).
\bigskip
In this paper we will use the following parametrization of the constants
$\a_k$ and $\ab_k$, fixing their dependence on $k$:
$$\a_k = {e^{i \phi ({2 k \over n}-1)} \over \sin{\pi k \over n} } \ , \qquad
\ab_k = {e^{i \vphi ({2 k \over n}-1)} \over \sin{\pi k \over n} }.
    \eqno{(\rm \secB.10)}$$
This is a suitable choice because it includes a large class of
interesting models.
\mn
For $\phi = \vphi = 0$ one obtains real
$\a_k = \ab_k = {1 \over \sin{\pi k \over n}}$. This leads to
models with a second order phase transition at $\lambda=1$ which
can be described by a parafermionic conformal field theory in the limit
$N \to \infty$ at criticality $\q{\fatzamA}\q{\alcaraz}$.
These so-called Fateev-Zamolodchikov-models $\q{\fateev}$
lead to extended conformal algebras ${\cal WA}_{n-1}$ where the simple
%%% Modification
% Avoid impression that Z_n parafermionic CFTs have W-algebras
% with simple fields of dimension 2, ..., n
fields have conformal dimension $2, \ldots, n$ for generic values of
the central charge $c$ $\q{\lykyanov}$. The spectrum of
the Hamiltonian (\secB.2) can be described by the first unitary minimal
model of the algebra ${\cal WA}_{n-1}$. For $n=3$ the symmetry algebra
is Zamolodchikov's well-known spin-three extended conformal algebra
$\q{\zam}$ at $c={4 \over 5}$.
\mn
Choosing $\phi = \vphi = {\pi \over 2}$ in (\secB.10) for the Hamiltonian
(\secB.2) yields the superintegrable chiral Potts model. For $n=3$ such
complex parameters in a spin chain Hamiltonian were first investigated in
detail by Howes, Kadanoff and denNijs $\q{\hkn}$. The integrability
of this chain was then recognized by von Gehlen and Rittenberg who also
generalized it to higher $\Zed_n$ $\q{\gehri}$. More precisely, the authors
of $\q{\gehri}$ showed that the $\Zed_n$-Hamiltonian (\secB.2) with
(\secB.10) at $\phi = \vphi = {\pi \over 2}$ is integrable for all values of
the inverse temperature $\la$ using the Dolan-Grady integrability
condition $\q{\dogra}$.  This particular kind of integrability is
called `superintegrability'
%%% Modification
% clarifying parenthesis added
(note that this terminology is not used entirely
consistent in the literature -- in contrast to us, some authors include
the generalized Yang-Baxter relations in the notion of superintegrability).
Ahn et al.\ $\q{\ahn}$ showed that the Hamiltonian (\secB.1) is still
integrable at $\phi = \vphi = {\pi \over 2}$ for
$\hat{\a}_k = \a_k = \ab_k = 1 - i \cot{\pi k \over n}$ and any
$\la$, $\hat{\la}$. Their argument used Onsager's algebra in order to
construct an infinite set of commuting conserved charges. Note that the
Hamiltonian (\secB.1) subject to the above constraints is {\it not
superintegrable} for general values of the parameters. Anyway,
one can introduce a further parameter $\hat{\la}$ into (\secB.2) without
spoiling integrability
\footnote{${}^{1})$}{
For $n=2$ and $\la = \hat{\la}$ this gives rise to extra symmetries of the
Hamiltonian -- one obtains an XY quantum chain $\q{\ahn}$ that is
invariant under an additional global $U(1)$ symmetry group. However,
one can easily check that for $n>2$ the Hamiltonian
(\secB.1) is not invariant under any non-trivial change of bases
$\Ga_j \to a \Ga_j + b \Xi_j$, $\Xi_j \to c \Ga_j + d \Xi_j$.
}.
\mn
The parametrization (\secB.10) also includes the family of integrable
%%% Modification
% 1. inappropriate name removed
% 2. range of citations enlarged
models discovered in $\q{\yang-\mccoyadv}$
which interpolates between the integrable cases at $\phi = \vphi = 0$,
$\la = 1$ and $\phi = \vphi = {\pi \over 2}$. The Hamiltonian
(\secB.2) is integrable if one imposes the additional constraint
$$\cos \vphi = \lambda \cos \phi    \eqno{(\rm \secB.11)}$$
on the parametrization (\secB.10).
For $\phi = \vphi = 0$ this yields $\la=1$ -- the conformally invariant
critical points. At $\phi = \vphi = 0$, the Hamiltonian
is self-dual, i.e.\ it is invariant under a duality-transformation such
that $H^{(n)}_N(\la) \cong \la H^{(n)}_N(\la^{-1})$. The Hamiltonian is also
self-dual on the superintegrable line $\phi=\vphi={\pi \over 2}$.
$H^{(n)}_N$ with the choices (\secB.10), (\secB.11) is in general
not self-dual any more
whereas particular choices yield a self-dual Hamiltonian. If we choose
for (\secB.10) $\phi=\vphi$ and neglect (\secB.11) $H^{(n)}_N$ will be
self-dual again. Therefore we choose to refer to (\secB.2) with (\secB.10) as
the general `chiral Potts model'. We will not consider the integrable case
where the additional constraint (\secB.11) is satisfied in detail.
\bigskip
We will now explain why we are going to focus on the Hamiltonian
(\secB.2) rather than considering the more general case (\secB.1).
For $\la = 0$ (\secB.1) is just a different representation
of (\secB.2). Thus, although we will certainly obtain different numerical
results, the main structures are unchanged by the extra term in (\secB.1).
In this paper we will use for example perturbation theory. The free
part of the Hamiltonian $H_0$ is the same in (\secB.2) and in (\secB.1):
$H_0 = - \sum_{j,k} \ab_k \si_j^k$. Only the potential $V$ is changed.
For (\secB.2) we have $V = - \sum_{j,k} \a_k \Ga_j^k \Ga_{j+1}^{n-k}$
whereas for (\secB.1) we have an extra term
$\hat{V} = - \sum_{j,k} \a_k \Ga_j^k \Ga_{j+1}^{n-k}
+ h \hat{\a}_k \Xi_j^k \Xi_{j+1}^{n-k}$ with $h := \la^{-1} \hat{\la}$.
Obviously, the action of $V$ and $\hat{V}$ on the eigenstates (\secB.9)
of charge and momentum is the same apart from different
numerical constants.
Furthermore, the extra term in (\secB.1) spoils duality. Thus, we will
not consider the Hamiltonian (\secB.1) explicitly any more. It is always
understood that our results apply to it with only minor modifications.
In particular, the quasiparticle picture we will derive for the
Hamiltonian (\secB.2) will hold for (\secB.1) as well.
\bigskip
Our main interest is the spectrum in the limit $N \to \infty$ of $H_N^{(n)}$.
Of course, we have to specify how the limit is to be taken.
In order to be able to study the spectrum in this limit we
concentrate on the $N$-dependence of the Hilbert spaces
$\HN = \D(H_N^{(n)})$. Consider the following embedding of Hilbert spaces:
$$\eqalign{
\HN & \to \H_{M} \qquad N < M \cr
\pstate{i_1 \ldots i_N}_P^{} &\mapsto
\pstate{i_1 \ldots i_N \underbrace{0 \ldots 0}_{
M-N \phantom{l} {\rm times}} }_P^{}. \cr
}  \eqno({\rm \secB.12})$$
This definition is motivated by the well-known fact that matrix
elements of (\secB.2) in momentum space are almost independent of $N$.
We will see in the following sections that this definition is indeed
useful.
\sn
Using the inclusion map (\secB.12) we can define the Hilbert space $\H$
as the closure of an inductive limit
$$\H := \{
\state{x} \mid \exists N: \quad \state{x} \in \HN
\}^{\widetilde{}}_{}.   \eqno({\rm \secB.13})$$
Furthermore, we shall not consider the limit of $H_N^{(n)}$ directly.
Instead, we shall subtract the groundstate energy $E_N^0$ first and then
consider the weak limit of the operator
$$\Delta H_N^{(n)} := H_N^{(n)} - E_N^0 \id.
\eqno({\rm \secB.14})$$
Similarly, we define $T$ to be the weak limit of $T_N$.
For each finite $N$ eqs.\ (\secB.7) and (\secB.9) imply that
the Hilbert space $\HN$ is graded into charge and momentum eigenspaces:
$$\HN = \bigoplus_P \bigoplus_{Q=0}^{n-1} \HN^{P,Q}.  \eqno({\rm \secB.15})$$
In the limit $N \to \infty$ the grading (\secB.15) translates into the fact
that $\Delta H^{(n)}$ and $T$ can be written in terms of the same
projection-valued measure $\{ \Pi_{\mu}^Q \}$:
$$T = \sum_{Q=0}^{n-1} \int e^{i P(\mu)} {\rm d}\Pi_{\mu}^Q \ , \qquad
\Delta H^{(n)} = \sum_{Q=0}^{n-1} \int \Delta E(\mu) {\rm d}\Pi_{\mu}^Q
     \eqno({\rm \secB.16})$$
with $0 \le P(\mu) < 2 \pi$. The $\{ \Pi_{\mu}^Q \}$ can be thought of as
infinite dimensional generalizations of projection operators onto
eigenspaces of charge $Q$ and momentum $P(\mu)$. Thus, (\secB.16) is
just the proper formulation of (\secB.15) in the infinite dimensional
case. The existence of the limits and projection valued measures in
(\secB.16) is not at all obvious. However, this is guaranteed by the
quasiparticle picture whereof a proof is presented in appendix {\appB}.
\mn
The definition in (\secB.14) is motivated by the fact that the smallest
eigenvalue of (\secB.2) has a leading term proportional to $N$ and the
excitation spectrum usually is defined with respect to this reference
eigenvalue. With the definition (\secB.14) the Hamiltonian $\Delta H_N^{(n)}$
is bounded from below. This automatically yields an operator
$\Delta H^{(n)}=\lim\limits_{N \to \infty} \Delta H_N^{(n)}$ with positive
spectrum and an eigenvector for eigenvalue $0$. Note that this definition
of the limit implies that any point where at finite $N$ eigenvalues exist
that are arbitrarily close to it belongs to the spectrum. In particular,
the spectrum forms a closed set.
\medskip
Before proceeding let us make a few further comments on our definition
of the limit. First note that $H_N^{(n)}$ is defined only
on $\D(H_N^{(n)}) = \HN \subset \H$. Of course, we could extend it
linearly (e.g.\ by $0$) to the complete Hilbert space $\H$. However,
it is easy to show that the limit $\Delta H^{(n)}$ does not depend on the
particular extension chosen as long as it is uniformly bounded for all
$N$. We will therefore not make use of any particular extension.
\sn
Secondly, it is convenient to let $H_N^{(n)}$ act on vectors in
$\D(H_N^{(n)}) = \HN$ which corresponds to choosing a particular
representative for a vector in the Hilbert space $\H$. This is useful
because $H_N^{(n)}$ naturally acts on chains of length $N$. However,
such a vector always has to be thought of as lying in $\H$ and,
in particular, in all $\H_M$ with $M \ge N$. Although the notation
might propose this, a limit in the chain length never has to be applied
to momentum eigenstates. Of course, other states than (\secB.9)
are not naturally embedded into $\H$ and therefore have to be expanded
in terms of them. This might lead to $N$-dependent coefficients
and a limit might have to be applied to the {\it coefficients}.
\sn
Finally, it is noteworthy that the Hilbert space $\H$ can be thought
of as a kind of universal tensor product. Any tensor product
of spaces $\HN$ and $\H_M$ can be naturally identified with
$\H_{N+M}$: $\HN \otimes \H_M \cong \H_{N+M}$. Therefore the
definition (\secB.13) yields an object that is closed with
respect to taking tensor products. Note that we have chosen
a particular topology on $\H$ which is not the one usually chosen
on the tensor algebra of a vector space. Still, this observation
is useful to guarantee the completeness of the construction
to be presented in section \secE.
\bn
\chapsubtitle{\secC.\ Generalities about Perturbation Theory}
\mn
In this section we review the general outline for perturbation theory
to all orders as presented in $\q{\baym}$
which directly applies to the degenerate case as well.
\sn
The Hamiltonian (\secB.2) can be written as
$$H = H_0 + \la V     \eqno({\rm \secC.1})$$
with $H_0 = - \sum_{j,k} \ab_k \si_j^k$,
$V = - \sum_{j,k} \a_k \Ga_j^k \Ga_{j+1}^{n-k}$.
The eigenstates for $H_0$ are obvious, thus we have solved:
$$H_0 \state{a} = E_{\state{a}}^{(0)} \state{a}.      \eqno({\rm \secC.2})$$
Now one can solve
$$H \state{a(\la)} = E_{\state{a}} \state{a(\la)}
    \eqno({\rm \secC.3})$$
for small $\la$ as follows:
Let $q_{\state{a}}$ be the projector onto the eigenspace of $H_0$ with
eigenvalue $E_{\state{a}}^{(0)}$. We can treat non-degenerate and
degenerate perturbation theory alike if we choose $\state{a}$ such
that
$$q_{\state{a}} V \state{a} = E_{\state{a}}^{(1)} \state{a}
     \eqno({\rm \secC.4})$$
with a constant $E_{\state{a}}^{(1)}$,
i.e.\ $q_{\state{a}} V q_{\state{a}}$ is to be chosen diagonal.
One also needs a regularized resolvent $g(z)$ of $H_0$:
$$g(z) := \left( \id - q_{\state{a}} \right) \left(z-H_0\right)^{-1}.
      \eqno({\rm \secC.5})$$
Then, the Wigner-Brillouin perturbation series
$$E_{\state{a}} = \sum_{\nu=0}^{\infty} \la^{\nu} E_{\state{a}}^{(\nu)}
\ , \qquad
\state{a(\la)} = \sum_{\nu = 0}^{\infty} \la^{\nu} \state{a, \nu}
      \eqno({\rm \secC.6})$$
is given by the following recurrence relations $\q{\baym}$:
$$\eqalign{
\state{a, 0} &= \state{a} \cr
\state{a, \nu} &= g(E_{\state{a}}^{(0)}) \left\{
       V \state{a, \nu-1} - \sum_{\mu = 1}^{\nu-1}
       \state{a, \nu-\mu} E_{\state{a}}^{(\mu)}
       \right\}, \cr
E_{\state{a}}^{(\nu+1)} &= \astate{a} V \state{a, \nu}. \cr
}       \eqno({\rm \secC.7})$$
Note that neither $\state{a(\la)}$ nor $\state{a, \nu}$ are in general
normalized although $\state{a}$ must be normalized to one.
Observe that the derivation of (\secC.7) does not rely on $H$
being hermitean. Therefore, (\secC.7) may also be applied
to diagonalizable but non-hermitean $H$.
\sn
The radius of convergence of the series (\secC.6) can be more
easily discussed in a different framework. Therefore, we postpone such a
discussion to section \secI.
\bigskip
There is one observation that makes explicit evaluation of high orders
for the $\Zed_n$-Hamiltonian (\secB.2) possible.
The energy-eigenvalues $E_{\state{a}}$ of $H_N^{(n)}$ do depend on the
chain length $N$. However, for the low lying gaps $\Delta E_{\state{a}}$
\footnote{${}^{2})$}{This will apply precisely to the fundamental
quasiparticle states to be discussed below.}
of $\Delta H_N^{(n)}$ (see (\secB.14) ) the coefficients for powers
of $\la$ become independent of $N$ up to order $\la^{N-2}$
(see e.g.\ $\q{\tang}$). Intuitively, this can be inferred from the
fact that (\secB.2) shows only nearest neighbour interaction and thus we
need $N-1$ powers in $V$ to bring us around a chain of length $N$.
Smaller powers in $V$ (or $\la$) do not feel that the length of the chain
is finite.
\bn
\chapsubtitle{\secD.\ High-temperature expansions}
\mn
In this section we study the low lying levels in the
spectrum of the $\Zed_3$-chiral Potts model perturbatively. In particular,
we calculate the dispersion relations of the lowest excitations in the
charge sectors $Q=1$ and $Q=2$. Some first results in this direction
have been presented in $\q{\weA}$ for the self-dual version of
this model. In this section we derive higher orders and admit general
$\phi \ne \vphi$
%%% Modification
% footnote added to explain limitations of length of perturbation series
\footnote{${}^{3})$}{Note that the main limitation of the length of most of
the series to be presented in this section is that we explicitly keep the
dependence on the parameters.}.
We also present some explicit results on the next excitations.
\sn
Perturbation expansions had already been used in $\q{\hkn}$, and were
again used e.g.\ in $\q{\hela}$ and $\q{\tang}$ in order to obtain
some results for spectra and order parameters on the superintegrable
line. Recently, low-temperature expansions have been applied in $\q{\han}$
to spectra and correlation functions for general values of the parameters.
Here, we focus on the high-temperature regime.
\medskip
For arbitrary $n$, $N$ the groundstate of the Hamiltonian (\secB.2)
in the limit $\la \to 0$ is given by:
$$\state{{\rm GS}} :=
\state{0 \ldots 0} \eqno({\rm \secD.1})$$
provided that $-{\pi \over 2} \leq \vphi \leq {\pi \over 2}$. For $n=3$
(\secD.1) will be the groundstate for the larger range
$-\pi \leq \vphi \leq \pi$ and for $n=4$ (\secD.1) is the groundstate for
$-{5 \pi \over 6} \leq \vphi \leq {5 \pi \over 6}$.
\mn
The first excited states at $\la = 0$ for $Q > 0$ and arbitrary $P$
are the states
$$\pstate{s^Q}_P := \pstate{Q 0 \ldots 0}_P^{}   \eqno({\rm \secD.2})$$
in the range $-{\pi \over 2} \leq \vphi \leq {\pi \over 2}$.
According to our definition of the space $\H$ in section \secB,
the states (\secD.2) give rise to proper eigenstates in
the limit of $\Delta H^{(n)}$. Thus, the corresponding gaps
$\Delta E_{Q,0} (P)$ belong to the point spectrum of $\Delta H^{(n)}$.
\mn
%%% Modification
% additional paragraph
More generally, we wish to argue later on that the complete spectrum can be
explained in terms of quasiparticles. At $\la = 0$, a single-particle
excitation corresponds to just one flipped spin (\secD.2). Due to the absence
of interactions $k$-particle states have $k$ flipped spins at $\la = 0$. For
$\la > 0$ one would have to take the interactions into account using
perturbation theory. Although we are in general not able to perform such a
computation directly, it may still be suggestive to think in terms of such
states. In fact, such a picture is quite traditional (see e.g.\ $\q{\camp}$).
\medskip
In the following we will use the abbreviations:
$$\cab := \bcos{{\vphi \over 3}} \ , \qquad
\chab := \bcos{{\pi - \vphi \over 3}} \ , \qquad
\Rab := 1 - 4 \cab^2 \ , \qquad
\cnab{r} := \bcos{{r \phi \over 3}}.
     \eqno{(\rm \secD.3)}$$
For $n=3$ we can calculate the groundstate energy per site $e_0$
which is defined by $E_N^0 = N e_0$ perturbatively:
$$\eqalign{
e_0 = &-{4 \over \sqrt{3}} \cab
      -{2 \la^2 \over 3 \sqrt{3} \cab} -{\ccab \la^3 \over 9 \sqrt{3} \cab^2}
       + {\sqrt{3} \over 81 \cab} \left \{ {1 \over 2 \cab^2 }
             +{4 \over \Rab } \right \} \la^4
       + {\sqrt{3} \ccab \over 81 \cab^2} \left \{ {3 \over 4 \cab^2 }
             +{4 \over \Rab } \right \} \la^5
         + \O(\la^6).  \cr
}  \eqno{(\rm \secD.4)}$$
Eq.\ (\secD.4) is independent of the chain length $N$ if $N>5$.
In order to convey some idea of the quality of such an expansion
we mention that for $\phi = \vphi = {\pi \over 2}$ and
$\la = {1 \over 2}$ the difference between (\secD.4) and the
result of a numerical diagonalization of the Hamiltonian
(\secB.2) performed with 12 sites is of magnitude $10^{-4}$.
Further comments on the accuracy of (\secD.4), in particular
at the boundary of the phase, can be found in $\q{\han}$
where the same expansion has been calculated for the
massive low-temperature phase.
\mn
Furthermore, for $n=3$ we obtain for the lowest $Q=1$ gap and $P=0$
using the states (\secD.2):
$$\eqalign{
\Delta E_{1,0}&(\phi, \vphi) = 4 \bsin{{\pi - \vphi \over 3}}
                - \la {4 \over \sqrt{3}} \cnab1
   -\la^2 { 2 \over 3 \sqrt{3} } {\Biggl \{ }
            {\cnab2 - 2 \over \cab }
          + {\cnab2 + 1 \over \chab }
            {\Biggr  \} } \cr
  &+ \la^3 {1 \over 9 \sqrt{3}}  {\Biggl \{}
    -{4 \cnab1 \over \cab \chab }
    +{\cnab1 \over \cab^2 }
    +{3 \cnab1 + \ccab \over \chab^2 }
    {\Biggr \} } \cr
  &-\la^4 {1 \over 27} {\Biggl \{ }
         {\cnab4 +
            1 \over \sqrt{3} \chab \cab^2 }
       + {\cnab4 -
            1 \over 2 \sqrt{3} \chab^2 \cab }
%%% Modification
% misprint corrected (a 4 deleted)
           + {\cnab4 + 4 \cnab2
             + 3 \over 2 \sqrt{3} \chab^3 }
           + {4 \cnab4 -12 \cnab2
             + 9 \over 2 \sqrt{3} \cab^3 } \cr
  &\qquad\qquad
   - {1 \over 3 \bsin{{\pi - \vphi \over 3}} \cab^2 }
   - {\cnab2 + 1 \over 3 \bsin{{\pi + \vphi \over 3}}
                                \chab^2 }
   + {\cnab2 - 2 \over 3 \bsin{{\pi + \vphi \over 3}} \cab^2 }
             {\Biggr \} } \cr
  &+\O(\la^5). \cr
}     \eqno({\rm \secD.5})$$
$\Delta E_{2,0}(\phi, \vphi)$ is given by
$\Delta E_{2,0}(\phi, \vphi) = \Delta E_{1,0}(-\phi, -\vphi)$.
\sn
For $n=3$ and general $P$ we obtain from the states (\secD.2) the following
perturbation expansion for the dispersion relation of the lowest $Q=1$
excitation:
$$\eqalign{
\EE_{1}&(P) :=
\Delta E_{1,0}(P, \phi, \vphi) = 4 \bsin{{\pi - \vphi \over 3}}
                - \la {4 \over \sqrt{3}} \bcos{P-{\phi \over 3}} \cr
  &- \la^2 {2 \over 3 \sqrt{3}} {\Biggl \{}
               { \bcos{{P+{2 \phi \over 3} }} + 1\over \chab }
             + { \bcos{{2 P-{2 \phi \over 3} }} - 2\over \cab }
                 {\Biggr \}} \cr
  &+ \la^3 {1 \over 9 \sqrt{3}} {\Biggl \{}
             - {2 \bcos{2 P + {\phi \over 3}} - 3 \bcos{P - {\phi \over 3}}
                  + 2 \bcos{3 P - \phi} - 2 \ccab
                        \over \cab^2 } \cr
  & \quad    -{2 \bcos{2 P + {\phi \over 3}} + 2 \bcos{P - {\phi \over 3}}
                   \over \cab \chab }
         +{\bcos{2 P + {\phi \over 3}} + 2 \bcos{P - {\phi \over 3}}+ \ccab
                   \over \chab^2 }
             {\Biggr \} } \cr
              &  +\O(\la^4) \cr
}     \eqno({\rm \secD.6})$$
and the lowest $Q=2$ excitation is given by:
$$\EE_{2}(P) :=
\Delta E_{2,0}(P, \phi, \vphi) = \Delta E_{1,0}(P, -\phi, -\vphi).
     \eqno({\rm \secD.7})$$
Eqs.\ (\secD.6) and (\secD.7) have already been presented in $\q{\lett}$
in a different form. Note that the agreement of (\secD.6) and (\secD.7)
with the results of a numerical diagonalization is usually
good as was discussed in detail in $\q{\lett}$.
\sn
In the previous section we mentioned that the $k$th orders of (\secD.5) --
(\secD.7) are independent of $N$ if $N \ge k+2$. In particular, this
implies the existence of the limits $N \to \infty$ of (\secD.5) and
(\secD.6) if the perturbative series converge at all.
\mn
In the derivation of (\secD.6) we have not assumed that
the Hamiltonian (\secB.2) is hermitean. Thus, we may admit
$\phi \in \Complex$. We have checked in a few cases that results
of a numerical diagonalization at $N=12$ sites are still in good
agreement with (\secD.6) also for complex $\phi$.
\mn
%%% Modification
% paragraph added in order to explain "short" length of series
We would like to mention that
it is no problem to compute further orders of the series (\secD.4), (\secD.5)
and (\secD.6). In fact, we have indeed done so (see $\q{\thesis}$) but
refrain from presenting the results because the next orders are very
complicated and not relevant for our purposes.
\medskip
Obviously, for $\phi=\vphi = {\pi \over 2}$ we have to perform degenerate
perturbation theory. The correct perturbative excited state for $Q=2$
and $P=0$ is for odd $N$:
$$\sqrt{{2 \over N+1}} {\bigg (}
\pstate{2 0 \ldots 0}_0 +
\pstate{1 1 0 \ldots 0}_0 + \ldots
\pstate{1 \underbrace{0 \ldots  0}_{{N-3 \over 2}} 1 0 \ldots 0}_0
{\bigg )} \eqno({\rm \secD.8a})$$
and for even $N$:
$$\sqrt{{2 \over N+1}} {\bigg (}
\pstate{2 0 \ldots 0}_0 +
\pstate{1 1 0 \ldots 0}_0 + \ldots
\pstate{1 \underbrace{0 \ldots  0}_{{N \over 2} - 2} 1 0 \ldots 0}_0 +
{1 \over \sqrt{2}}
\pstate{1 \underbrace{0 \ldots  0}_{{N \over 2} - 1} 1 0 \ldots 0}_0
{\bigg )}. \eqno({\rm \secD.8b})$$
With this state we obtain for $N > 9$:
$$\Delta E_{2,0}\left({\textstyle {\pi \over 2}, {\pi \over 2}}\right)
 = 4 (1 - \la) + \O(\la^9)   \eqno({\rm \secD.9})$$
as expected. In fact (\secD.9) has been proven exactly $\q{\baxterB}$
using different methods but
previous perturbative calculations were restricted to the non-degenerate
case $\Delta E_{1,0}$ at $\phi=\vphi = {\pi \over 2}$. This demonstrates
the universality of the approach to perturbation expansions outlined
in section \secC.
\medskip
Also for the higher excitations we must apply degenerate
perturbation theory. The next simplest case are the states where
two spins are different from zero. For general $P$,
$-{\pi \over 2} < \vphi < {\pi \over 2}$ the space of the excitation
with one spin flipped into charge sector $Q_1$ and another one flipped
into charge sector $Q_2$ is spanned by the states
$$\pstate{t^{Q_1,Q_2}_j}_P :=
\pstate{Q_1 \underbrace{0 \ldots 0}_{j-1 \ {\rm times}} Q_2 0 \ldots 0}_P \ ,
\qquad 1 \le j \le
\cases{N-1, & if $Q_1 \ne Q_2$;\cr
   \left\lb{N \over 2}\right\rb, & if $Q_1 = Q_2$.\cr}
 \eqno({\rm \secD.10})$$
Obviously, we will have to consider two cases: $Q_1 \ne Q_2$ and
$Q_1 = Q_2$.
\sn
Let us first consider $Q_1 \ne Q_2$. For $n=3$ we can choose
$Q_1 = 1$, $Q_2 = 2$. Then we may omit the upper indices of $t$
because they are uniquely fixed: $\pstate{t_j}_P :=
\pstate{t^{1, 2}_j}_P$.
Now, the potential $V$ acts in the space (\secD.10) as:
$$\eqalign{
q r(V) \pstate{t_1}_P =&
-{2 \over \sqrt{3}} \left(
e^{-i(\phit - P)} \pstate{t_{N-1}}_P
+ 2 \cPh e^{-i(\phit + \Ph)} \pstate{t_2}_P \right)
\cr
q r(V) \pstate{t_j}_P =&
-{2 \over \sqrt{3}} \left(
2 \cPh e^{i(\phit + \Ph)} \pstate{t_{j-1}}_P
+ 2 \cPh e^{-i(\phit + \Ph)} \pstate{t_{j+1}}_P \right) \cr
& \hskip 5cm
\qquad 1 < j < N-1 \cr
q r(V) \pstate{t_{N-1}}_P =&
-{2 \over \sqrt{3}} \left(
2 \cPh e^{i(\phit + \Ph)} \pstate{t_{N-2}}_P
+ e^{i(\phit - P)} \pstate{t_{1}}_P \right)
\cr
} \eqno({\rm \secD.11})$$
where $q$ is the projector onto the space (\secD.10).
Although it is not difficult to diagonalize (\secD.11) numerically
for comparably long chains (e.g.\ $N=100$), we did not succeed in
obtaining a closed expression for the eigenvalues or eigenvectors.
\mn
In the second case, i.e.\ $n=3$ and $Q_1 = Q_2$ let us again
simplify notation by setting $\pstate{t^{+}_j}_P :=
\pstate{t^{1, 1}_j}_P$ and $\pstate{t^{-}_j}_P :=
\pstate{t^{2, 2}_j}_P$. Furthermore, introduce the abbreviation
$W$ by:
$$-{4 \over \sqrt{3}}  \bcos{\textstyle \Ph \mp \phit} W
\pstate{t^{\pm}_j}_P :=
q r(V) \pstate{t^{\pm}_j}_P.
 \eqno({\rm \secD.12})$$
In the case of two identical excitations we will also have to
distinguish between even and odd momenta in terms of lattice
sites. It is therefore convenient to introduce a further
abbreviation $\delta_P^N$ encoding this distinction:
$$\delta_P^N := 0\ , \quad {\rm if} \ {P N \over 2 \pi} \ {\rm odd};
\qquad \qquad
\delta_P^N := 1\ , \quad {\rm if} \ {P N \over 2 \pi} \ {\rm even}.
  \eqno({\rm \secD.13})$$
The action of the potential $V$ now is
$$\eqalign{
W \pstate{t^{\pm}_1}_P =& \left(
e^{-i \Ph} \pstate{t^{\pm}_2}_P \right)
\cr
W \pstate{t^{\pm}_j}_P =& \left(
e^{i \Ph} \pstate{t^{\pm}_{j-1}}_P
+ e^{-i \Ph} \pstate{t^{\pm}_{j+1}}_P \right)
\qquad 1 < j < \left\lb{\textstyle {N \over 2}}\right\rb-1 \cr
W \pstate{t^{\pm}_{\lb{N \over 2}\rb-1}}_P =&
\cases{
\left(e^{i \Ph} \pstate{t^{\pm}_{{N-5 \over 2}}}_P
+ e^{-i \Ph} \pstate{t^{\pm}_{{N-1 \over 2}}}_P \right),
& if $N$ odd; \cr
\left(e^{i \Ph} \pstate{t^{\pm}_{{N \over 2}-2}}_P
+ \delta_P^N \sqrt{2} e^{-i \Ph} \pstate{t^{\pm}_{{N \over 2}}}_P \right),
& if $N$ even \cr
} \cr
W \pstate{t^{\pm}_{\lb{N \over 2}\rb}}_P =&
\cases{
\left(e^{i \Ph} \pstate{t^{\pm}_{{N-3 \over 2}}}_P
- (-1)^{\delta_P^N} \pstate{t^{\pm}_{{N-1 \over 2}}}_P \right),
& if $N$ odd; \cr
\delta_P^N \sqrt{2} e^{i \Ph} \pstate{t^{\pm}_{{N \over 2}-1}}_P,
& if $N$ even. \cr
} \cr
} \eqno({\rm \secD.14})$$
At first sight (\secD.14) looks much more complicated than (\secD.11).
This is however misleading and the matrix $W$ can be diagonalized explicitly.
In order to do so, we exploit a connection to graph theory
(see e.g.\ $\q{\jones}$). In this section, we restrict
to a graphical representation of (\secD.11) and (\secD.14) --
the calculation is spelled out in detail in appendix {\appA}.
\mn
Each vector $\pstate{t_j}_P$
will be symbolized as a `$\bullet$' with the index written above.
The action of the potential $V$ is symbolized by lines, with the
square of the matrix elements (up to an isomorphism to be presented
in appendix \appA) attached to them. Assume first that
we could distinguish the two flips we make. Then the graphical
representation for the action of the potential $V$ (or $W$)
would be
$$\bu{1}\mskp\Hr{1}\mskp\bu{2}\cdots\cdots\mkern-12mu
\bu{N - 2}\mkern-12mu
\mskp\Hr{1}\mskp\mkern-12mu
\bu{N - 1} = \left({\cal A}_{N-1}\right).
  \eqno({\rm \secD.15})$$
Here `$\left({\cal L}_k\right)$' denotes the incidence matrix
derived from the Cartan matrix of a Lie algebra ${\cal L}_k$.
However, the states $\pstate{t^{\pm}_j}_P$ and $\pstate{t^{\pm}_{N-j}}_P$
are proportional to each other and must therefore be identified.
Furthermore,
it turns out that for $N$ even and ${N P \over 2 \pi}$ odd
$\pstate{t^{\pm}_{N \over 2}}_P  = 0$ vanishes identically. This
already splits the graph (\secD.15) into two disjoint parts.
Therefore, a graphical representation of (\secD.14) is given by:
$$\eqalign{
W \cong \bu{1}\mskp\Hr{1}\mskp\bu{2}\cdots\cdots\mkern-12mu
\bu{{N\over 2} - 2}\mkern-12mu
\mskp\Hr{1}\mskp\mkern-12mu
\bu{{N\over 2} - 1} & = \left({\cal A}_{{N\over 2} -1}\right)
\qquad {\rm for} \ N \ {\rm even} , \ {N P \over 2 \pi} \ {\rm odd} \cr
W \cong \bu{1}\mskp\Hr{1}\mskp\bu{2}\cdots\cdots\mkern-12mu
\bu{{N-3\over 2}}\mkern-12mu
\mskp\Hr{1}\mskp\mkern-12mu
\bu{{N-1\over 2}}\mkern-21mu\bigcirc \ {\scriptstyle 1}&
 = \left({\cal T}_{{N-1\over 2}}\right)
\qquad \ {\rm for} \ N \ {\rm odd} \cr
W \cong \bu{1}\mskp\Hr{1}\mskp\bu{2}\cdots\cdots\mkern-12mu
\bu{{N\over 2} - 2}\mkern-12mu\mskp\Hr{1}\mkern-12mu
\mskp\bu{{N\over 2} - 1}\mkern-14mu\mskp\Hrr{2}
\mkern-4mu\mskp\bu{{N\over 2}} &
 = \left({\cal B}_{{N\over 2}}\right)
\qquad \ \ \ {\rm for} \ N \ {\rm even} ,
 \ {N P \over 2 \pi} \ {\rm even}. \cr
}  \eqno({\rm \secD.16})$$
Fortunately, all the graphs (\secD.16) have norm less or equal to $2$
\footnote{${}^{4})$}{$\left({\cal T}_k\right)$ is the Tadpole graph.}.
Because the eigenvalues of such graphs are classified
$\q{\jones}$ we can derive the first order explicitly.
\medskip
In the case of (\secD.11) the situation is different. In order to
simplify the discussion consider the case $P = \phi = 0$. Then one
can represent (\secD.11) as
$$V \approx
\cdots\cdots\mkern-12mu\bu{N-1}\mkern-12mu\mskp\Hr{1}
\mskp\bu{1}\mskp\Hrf{4}\mskp\bu{2}\cdots\cdots\mkern-12mu
\bu{N - 2}\mkern-12mu
\mskp\Hrf{4}\mskp\mkern-12mu
\bu{N - 1}\mkern-12mu
\mskp\Hr{1}\cdots\cdots
\ .  \eqno({\rm \secD.17})$$
Note that instead of drawing a closed diagram we have represented
part of it twice. It is easy to see that the norm of (\secD.17) is
larger than 3 (it tends to 4 for $N \to \infty$). The absence
of explicit expressions for the eigenvalues of such graphs prevented
us from deriving an explicit expression for the first order
of two-particle states in the $Q=0$ sector.
\medskip
The result of the calculation in appendix {\appA}
for the eigenvectors of the matrix $W$ as given by (\secD.14) is:
$$\eqalign{
\pstate{\tau^{\pm}_k}_P := {2 \over \sqrt{N}}
\biggl\{& \sum_{j=1}^{\lb{N \over 2}\rb - 1}
      \bsin{{(2 k - \delta_P^N) \, j \, \pi \over N}}
      e^{-i \Ph (j-1)} \pstate{t^{\pm}_j}_P \cr
&+    {\sqrt{2} \over \sqrt{3 + (-1)^N }}
      \bsin{{(2 k - \delta_P^N) \, \lb{N \over 2}\rb \, \pi \over N}}
      e^{-i \Ph (\lb{N \over 2}\rb-1)} \pstate{t^{\pm}_{\lb{N \over 2}\rb}}_P
\biggr\}. \cr
}  \eqno({\rm \secD.18})$$
\mn
The final result for the first order expansion of the energy for these
excitations is for $N \ge 3$:
$$\eqalign{
\Delta E_{2,k}(P,\phi,\vphi) &= 8 \bsin{{\pi - \vphi \over 3}}
        - \la {8 \over \sqrt{3}} \bcos{\Ph - \phit}
                                 \bcos{{(2 k-\delta_P^N) \pi \over N}}
        + \O(\la^2) \ , \cr
\Delta E_{1,k}(P,\phi,\vphi) &= \Delta E_{2,k}(P,-\phi,-\vphi) \ ,
\qquad  1 \le k \le \left\lb{N+\delta_P^N-1 \over 2}\right\rb. \cr
}  \eqno({\rm \secD.19})$$
For remarks on the second order see appendix {\appA}.
\bn
\chapsubtitle{\secE.\ Evidence for quasiparticle spectrum}
\mn
In this section we present an argument using perturbation theory
that the spectrum of the $\Zed_n$-Hamiltonian (\secB.2) can be interpreted
in terms of quasiparticles for a wide range of parameters.
In the case of $\Zed_3$, the dispersion
relations of the two fundamental particles with $Q=1$ and $Q=2$
are given by (\secD.6) and (\secD.7).
\medskip
The results in $\q{\weA-\kedem}$ suggest that we may expect a
quasiparticle spectrum. More precisely, all excitation energies
$\Delta E_{Q,r}(P, \phi, \vphi)$ should satisfy
$$\Delta E_{Q,r}(P, \phi, \vphi) = \sum_{k=1}^{m_r} \EE_{Q^{(k)}}(P^{(k)})
\, , \quad
P = \sum_{k=1}^{m_r} P^{(k)} \hbox{ mod } 2 \pi \, , \quad
Q = \sum_{k=1}^{m_r} Q^{(k)} \hbox{ mod } n
     \eqno({\rm \secE.1})$$
where $\EE_1(P), \ldots, \EE_{n-1}(P)$ are the energies of the
$n-1$ fundamental quasiparticles.
Additionally, the fundamental quasiparticles seem to satisfy a Pauli
principle, i.e.\ $Q^{(i)} = Q^{(j)}$ implies $P^{(i)} \ne P^{(j)}$.
In particular, for $n=3$, (\secD.6) and
(\secD.7) are the dispersion relations of the fundamental $Q=1$
and $Q=2$ quasiparticles and all other states can be obtained
by composition under the assumption that energy, momentum and charge
are additive.
\mn
Although the result (\secE.1) may not be very surprising
it should be clear to the reader that a particle interpretation is not
directly incorporated into the Hamiltonian (\secB.2).
\medskip
Before presenting a proof of (\secE.1) we would like to add a remark
on Figs.\ 2--4 of $\q{\lett}$:
In the limit $N \to \infty$ the eigenvalues seem to become dense such that
we may expect to interpret the energy bands as continuous spectrum
in the weak limit of the Hamiltonian. Note that according to our
definition, the single particle excitations (\secD.2) lead to point
spectrum. One also observes that the energy bands are filled from the
interior such that their boundaries do not belong to the spectrum
for any finite $N$. However, we have pointed out in section {\secB}
that the spectrum is closed in the infinite chain limit. Thus, the
boundaries of the energy bands will belong to the spectrum in this
infinite $N$ limit. It is worthwhile noting that the normalization
factors ${2 \over \sqrt{N}}$ for the two-particle states
in (\secD.18) demonstrate that these
states tend to zero for $N \to \infty$ and will therefore
not give rise to proper eigenvectors. This confirms that with
our definition of the limit composite particle states belong to
the continuous spectrum.
\medskip
Before proceeding with the general discussion let us first look a
little closer at the two-particle states. First note that the derivation
of the energy of two identical particles to first order in $\la$ does
essentially not depend on the number $n$ of states involved. It is therefore
straightforward to generalize (\secD.19) to arbitrary $n$:
$$\eqalign{
\Delta E_{2 Q,k}(P,\phi,\vphi) = &
       2 \left(\sum_{k=1}^{n-1} \ab_k (1 - \om^{Q k}) \right)
        - 4 \la {\bcos{\Ph - \left(1-{2 Q \over n}\right) \phi}
                    \bcos{{(2 k-\delta_P^N) \pi \over N}}
                 \over \bsin{\pi Q \over n} } \cr
       &+ \O(\la^2) \ ,
\qquad  1 \le k \le \left\lb{N+\delta_P^N-1 \over 2}\right\rb \cr
}  \eqno({\rm \secE.2})$$
where $\delta_P^N$ was defined in (\secD.13): $\delta_P^N = 0$ if
${P N \over 2 \pi}$ odd and $\delta_P^N := 1$ for ${P N \over 2 \pi}$ even.
Comparing (\secE.2) with the first order expansion for the single-particle
states eq.\ (39) of $\q{\weA}$ one observes that this first order expansion
of the two-particle excitations is in agreement with the quasiparticle rule
(\secE.1). Up to first order the composite particle states satisfy
either $2 \Delta E_{Q,0}(P, \phi, \vphi) <
\Delta E_{2Q,k} (2 P, \phi, \vphi) <
2 \Delta E_{Q,0}(P+2 \pi, \phi, \vphi)$ or
$2 \Delta E_{Q,0}(P, \phi, \vphi) >
\Delta E_{2Q,k} (2 P, \phi, \vphi) >
2 \Delta E_{Q,0}(P+2 \pi, \phi, \vphi)$ depending on which one of the
single particle energies is larger. Thus, the two-particle states
do indeed lie inside the energy band of two single-particle states
and the boundaries are not included. Even more, we can see from
(\secE.2) that the two-particle states become dense
in this energy band for $N \to \infty$.
\medskip
Let us now present a more abstract argument which ensures the validity of
(\secE.1). The interaction in the Hamiltonian (\secB.1) is very short
ranged -- in fact, only among nearest neighbours. In the massive
high-temperature phase there is no spontaneous order and the correlation
length is finite. Thus, if one puts two excitations of `short' chains with
a sufficient separation on a longer chain, the interaction will be
negligible. For example, putting one single-particle
excitation one the left half of the chain and another on the right
half will approximate a two-particle excitation.
\mn
We make this derivation of the quasiparticle interpretation of the spectrum
more precise using perturbative arguments.
%%% Modification
% this argument has partially shifted to section 4
According to the remarks at the beginning of section {\secD}
the quasiparticle spectrum with flat dispersion curves is
easily verified for $\la=0$.
%It is easy to see that for $\la = 0$ we do have a quasiparticle
%spectrum with flat dispersion curves. At $\la = 0$, a single-particle
%excitation corresponds to just one flipped spin. Due to the absence
%of interactions $k$-particle states have $k$ flipped spins at $\la = 0$.
%Therefore, it is very useful to think in terms of states (see e.g.\
%(\secD.2), (\secD.10)) keeping in mind that for $\la > 0$ one has to
%take the interactions into account using perturbation theory.
%\mn
In this section we sketch a proof that the quasiparticle
picture remains valid for $\la > 0$ -- we just present
the main ideas. A modified rigorous version of the proof is spelled
out in appendix {\appB}.
\sn
First, we notice that
$$\eqalign{
\Delta H_{N+M}^{(n)} &= \Delta H_N^{(n)} \otimes \id
                   + \id \otimes \Delta H_M^{(n)}
                   + \O(\Delta H_{N,M}) \ , \cr
T_{N+M}        &= \left\{ \id + \O(T_{N,M}) \right\} T_N \otimes T_M \cr
}    \eqno({\rm \secE.3})$$
where `$\O(\Xi_{N,M})$' denotes an operator acting only at sites
$0$, $N-1$, $N$ and $N+M-1$. One of the main steps of the proof
is to show that these boundary operators vanish in the limit
$N, M \to \infty$. It should be clear to the reader that the
coproduct rule (\secE.3) is going to be crucial for the derivation
of the quasiparticle picture (\secE.1). In particular, our
proof cannot be easily modified to accommodate more complicated
selection rules and will therefore be specific for $\Zed_n$-spin
quantum chains
\sn
If we can build a composite state of any two states
we have to show that energy, charge and momentum behave additive
under this composition and that we can construct all states.
Then, the quasiparticle structure follows by induction.
\sn
Composite particle states are expected to give rise to continuous
spectrum. This is a technical complication in the argument we are
going to give because it is not possible to use eigenstates but
we must show that the resolvent is unbounded.
However, for each finite $N$ the Hamiltonian has a complete set
of eigenstates. We have already argued in section {\secB} that
the resolvent becomes unbounded for any energy if it can be
approximated by eigenvalues of $\Delta H_N^{(n)}$. This in turn
can be ensured by providing a sequence of vectors
$\pstate{k; E}_{P}^{}$ that approximate eigenvectors of
$\Delta H_N^{(n)}$ to eigenvalue $E$ for $N$ large. Thus, we
would have to take two limits simultaneously. However, a standard
argument shows that it is no loss of generality to restrict to
the diagonal sequence $k=N$.
\sn
The assumption in the induction is that we can choose two sequences of states
$\pstate{N; E_1}_{P_1}^{} \in \H_N$ and
$\pstate{M; E_2}_{P_2}^{} \in \H_M$ such that in the weak limits
of $\Delta H^{(n)}$ and $T$ they give rise to unbounded resolvents
at $E_k$, $e^{i P_k}$:
$$\eqalign{
\lim_{N \to \infty} (\Delta H_N^{(n)} - E_1) \ \pstate{N; E_1}_{P_1}^{}
     &= 0 \, , \qquad
\lim_{N \to \infty} (T_N - e^{i P_1}) \ \pstate{N; E_1}_{P_1}^{} = 0 \, , \cr
\lim_{M \to \infty} (\Delta H_M^{(n)} - E_2) \ \pstate{M; E_2}_{P_2}^{}
     &= 0 \, , \qquad
\lim_{M \to \infty} (T_M - e^{i P_2}) \ \pstate{M; E_2}_{P_2}^{} = 0. \cr
}    \eqno({\rm \secE.4})$$
We know that such sequences of states exist at least for the
single-particle states -- the perturbative series for $\pstate{s^Q}_P$.
\sn
The second major step in the proof is to consider now the state
$\pstate{N; E_1}_{P_1}^{} \otimes \pstate{M; E_2}_{P_2}^{} \in \H_{N+M}$.
{}From (\secE.3) one has
$$\eqalign{
\Delta H_{N+M}^{(n)} ( \pstate{N; E_1}_{P_1}^{}
    \otimes \pstate{M; E_2}_{P_2}^{} )
 =& (\Delta H_N^{(n)} \pstate{N; E_1}_{P_1}^{})
    \otimes \pstate{M; E_2}_{P_2}^{} \cr
  &+ \pstate{N; E_1}_{P_1}^{} \otimes (\Delta H_M^{(n)}
    \pstate{M; E_2}_{P_2}^{}) \cr
  &+ \O(\Delta H_{N,M}) (\pstate{N; E_1}_{P_1}^{}
     \otimes \pstate{M; E_2}_{P_2}^{})     \ , \cr
T_{N+M} ( \pstate{N; E_1}_{P_1}^{} \otimes \pstate{M; E_2}_{P_2}^{} )
 =& (T_N \pstate{N; E_1}_{P_1}^{}) \otimes (T_M \pstate{M; E_2}_{P_2}^{}) \cr
  &+ \O(T_{N,M}) (T_N  \pstate{N; E_1}_{P_1}^{})
             \otimes (T_M \pstate{M; E_2}_{P_2}^{}).   \cr
}    \eqno({\rm \secE.5})$$
The vanishing of the boundary terms in (\secE.5) can be shown using e.g.\
perturbative arguments. The crucial point in the argumentation is that
the momentum eigenstates have normalization factors $N^{-{1 \over 2}}$,
$M^{-{1 \over 2}}$. Any operator acting only at boundaries yields only
a finite part of these states in contrast to the operators $T_N$ and
$\Delta H_{N}^{(n)}$ which act
on the complete chain and yield complete momentum eigenstates.
The finite pieces of momentum eigenstates are suppressed by the
normalization factors $N^{-{1 \over 2}}$ in the infinite chain length
limit.  For example, for the translation operator $T_N$
it is easy to verify explicitly that the boundary terms tend to zero
at $\la=0$ using precisely this argument. The argumentation for
the Hamiltonian is analogous but slightly more complicated.
A similar perturbative argument has already been presented in $\q{\han}$
in order to show the vanishing of the $Q$-dependence in the low-temperature
regime.
\sn
These rather technical details are spelled out in appendix {\appB}.
\sn
We have shown that the boundary operators $\O(\Delta H_{N,M})$ and
$\O(T_{N,M})$ vanish as $N$, $M$ go to infinity. Thus, in this limit
$$\eqalign{
\lim_{N, M \to \infty} (\Delta H_{N+M}^{(n)} - (E_1 + E_2))
    ( \pstate{N; E_1}_{P_1}^{} \otimes \pstate{M; E_2}_{P_2}^{} ) &= 0
          \ , \cr
\lim_{N, M \to \infty} (T_{N+M} - e^{i (P_1 + P_2)})
    ( \pstate{N; E_1}_{P_1}^{} \otimes \pstate{M; E_2}_{P_2}^{} ) &= 0 \cr
}    \eqno({\rm \secE.6})$$
holds.
This shows that energy $E$ and momentum $P$ are additive -- the additivity
of the charge $Q$ is obvious. One can always build a basis for the
space $\H_{K} = \otimes^K \H_{1}$ by considering tensor products of
basis vectors in $\H_N$ and $\H_M$ with $N+M = K$. This is precisely what
we have done. Thus, the above procedure does indeed yield the complete
spectrum.
\mn
One should be careful about the requirements that enter in our proof
of the quasiparticle picture in order not to mistake it for more general
than it is. Note that the vanishing of boundary terms is a crucial
part of the proof. However, boundary terms are substantial for conformally
invariant systems with long ranged correlations. Also in the low-temperature
phase boundary terms play an important r\^ole because the free part of the
Hamiltonian depends on the difference of neighbouring spins
%%% Modification
% more careful with reference
(see also $\q{\han}$).
Thus, our proof applies neither to critical points where one might have
conformal invariance nor to the low-temperature phase. Furthermore, we
have used the explicit form (\secB.1) of the Hamiltonian (for example
for the selection rules in (\secE.1)).
\sn
It should be clear to the reader that our argument relies on a perturbation
series for the single-particle states and is valid only if this series
is convergent. We will discuss
the radius of convergence for the $\Zed_3$-chain in more detail in section
{\secI}. At this place we would just like to mention that this perturbative
argument cannot be applied to massless incommensurate phases
because the main limitations on the convergence come from level
crossings which are characteristic for massless incommensurate phases.
\sn
Note that we have not assumed the Hamiltonian to be hermitean. In particular,
the quasiparticle picture should also be valid for $\phi \in \Complex$
as long as the single-particle excitations exist and converge.
This is indeed supported by numerical calculations $\q{\yildirim}$.
\bigskip
The argument proving the quasiparticle structure can be refined in order
to give an upper estimate for the rate of convergence in $N$ of the
energy of a $k$-particle state. As an approximation to a $k$-particle state
for $k N$ sites, total energy $E_{\rm tot}$ and total momentum $P$
we may take the $k$-fold tensor product of single-particle states
$$\pstate{k N; E_{\rm tot}}_{P}
  := \pstate{N; E_1}_{P_1} \otimes \ldots \otimes \pstate{N; E_k}_{P_k}
    \eqno({\rm \secE.7})$$
with $E_{\rm tot} = \sum_{l=1}^k E_l$, $P = \sum_{l=1}^k P_l$.
Now, the deviation from the limit $N \to \infty$ is given by:
$$\eqalign{
{}_{P}\apstate{k N; E_{\rm tot}} \Delta H_{k N}^{(n)}
  \pstate{k N; E_{\rm tot}}_{P} - E_{\rm tot}
&= \prod_{l=1}^k {}_{P_l} \apstate{N; E_l} \O(\Delta H_N)
   \pstate{N; E_l}_{P_l} \cr
&= \O(N^{-k}).
}    \eqno({\rm \secE.8})$$
$\O(\Delta H_N)$ is some operator that acts only at sites $1$ and $N$.
The first equality simply uses the definition of the scalar product in
tensor products. The last equality is more profound and
due to the fact that operators acting only at boundaries
of the chain are suppressed by $N^{-1}$ due to the normalization factor in
the finite fourier transformation for momentum eigenstates.
This shows that the deviation of the energy of a $k$-particle state
($k > 1$) from the limit is
at most of order $N^{-k}$ for $N \to \infty$. Of course, one might find
better approximations for the eigenstates and the convergence
could be faster. Thus, the $N$-dependence of some energy eigenvalue
gives only a lower bound on the number $k$ of particles involved.
\smallskip
This general argument is confirmed by our results for the two-particle
states. Expanding $\cos(x) = 1 - {1 \over 2} x^2 + \O(x^4)$ we can read
off from (\secE.2) that the first order correction of the
$k$th two-particle state with respect to the boundary of the energy
band behaves as $N^{-2}$. This is precisely what we expect from the
general considerations.
\smallskip
This argument shows in particular that in a finite-size system the
{\it energy} of any state remains unchanged to order ${1 \over N}$.
According to the argument presented at the end of section {\secC}
the energies of the fundamental particle states have to converge
exponentially in $N$ and the energies of composite particle states
have corrections at most of order ${1 \over N^2}$. Thus, the only
modification in (\secE.1) at order ${1 \over N}$ in the massive high
temperature phase is a discretization of the momentum (and possible
minor modifications of the Brillouin zones and selection rules
$\q{\kedem}$).
\bigskip
Note that the proof of the vanishing of boundary terms as sketched above
and presented in detail in appendix {\appB} also directly applies to the
Hamiltonian (\secB.2) itself. So far, we have restricted ourselves to
periodic boundary conditions $\Ga_{N+1} = \Ga_1$. However, one could also
impose toroidal boundary conditions: `Cyclic' boundary conditions
$\Ga_{N+1} = \om^{-R} \Ga_1$ or `twisted' boundary conditions
$\Ga_{N+1} = \om^{-R} \Ga_1^{+}$. Even `free' boundary conditions
$\Ga_{N+1} = 0$ are well-known in the literature. Our argument shows
that all these different choices lead to the same spectrum in the
limit $N \to \infty$. In particular, our results are valid for all
choices of boundary conditions and one is free to choose those which
seem most appropriate, e.g.\ one can leave the ends of the chain open
instead of the unnatural end-identification for a realistic physical
system.
\sn
Again, this observation for the massive high-temperature phase is to
be contrasted with other situations. In particular, at the second order
phase transition $\phi = \vphi = 0$, $\la=1$ the correlation length
becomes infinite and the boundary terms are very important
$\q{\cardyB-\automos}$. Even in the massive low-temperature phase one
observes long range order and boundary terms cannot be neglected $\q{\han}$.
\bigskip
So far, we have not addressed the question of whether the fundamental
particles satisfy a Pauli principle or not -- note that the above discussion
is intrinsically insensitive to a Pauli principle because the limit
was defined such that the spectrum forms a closed set. Nevertheless, for
the special case $n=3$ and $\phi = \vphi = {\pi \over 2}$, eq.\ (\secE.1)
was obtained in $\q{\dkcoy}$ supplemented with the Pauli principle mentioned
below (\secE.1). Fortunately, due to (\secE.2), we have some control over the
finite-size dependence of the scattering states of two identical particles in
the general case. Up to first order in $\la$ these finite-size effects
do essentially neither depend on the charge $Q$ nor on the number of states
$n$. Therefore, the nature of the fundamental excitations can be determined
by looking at one particular choice of $Q$ and $n$. However, for $n=2$ one
obtains the Ising model where it is well-known that the excitation spectrum
can be explained in terms of one fundamental {\it fermion} (see e.g.\
$\q{\kogut}$). This indicates that the fundamental excitations for general
$n$ should be regarded as fermions. In particular, for a scattering
state of two identical excitations $i$ and $j$ the momenta must satisfy
$P_i \ne P_j$. In a scattering state of two {\it different} fundamental
particles these two fundamental particles can easily be distinguished
because they carry different $\Zed_n$-charges. Therefore, two different
particles should not be subject to a Pauli principle (like it is the
case for two different non-interacting fermions).
\bn
\chapsubtitle{\secG.\ Correlation functions}
\mn
In recent papers a systematic investigation of the correlation
functions of the $\Zed_3$-chiral Potts model in the massive phases
has been started. First, a non-vanishing wave vector
has been predicted in $\q{\gehlenph}\q{\krallm}$ for the massive
high-temperature phase and its critical exponent was calculated
from level crossings. Next, perturbative calculations for the
massive {\it low-temperature phase} analogous to those
to be presented below have been reported in $\q{\han}$.
We also studied the correlation function for the operator
$\Ga$ in the massive high-temperature phase numerically
in $\q{\lett}$ and were able to demonstrate an oscillation.
In $\q{\lett}$ some of the results to be presented below we
already cited without derivation.
Note also that for the massless phases around $\la \sim 1$ of the
$\Zed_3$-chain correlation functions have been derived in $\q{\albcoy}$
borrowing results from conformal field theory.
\sn
In this section we study correlation functions for the $\Zed_3$-chiral
Potts model perturbatively.
Before defining correlation functions, we first note that the
two-point functions are translationally invariant because
the groundstate $\vac$ is translationally invariant:
$$\eqalign{
\avac \Ga_{x+r}^{+} \Ga_r \vac &= \avac \Ga_{x+1}^{+} \Ga_1 \vac \ , \cr
\avac \si_{x+r}^{+} \si_r \vac &= \avac \si_{x+1}^{+} \si_1 \vac . \cr
}    \eqno({\rm \secG.1})$$
Thus, it makes sense to define the correlation function for an operator
$\Xi$ by the following expression:
$$C_{\Xi}(x) := {\avac \Xi_{x+1}^{+} \Xi_1 \vac \over \normvac}
              - {\avac \Xi_{x+1}^{+} \vac \avac \Xi_1 \vac \over
                   \normvac^2}
    \qquad 0 \le x < {N \over 2}    \eqno({\rm \secG.2})$$
where $\vac$ is the eigenvector of the Hamiltonian to lowest energy.
Here, we do not assume that $\vac$ is normalized to one and have therefore
included the proper normalization factors in (\secG.2). The correlation
functions of the operators $\Ga_x$ and $\si_x$ have the property
$$\eqalign{
C_{\Ga}(-x) &= C_{\Ga}(x)^{*} \ , \cr
C_{\si}(-x) &= C_{\si}(x)^{*} = C_{\si}(x) \cr
}    \eqno({\rm \secG.3})$$
such that it makes sense to restrict to positive $x$. Note that
(\secG.3) follows by complex conjugation using (\secG.1).
Explicit calculations show the validity of (\secG.1) and (\secG.3) as well.
\medskip
For simplicity we will first neglect the correction term for the uncorrelated
part as well as the normalization in (\secG.2) and consider the following
expression:
$$c_{\Xi}(x) := \avac \Xi_{x+1}^{+} \Xi_1 \vac
    \qquad 0 \le x < {N \over 2}.    \eqno({\rm \secG.4})$$
The operator $\Xi$ for the $\Zed_3$-chiral Potts model can be
either $\Gamma$ or $\sigma$. For $n>3$ also different powers of
these operators may be interesting.
\medskip
One can use the quasiparticle picture which we have already derived
in order to rewrite a correlation function $C_{\Xi}(x)$ as follows:
$$\eqalign{
C_{\Xi}(x)&={\sum_{n=0}^{\infty} \int_0^{2 \pi}
               \left(\prod_{i=1}^n {\rm d}p_i\right) \,
               \avac \Xi_{x+1}^{+} \state{p_1,\ldots,p_n}
               \astate{p_1,\ldots,p_n} \Xi_1 \vac
               \over \normvac}
         - {\abs{\avac \Xi_1 \vac}^2 \over \normvac^2} \cr
    &= \sum_{n=1}^{\infty} \int_0^{2 \pi}
        \left(\prod_{i=1}^n {\rm d}p_i\right) \,
        e^{i x \left(\left(\sum_{j=1}^n p_j \right) - P_{\vac}\right)} \,
                 { \abs{\astate{p_1,\ldots,p_n} \Xi_1 \vac}^2
                 \over \normvac} \cr
}    \eqno({\rm \secG.5})$$
where we have inserted a complete set of normalized $n$-particle states
$\state{p_1,\ldots,p_n}$. Representations similar to (\secG.5) have been
used in quantum field theory for a long time (see e.g.\ $\q{\lehmann}$)
and are well-known to be useful for the evaluation of correlation functions
of statistical models (see e.g.\ $\q{\mussardo}$). According to (\secG.5)
one could compute the correlation function $C_{\Xi}(x)$ by computing its
`form factors' $\astate{p_1,\ldots,p_n} \Xi_1 \vac$, but one can even derive
interesting results without doing so.
Clearly, if the groundstate $\vac$ has non-zero momentum $P_{\vac} \ne 0$ we
expect an oscillatory contribution to the correlation function.
However, one can read off from (\secG.5) that an oscillatory contribution
is also to be expected if $P_{\vac} = 0$ but the model breaks parity which
precisely applies to the massive high-temperature phase of the
chiral Potts model. The correlation functions of massive models
in general have an exponential decay, i.e.\
$C_{\Xi}(x) = e^{-{x \over \xi}} f_{\Xi}(x)$ where $f_{\Xi}(x)$ is
some bounded function. According to (\secG.5) we also expect an
oscillatory contribution of the form $e^{i {2 \pi x \over L}}$.
In summary, we expect correlation functions of the approximate form
$$C_{\Xi}(x) \sim  e^{-{x \over \xi}+i {2 \pi x \over L}}.
    \eqno({\rm \secG.6})$$
$\xi$ is called `correlation length' and $L$ is the `oscillation length'
($L^{-1}$ is the `wave vector').
\mn
More precisely, for the $\Zed_3$-chiral Potts model the operator
$\Ga_1$ creates $Q=1$-single-particle excitations from the groundstate.
The dispersion relations of these particles clearly violate parity.
Therefore we expect that $C_{\Ga}(x)$ is of the form (\secG.6).
The action of the operator $\si_1$ is much less spectacular.
In particular, it leaves the charge sector $Q=0$ invariant and thus
it need not necessarily have an oscillatory contribution. In fact,
from (\secG.3) we see that $C_{\si}(x)$ should be real which in
view of (\secG.6) implies the absence of oscillations.
\mn
Symmetries of the Hamiltonian translate into symmetries of the form
factors. In certain cases these symmetries are already sufficient to
compute the oscillation length $L$. In appendix {\appC} we demonstrate
this in a few cases for the correlation function $C_{\Ga_{}^Q}(x)$ of
%%%% Modification
% Comment on symmetry needed
the $\Zed_n$-chiral Potts model. For ${\rm Re}(\phi)=\pi$ one observes
a shifted parity symmetry $\q{\yildirim}$ that can be derived e.g.\ along
the lines of $\q{\mccoyadv}$. Using this symmetry one can show
(see appendix~{\appC}) that
$$C_{\Ga_{}^Q}^{}(x) = e^{2 \pi i x \over L} f_{Q,r}(x)
    \eqno({\rm \secG.7a})$$
with
$$\eqalign{
f_{Q,r}(x) \in \Real & \qquad \forall x \, , \cr
L = \infty & \qquad \hbox{for } \phi = r \pi, r \in \Zed \ \hbox{ or } \
                                \vphi \in \Real, {\rm Re}(\phi) = 0 \, , \cr
L = {2n \over n-2Q} \, & \qquad \hbox{for }\vphi\in\Real, {\rm Re}(\phi)=\pi
                      \hbox{ and } 0 < Q < n \, . \cr
}    \eqno({\rm \secG.7b})$$
%%% Modification
% considerable changes - the following paragraph becomes unnecessary
%Strictly speaking, the symmetry of the Hamiltonian needed to obtain
%$L = {2n \over n-2Q}$ is shown in appendix {\appC} only for $Q$ invertible
%%%% Modification
%% parenthesis added
%in $\Zed_n$ (those $Q$ whose greatest common divisor with $n$ is 1)
%but there are indications that it is indeed true for all the
%cases claimed in (\secG.7b) $\q{\yildirim}$ \footnote{$^{5})$}{
%In fact, the symmetries of the Hamiltonian needed in order to show (\secG.7)
%were first observed numerically in the non-hermitean integrable case by
%K.\ Yildirim $\q{\yildirim}$ which inspired our proof in appendix \appC.
%}.
\medskip
Let us now turn to the explicit computation of correlation functions
for the $\Zed_3$-chain.
In order to be able to calculate the correlation functions we need
to know the groundstate $\vac$. We will calculate it from
the free ground state $\state{{\rm GS}}$ using the perturbation
expansion (\secC.7).
We should stress again that although we assume the free groundstate
$\state{{\rm GS}}$ to be normalized to
$1$, this is not necessarily true for the complete state $\vac$.
The expansion of the groundstate $\vac$ provides us with an
expansion for the correlation functions in powers of $\la$
$$c_{\Xi}(x) = \sum_{\nu=0}^{\infty} \la^{\nu} c_{\Xi}^{(\nu)}(x)
    \eqno({\rm \secG.8})$$
where we again neglect an irrelevant overall normalization factor which
depends on $\la$. Note that according to (\secC.7) a $k$th order
expansion of the groundstate yields a $k+1$th order expansion
of the groundstate energy as a byproduct.
\medskip
Using the state (\secD.1) one can calculate for the $\Zed_3$-chiral
Potts model in the high-temperature phase the first orders in
$\la$ for $c_{\Ga}(x)$:
$$\eqalign{
c_{\Ga}^{(0)}(x) =& \delta_{x,0} \ , \qquad \qquad \qquad
c_{\Ga}^{(1)}(x) = \delta_{x,1}
                     {e^{i {\phi \over 3}} \over 3 \cab } \ , \cr
c_{\Ga}^{(2)}(x) =& {1 \over 6 \cab^2} {\Biggl \{}
                     \delta_{x,0} {N \over 3}
                   + \delta_{x,1} {e^{- i {2 \phi \over 3}} \over 2}
                   + \delta_{x,2} e^{i {2 \phi \over 3}}
                    {\Biggr \}} \ . \cr
}    \eqno({\rm \secG.9})$$
In order to save place we present higher orders only in the final,
properly normalized form (\secG.15).
\sn
For the first orders of $c_{\si}(x)$ we obtain
$$\eqalign{
c_{\si}^{(0)}(x) &= 1 \ , \qquad \qquad \qquad
c_{\si}^{(1)}(x) = 0 \ , \cr
c_{\si}^{(2)}(x) &= {1 \over 3 \cab^2} \left \{
                     \delta_{x,0}
                   + {\delta_{x,1} \over 4}
                   + {N-6 \over 6}
                    \right \} \ . \cr
}    \eqno({\rm \secG.10})$$
Again, we have postponed presentation of higher orders to the
final, properly normalized result (\secG.14).
\sn
Let us now discuss the correction terms in (\secG.2). The
operator $\Gamma_x$ creates charge such that charge conservation implies
$\avac \Ga_x^{+} \vac = \avac \Ga_x \vac = 0$ for all $x$. Thus
$$C_{\Ga}(x) = {c_{\Ga}(x) \over \normvac}.    \eqno({\rm \secG.11})$$
The corrections for the operator $\si$ are more complicated.
Using the expansion (\secC.7) for the groundstate one obtains
independent of $x$
$$\eqalign{
\avac & \si_x^{+} \vac  = \avac \si_x \vac^{*} \ , \cr
\avac & \si_x \vac  =
  1 + \la^2 {N-3 \over 18 \cab^2}
      + \la^3 {(N-3) \ccab \over 54 \cab^3} \cr
    & + \la^4 {1 \over 81 \cab^2} \left\{
            {9 \ i \ \bsin{{2\vphi \over 3}}
          +  9 - 4 N + (28 N - 90) \cab^2 \over 3 \Rab^2}
          + {(N-3)^2 \over 8 \cab^2}
       \right\} \cr
    & + \O(\la^5)
    \qquad \qquad \forall x. \cr
}    \eqno({\rm \secG.12})$$
In order to be able to evaluate (\secG.2) we have to
divide (\secG.12) by the norm squared of $\vac$ before we subtract it.
We apply
$\left( 1 + \sum_{\nu=1}^{\infty} a_{\nu} \la^{\nu} \right)^{-1}
 = \sum_{\mu = 0}^{\infty} \left( -\sum_{\nu=1}^{\infty} a_{\nu} \la^{\nu}
          \right)^{\mu}$
to the norm of $\vac$
$$\eqalign{
\normvac = & 1 + \la^2 {N \over 18 \cab^2}
    + \la^3 {N \ccab \over 54 \cab^3}
    + \la^4 {N \over 81 \cab^2} \left\{
            {4 (7 \cab^2 - 1) \over 3 \Rab^2 }
          + {N \over 8 \cab^2 }
      \right\}
    + \O(\la^5) \cr
}    \eqno({\rm \secG.13})$$
and obtain a normalized expression for the one-point function
(\secG.12). It is not surprising that up to the order calculated
%%% Modification
% equality true only for absolute values
one has the equality $\abs{{\avac \si_x \vac \over \normvac}}
= \abs{C_{\Ga}^{lt}(1)}$ at the dual point in the low-temperature
phase. In fact, this is to be expected from the
proof of duality presented in the appendix of $\q{\han}$ (eq.\ (A.5) ).
\sn
Inserting (\secG.12) and (\secG.13) into (\secG.10) leads to
$$\eqalign{
C_{\si}^{(0)}(x) =&
C_{\si}^{(1)}(x) = 0 \, , \quad
C_{\si}^{(2)}(x) = {1 \over 3 \cab^2} \left \{
                     \delta_{x,0}
                   + {\delta_{x,1} \over 4}
                    \right \} \, , \quad
C_{\si}^{(3)}(x) = {\ccab \over 9 \cab^3} \left \{
                     \delta_{x,0}
                   + {\delta_{x,1} \over 4}
                    \right \} \, , \cr
C_{\si}^{(4)}(x) =& {1 \over 27 \cab^2} {\Biggl \{}
                   - \delta_{x,0} \left (
                           {2 (1 - 10 \cab^2)
                               \over \Rab^2 }
                         + {3 \over 2 \cab^2}
                                  \right )
                   + \delta_{x,1} \left (
                           {1 + 20 \cab^2
                               \over 3 \Rab^2 }
                         - {1 \over \cab^2}
                                  \right ) \cr
                  & \phantom{{1 \over 27 \cab^2} {\Biggl \{} }
                   + \delta_{x,2} \left (
                           {2( 1 + 2 \cab^2)
                               \over 3 \Rab^2 }
                         + {1 \over 16 \cab^2}
                                  \right )
                    {\Biggr \}}. \cr
}    \eqno({\rm \secG.14})$$
Note that also the $N$-dependence in (\secG.9) is due to the $N$-dependence
(\secG.13) of the norm of $\vac$. If we normalize $\vac$ properly to $1$
we have
$$\eqalign{
C_{\Ga}^{(0)}(x) =& \delta_{x,0} \ , \qquad \qquad
C_{\Ga}^{(1)}(x) =  \delta_{x,1}
                     {e^{i {\phi \over 3}} \over 3 \cab } \ , \cr
C_{\Ga}^{(2)}(x) =& {1 \over 6 \cab^2} {\Biggl \{}
                     \delta_{x,1} {e^{- i {2 \phi \over 3}} \over 2}
                   + \delta_{x,2} e^{i {2 \phi \over 3}}
                    {\Biggr \}} \ , \cr
C_{\Ga}^{(3)}(x) =& {1 \over 54 \cab} {\Biggl \{}
                   - \delta_{x,1} \ e^{i {\phi \over 3}} \left(
                           {1 \over \cab^2}
                         + {8 \over \Rab}
                            \right)
                   + \delta_{x,2} \ e^{-i {\phi \over 3}} \left(
                           {2 \over \cab^2}
                         - {8 \over \Rab}
                            \right)
                   + \delta_{x,3} {5 e^{i \phi} \over \cab^2}
                    {\Biggr \}} \cr
}    \eqno({\rm \secG.15a})$$
which obviously is $N$-independent. Finally, in this case
we obtain for the fourth order
$$\eqalign{
C_{\Ga}^{(4)}&(x) = {1 \over 81 \cab^2} {\Biggl \{}
            - \delta_{x,1} \left( e^{i {4 \phi \over 3}}
             + 4 e^{-i {2 \phi \over 3}} \right)
                  \left( {9 \over 16 \cab^2}
           + {3 \over \Rab}
                  \right) \cr
          &+ \delta_{x,2} \left( {8 e^{i {2 \phi \over 3}}
          (19 \cab^2 - 4) \over 3 \Rab^2 }
           + {3 e^{-i {4 \phi \over 3}} - 20 e^{i {2 \phi \over 3}}
                  \over 8 \cab^2}
           - {3 e^{-i {4 \phi \over 3}} \over \Rab}
                  \right) \cr
          &+ \delta_{x,3} \left(
           {40 \cab^2 - 7 \over \Rab^2 }
               + {9 \over 4 \cab^2 }
                  \right)
           + \delta_{x,4} {35 e^{i {4 \phi \over 3}} \over 8 \cab^2}
            {\Biggr \}} \cr
}    \eqno({\rm \secG.15b})$$
which is also $N$-independent. More precisely, $C_{\Ga}^{(k)}(x)$ and
$C_{\si}^{(k)}(x)$ are independent of $N$ if $N > 2 k$ and $x \le k$.
\medskip
$C_{\si}(x)$ is real and positive for all values of $\phi$ and $\vphi$
up to the order calculated. However, it is not easy to read off from
(\secG.14) what might be the form for large $x$. Thus, we specialize
to $\phi = \phi = {\pi \over 2}$ and calculate two further orders for
$C_{\si}(x)$:
$$\eqalign{
C_{\si}^{(0)}(x) =& C_{\si}^{(1)}(x) = C_{\si}^{(3)}(x)
                 = C_{\si}^{(5)}(x)= 0 \ , \cr
C_{\si}^{(2)}(x) =& {1 \over 9} \left \{
                     4 \delta_{x,0} + \delta_{x,1} \right \} \ , \cr
C_{\si}^{(4)}(x) =& {1 \over 81} \left \{
                     5 \delta_{x,0} + 2 \delta_{x,2} \right \} \ , \cr
C_{\si}^{(6)}(x) =& {1 \over 6561} \left \{
                     190 \delta_{x,0} - 13 \delta_{x,1} + 38 \delta_{x,2}
                     + 60 \delta_{x,3} \right \}. \cr
}    \eqno({\rm \secG.16})$$
As a byproduct we verified in this case two further orders of the equality
%%% Modification
% equality true only for absolute values
$\abs{{\avac \si_x \vac \over \normvac}} = \abs{C_{\Ga}^{lt}(1)}$ at the dual
point in the low-temperature phase.
\smallskip
$C_{\Ga}(x)$ in general has a non-vanishing imaginary part and therefore
is worth while being considered in more detail. Thus, we specialize
again to the superintegrable case $\phi = \vphi = {\pi \over 2}$ and
obtain after calculating two further orders
$$\eqalign{
C_{\Ga}^{(0)}(x) =& \delta_{x,0} \ , \qquad \qquad
C_{\Ga}^{(1)}(x) =  \delta_{x,1} \left({1 \over 3} + i {\sqrt{3} \over 9}
            \right) \ , \cr
C_{\Ga}^{(2)}(x) =& {1 \over 18} \left\{\delta_{x,1} +2 \delta_{x,2} \right\}
      + i {\sqrt{3} \over 18} \left\{ -\delta_{x,1} + 2 \delta_{x,2}
      \right\} \ , \cr
C_{\Ga}^{(3)}(x) =& {1 \over 81} \left\{ 4 \delta_{x,1} + 10 \delta_{x,2}
            \right\}
      + i {\sqrt{3} \over 243} \left\{ 4 \delta_{x,1} - 10 \delta_{x,2}
             + 20 \delta_{x,3} \right\} \ , \cr
C_{\Ga}^{(4)}(x) =& {1 \over 1458} \left\{ 27 \delta_{x,1} + 18 \delta_{x,2}
              + 210 \delta_{x,3} - 70 \delta_{x,4}\right\}
      + i {\sqrt{3} \over 1458} \left\{ -27 \delta_{x,1} + 18 \delta_{x,2}
             + 70 \delta_{x,4} \right\}
      \hskip 2pt ,  \cr
C_{\Ga}^{(5)}(x) =& {1 \over 2187} \left\{ 45 \delta_{x,1} + 108 \delta_{x,2}
         + 252 \delta_{x,4}- 126 \delta_{x,5} \right\} \cr
      &+ i {\sqrt{3} \over 2187} \left\{ 15 \delta_{x,1} - 36 \delta_{x,2}
             - 14 \delta_{x,3} + 84 \delta_{x,4} + 42 \delta_{x,5} \right\}
      \ , \cr
C_{\Ga}^{(6)}(x) =& {1 \over 39366} \left\{381 \delta_{x,1} +214 \delta_{x,2}
              + 2314 \delta_{x,3} + 784 \delta_{x,4} + 2310 \delta_{x,5}
              - 1848 \delta_{x,6} \right\} \cr
      &+ i {\sqrt{3} \over 39366} \left\{-381 \delta_{x,1} +214 \delta_{x,2}
             - 784 \delta_{x,4} + 2310 \delta_{x,5} \right\}. \cr
}    \eqno({\rm \secG.17})$$
Of course, we still have to calculate the sum (\secG.8). Thus, changes
of signs in individual orders need not necessarily turn up in the final
result. In fact, it turns out that the imaginary part of $C_{\Ga}(x)$
is always positive up to order 6 because the smallest orders are positive
and they dominate the others. However, for sufficiently small $\la$ the
real part does indeed change signs around $x=4$. Although we are not able
to verify if it becomes positive again around $x=12$ (which would need
more than the double of the orders which we have calculated) it
is in good agreement with the expected form (\secG.6).
Therefore we fit (\secG.17) by a complex exponential function.
In summary, (\secG.17) indicates that
$$\eqalignno{
C_{\Ga}(x) &= a \ e^{\left({2 \pi i \over L} - {1 \over \xi_{\Ga}}\right) x}
            + (1-a) \delta_{x,0} \ , &({\rm \secG.18a})\cr
C_{\si}(x) &= p \ e^{- {x \over \xi_{\si}} } + q \delta_{x,0}
             &({\rm \secG.18b})\cr
}$$
such that $C_{\Ga}(x)$ is of the form (\secG.6) for $x > 0$.
In (\secG.18) we have also taken into account that from (\secG.16)
${C_{\si}(0) \over C_{\si}(1)} \approx 4$ independent of the correlation
length $\xi_{\si}$.
\sn
If (\secG.18a) is the correct form for $C_{\Ga}(x)$ we infer from (\secG.17)
that $L$ is about $14$ for small $\la$. We can also see from
the higher orders that $L$ increases with increasing $\la$ such that
it might well be singular at $\la = 1$. The correlation length $\xi$ tends
to zero as $\la \to 0$. This implies that -- after proper re-normalization
of the Hamiltonian -- the mass gap becomes infinite at $\la = 0$. It has
already been observed in $\q{\weA}$ that there are physical
reasons to divide (\secB.2)  by $\sqrt{\la}$ which would have exactly the
effect of infinite mass at $\la = 0$. Fits to (\secG.18) for
$\la \in \{ {1 \over 4}, {1 \over 2}, {3 \over 4} \}$ in the superintegrable
case are given by the values in the following table:
%%%%%%%%%%%%%%%%%%%%%%%%%%%%%%%%%%%%%%%%%%
% Table
\mn
\centerline{\vbox{
\hbox{
\vrule \hskip 1pt
\vbox{ \offinterlineskip
\def\tablespace{height2pt&\omit&\vl&\omit&&\omit&&\omit&\vl&\omit&&
                          \omit&&\omit&\vl&\omit&&\omit&\cr}
\def\tablerule{ \tablespace
                \noalign{\hrule}
                \tablespace        }
\hrule
\halign{&\vrule#&
  \strut\hskip 4pt\hfil#\hfil\hskip 4pt\cr
\tablespace
\tablespace
& $\la$  &\vl& $\xi_{\Ga}$ && $a$       && $L$      &\vl&
          $\xi_{\si}$ && $p$       && $q$       &\vl&
            $P_{\min}$ && ${L P_{\min} \over 2 \pi}$ &
                                           \cr \tablespace \tablerule
& $0.25$ &\vl& $0.55(3)$   && $0.55(5)$ && $14.3(2)$ &\vl&
          $0.25(2)$   && $0.35(4)$ && $0.32(4)$ &\vl&
            $0.471$     && $1.07(2)$ & \cr \tablespace
& $0.50$ &\vl& $0.9(1)$    && $0.59(3)$ && $16.5(8)$ &\vl&
          $0.38(4)$   && $0.35(3)$ && $0.24(3)$ &\vl&
            $0.401$     && $1.05(5)$ & \cr \tablespace
& $0.75$ &\vl& $1.5(6)$    && $0.64(3)$ && $18.3(8)$ &\vl&
          $0.55(6)$   && $0.36(2)$ && $0.09(2)$ &\vl&
            $0.308$     && $0.90(4)$ & \cr \tablespace
}
\hrule}\hskip 1pt \vrule}
\hbox{Table 1: Parameters for the correlation functions
            (\secG.18) at $\phi = \vphi = {\pi \over 2}$ }}
}
\mn
%%%%%%%%%%%%%%%%%%%%%%%%%%%%%%%%%%%%%%%%%%
The estimates in table 1 have been obtained as follows.
First, $\xi_{\Ga}$ has been estimated by calculating
$\re(\ln({C_{\Ga}(x) \over C_{\Ga}(x+1) }))^{-1}$ and averaging
over $x$. Next, the zero of $\re(e^{x \over \xi_{\Ga}} C_{\Ga}(x))$
has been estimated by linear interpolation for two neighbouring
values and ${L \over 4}$ was obtained by averaging. Finally,
$a$ was estimated such that the difference
$$\re(C_{\Ga}(x)) -
a e^{-{x \over \xi_{\Ga}}} \bcos{{2 \pi x \over L}}
    \eqno({\rm \secG.19})$$
is minimal for $x=1,2$. That this procedure yields reasonable
fits is demonstrated by Fig.~1 which shows the stretched correlation
function $e^{x \over \xi_{\Ga}} C_{\Ga}(x)$ in comparison to the
fits. The `error bars' are not really error bars but given by
%%% Modification
% wrong sign in exponent changed
$a e^{x - 6 \over \xi_{\Ga}}$ which gives an idea how much the
values have actually been stretched and what might be the contribution
of the next orders in the perturbation expansion. The agreement
for all $x$ not only in the real part but also in the imaginary part
is convincing.
\mn
Table 2 shows the values $C_\Ga^{\rm pert.}(x)$ corresponding to Fig.\ 1.
This table also contains the numerical results for the correlation
function $C_\Ga^{\rm num.}(x)$ which were obtained in $\q{\lett}$
for $N=12$ sites at $\phi = \vphi = {\pi \over 2}$, $\la = {1 \over 2}$.
%%%%%%%%%%%%%%%%%%%%%%%%%%%%%%%%%%%%%%%%%%
% Table
\mn
\centerline{\vbox{
\hbox{
\vrule \hskip 1pt
\vbox{ \offinterlineskip
\def\tablespace{height2pt&\omit&\vl&\omit&&\omit&&\omit&&\omit&&
                          \omit&&\omit&&\omit&\cr}
\def\tablerule{ \tablespace
                \noalign{\hrule}
                \tablespace        }
\hrule
\halign{&\vrule#&
  \strut\hskip 2pt\hfil#\hfil\hskip 2pt\cr
\tablespace
& $x$  &\vl& \hskip 2pt 0 \hskip 2pt && 1 && 2 && 3 && 4 && 5 && 6 &
                                                      \cr \tablerule
&$C_{\Ga}^{\rm pert.}(x)$  &\vl& $1$ &&
          $\scriptstyle .18868+.07384 i$ &&
          $\scriptstyle .04561+.03980 i$ &&
          $\scriptstyle .00992+.01747 i$ &&
          $\scriptstyle .00091+.00674 i$ &&
          $\scriptstyle -.00088+.00263 i$ &&
          $\scriptstyle -.00074$ & \cr \tablerule
&$C_{\Ga}^{\rm num.}(x)$   &\vl& $1$ &&
          $\scriptstyle .18881+.07385 i$ &&
          $\scriptstyle .04587+.03967 i$ &&
          $\scriptstyle .01004+.01737 i$ &&
          $\scriptstyle .00126+.00679 i$ &&
          $\scriptstyle -.00056+.00224 i$ &&
          $\scriptstyle -.00080$ & \cr \tablespace
}
\hrule}\hskip 1pt \vrule}
\hbox{Table 2: Perturbative results (\secG.17)
            and numerical results at $N=12$ sites}
\hbox{\phantom{Table 2:}
            for the correlation function $C_{\Ga}(x)$
            at $\phi = \vphi = {\pi \over 2}$, $\la = {1 \over 2}$ }}
}
\mn
%%%%%%%%%%%%%%%%%%%%%%%%%%%%%%%%%%%%%%%%%%
The agreement between the results of both methods is good. This
shows that on the one hand higher orders are indeed negligible in
(\secG.17) for $x < 7$ and on the other hand that the finite chain
length does not considerably affect the correlation function $C_{\Ga}(x)$.
\medskip
Let us now discuss the implications of (\secG.15) under the
assumption that (\secG.18a) is the correct form for general values
of the chiral angles.
{}From the leading orders in (\secG.15a) we read off the following identity
for the ratio of $C_{\Ga}(1)$ and $C_{\Ga}(2)$:
$${C_{\Ga}(2) \over C_{\Ga}(1)} = {
{e^{i{2 \phi \over 3}} \over 6 \cab^2} \la^2 + \O(\la^3) \over
{e^{i{\phi \over 3}} \over 3 \cab} \la + \O(\la^2) }
= {e^{i{\phi \over 3}} \over 2 \cab} \la + \O(\la^2).
    \eqno({\rm \secG.20a})$$
On the other hand we immediately obtain from (\secG.18a)
$${C_{\Ga}(2) \over C_{\Ga}(1)}
= e^{-{1 \over \xi_{\Ga}}} e^{{2 \pi i \over L}}.
    \eqno({\rm \secG.20b})$$
Comparison of (\secG.20a) and (\secG.20b) leads to
$$L = {6 \pi \over {\rm Re}(\phi)} \ , \qquad \qquad
\xi_{\Ga} = -{1 \over \ln{\left(
{\la \over 2 \bcos{{\vphi \over 3}} }
\right)}  - {{\rm Im}(\phi) \over 3} }
    \eqno({\rm \secG.21})$$
for small values of $\la$. It is noteworthy that we obtain the same
%%% Modification
% L inserted
result for the oscillation length $L$ if we apply a similar argument to
${C_{\Ga}(x_1) \over C_{\Ga}(x_2)}$ in lowest non-vanishing order with
$x_1,x_2 \in \{1,2,3,4\}$. At $\phi = {\pi \over 2}$ (\secG.21) yields
the approximations $L = 12$, $\xi_{\Ga} = 0.52$, $0.80$, $1.2$
for $\la = 0.25$, $0.50$, $0.75$. The agreement with the numbers of table 1
is very good. Thus, for very high temperatures the oscillation length $L$
is proportional to the inverse chiral angle $\phi^{-1}$. In particular,
the oscillation vanishes smoothly for $\phi \to 0$. In $\q{\weA}$ it
was shown that for very high temperatures the minimum of the dispersion
relation of the fundamental particles is also proportional to $\phi$.
More precisely, we read off from (\secD.6) that the minimum of
the dispersion relation is in first order perturbation theory at
$P_{\min} = {{\rm Re}(\phi) \over 3}$. Thus, we obtain from (\secG.21)
for very high temperatures
$$P_{\min} L\mid_{\la \to 0} = 2 \pi \qquad \forall \phi, \vphi.
     \eqno({\rm \secG.22})$$
Furthermore, the second order in (\secD.6) shows that the minimal
momentum $P_{\min}$ decreases with increasing $\la$ (compare also
$\q{\lett}$). Similarly,
we read off from (\secG.15) that the inverse oscillation length $L^{-1}$
%%% Modification
% inverse temperature $\la$
also decreases with increasing inverse temperature $\la$. Thus, (\secG.22) has
a chance to be valid for all values of $\la$ in the massive
high-temperature phase. Indeed, using the values of $P_{\min}$
given in table 8 of $\q{\weA}$ we see that $P_{\min} L = 2 \pi$
holds quite accurately for $\la = 0.25, 0.5, 0.75$ at
$\phi = \vphi = {\pi \over 2}$ (compare table 1).
Using numerical methods we have checked in $\q{\lett}$ that
$P_{\min} L = 2 \pi$ is indeed valid within the numerical accuracy for
general values of the parameters.
%\mn
The identity $P_{\min} L = 2 \pi$ can e.g.\ be derived from the
form factor expansion (\secG.5) if the Hamiltonian has suitable symmetries
as is demonstrated in appendix {\appC} for certain special cases.
However, it may well be that in general this relation is not exact but
an excellent approximation.
\medskip
Note that even at $\phi = \vphi = {\pi \over 2}$ the correlation lengths
$\xi_{\Ga}$ and $\xi_{\si}$ are clearly different. Furthermore,
$\xi_{\si}$ coincides with its dual in the low-temperature phase
whereas $\xi_{\Ga}$ does not (see $\q{\han}$). Recall that
for the correlation function $C_{\si}(x)$ only the spectrum
in the charge sector $Q=0$ is relevant but $C_{\Ga}(x)$
comes from the $Q=1$ sector. Using (\secG.5) this explains
the agreement of $\xi_{\si}$ with $\xi$ in the low-temperature phase;
in this phase all charge sectors have a spectrum that is identical with
the spectrum in the $Q=0$-sector at the dual point in the high-temperature
phase $\q{\han}$.
\bn
\chapsubtitle{\secH.\ The parity conserving Potts case}
\mn
So far, we have studied correlation functions for general values of
the parameters and for the superintegrable case. In this section
we discuss the standard parity-conserving $\Zed_3$ case in more
detail and compare the correlation length to the inverse mass gap.
We also examine the dispersion relation of the particle-/anti-particle
pair closer for this special case and show that, for general $\la$, there
is no simple relation between the square of the energy and the momentum like
the Klein-Gordon equation.
\mn
First, we note that for $\phi = \vphi = 0$ eq.\ (\secD.5) simplifies
considerably and we can calculate even higher orders:
$$\eqalign{
m(\la) := &
\Delta E_{1,0}(0, 0) =
\Delta E_{2,0}(0, 0) = {1 \over \sqrt{3}} \Bigl(
          6 - 4 \la - 2 \la^2 + \la^3 \cr
&{\textstyle
            - {179 \over 162} \la^4
            + {1099 \over 1458} \la^5 - {15865 \over 26244} \la^6
            + {163717 \over 629856} \la^7 - {4564375 \over 68024448} \la^8
          }\Bigr) + \O(\la^9). \cr
}    \eqno({\rm \secH.1})$$
In this case the mass gap is located at zero momentum. Therefore,
we defined (\secH.1) as `$m(\la)$'. The quality of the approximation
(\secH.1) can be judged from the values $m(\la)^{\rm pert.}$
%%%% Modification
% Tablenumber corrected
in table 3.
%%%%%%%%%%%%%%%%%%%%%%%%%%%%%%%%%%%%%%%%%%
% Table
\mn
\centerline{\vbox{
\hbox{
\vrule \hskip 1pt
\vbox{ \offinterlineskip
\def\tablespace{height2pt&\omit&&\omit&&\omit&&\omit&\cr}
\def\tablerule{ \tablespace
                \noalign{\hrule}
                \tablespace        }
\hrule
\halign{&\vrule#&
  \strut\hskip 4pt\hfil#\hfil\hskip 4pt\cr
\tablespace
\tablespace
&$\la$ && $m(\la)^{\rm num.}$  && $m(\la)^{\rm pert.}$
           && $(1-\la)^{-{5\over 6}} \, m(\la)^{\rm num.}$
                                                 & \cr \tablespace \tablerule
&$0$   && $2 \sqrt{3}$         &&$2 \sqrt{3}$ &&
                                       $3.4641016151377544$ & \cr \tablespace
&$0.1$ && $3.22213207613437(1)$&&$3.22213207621$ &&
                                      $3.51782783078230(1)$ & \cr \tablespace
&$0.3$ && $2.67861152211(9)$   &&$2.67861277$  &&
                                          $3.6057424206(1)$ & \cr \tablespace
&$0.5$ && $2.0620838(7)$       &&$2.06219$    &&$3.674216(1)$ &\cr\tablespace
&$0.6$ && $1.725364(1)$        &&$1.72591$    &&$3.702520(2)$ &\cr\tablespace
&$0.7$ && $1.3665(2)$          &&$1.36879$    &&$3.7269(5)$ & \cr \tablespace
&$0.8$ && $0.96507(6)$         &&$0.98796$    &&$3.6901(2)$ & \cr \tablespace
&$0.9$ && $0.5373(1)$          &&$0.57927$    &&$3.6606(7)$ & \cr \tablespace
&$0.95$&& $0.300(1)$           &&$0.36274$    &&$3.64(1)$   & \cr \tablespace
&$0.975$&&$0.166(2)$           &&$0.25105$    &&$3.59(4)$   & \cr \tablespace
&$1$   && $0.001(6)$           &&$0.13691$    && \omit      & \cr \tablespace
}
\hrule}\hskip 1pt \vrule}
%%% Modification
% Tablenumber corrected
\hbox{Table 3: Mass gap $m(\la)$ at $\phi = \vphi = 0$}}
}
\mn
%%%%%%%%%%%%%%%%%%%%%%%%%%%%%%%%%%%%%%%%%%
For comparison we have also included values for $m(\la)^{\rm num.}$
obtained from a numerical diagonalization of the Hamiltonian
at $4 \le N \le 12$ sites with extrapolation. The numerical values
are more precise but the calculations for each value of $m(\la)^{\rm num.}$
are quite complicated whereas for $m(\la)^{\rm pert.}$ one just has
to evaluate (\secH.1).
One observes that the approximation (\secH.1) is accurate even near
the critical point $\la =1$. However, we should mention that (\secH.1)
is an alternating sum and gives only good approximations
if an even number of orders is used. For example using only 7th
order one obtains a value of $0.18$ for $m(1)$. Even more, for $\la > 0.7$ the
approximations $m(\la)^{\rm pert.}$ are more accurate if one truncates the
sum (\secH.1) to 6th order (the 6th order approximation for $m(1)$ is
$0.026$ !). This demonstrates that close to the phase transition
$\la = 1$ this perturbation series is slowly convergent.
\mn
It is well-known that the critical exponent for $m(\la)$ at $\la=1$
equals ${5 \over 6}$.
%%% Modification
% Line of argumentation changed to avoid impression that
% we wish to compute a critical exponent
The series eq.\ (\secH.1) can be used to verify this critical
exponent with a DLog-Pad\'e analysis. In fact, this check has
already been performed in $\q{\hkn}$. In order to test the
critical behaviour throughout
the massive high-temperature phase we have also included the
%%% Modification
% Tablenumber corrected
values for $(1-\la)^{-{5\over 6}} \, m(\la)^{\rm num.}$ in table 3
\footnote{${}^{5})$}{Corresponding values for
$\la \in \{0, 0.25, 0.5, 0.75, 0.9\}$ have already been given in $\q{\weA}$.
The values presented here are however more accurate for $\la$ close
to the critical value $1$ (the estimated errors are realistic
as one can conclude from the fact that the ratios become constant
around $\la = 1$ within the estimated errors).
}.
Taking the critical exponent ${5 \over 6}$ as input we conclude
from the fact that this ratio yields a very slowly varying function
that the normalization of the Hamiltonian (\secB.2) is indeed meaningful
even far away from the critical region.
Note that this agreement between mass gap and extrapolation from the
critical point has already been observed in $\q{\weA}\q{\gehkal}$.
\bigskip
Next we will discuss the correlation function $C_{\Ga}(x)$ for
$\phi = \vphi = 0$.
If a statistical system has an isotropic field theory as limit, the
correlation length is related in this limit to the inverse of the smallest
gap between the ground state and the first excitation $\q{\kogut}$.
Therefore, one expects a relation $\xi \sim m(\la)^{-1}$ $\q{\kogut}$.
Note that for small values of $\la$ we expect a different behaviour
according to (\secG.21): $\xi_{\Ga} \sim \ln\left({\la \over 2}\right)^{-1}$.
We will now study these two relations more closely
by considering the correlation function $C_{\Ga}(x)$.
First, we specialize (\secG.15) to $\phi = \vphi = 0$ and
calculate two further orders. This leads to:
$$\eqalign{
C_{\Ga}^{(0)}(x) =& \delta_{x,0} \ , \qquad \qquad \qquad \qquad \qquad
C_{\Ga}^{(1)}(x) =  {\delta_{x,1} \over 3} \ , \cr
C_{\Ga}^{(2)}(x) =& {1 \over 12} \left\{\delta_{x,1}
                    + 2 \delta_{x,2} \right\} \ , \, \qquad \qquad
C_{\Ga}^{(3)}(x) =  {1 \over 162} \left\{5 \delta_{x,1}
                      + 14 \delta_{x,2} + 15 \delta_{x,3} \right\} \ , \cr
C_{\Ga}^{(4)}(x) =& {1 \over 11664} \left\{315 \delta_{x,1}
                      + 478 \delta_{x,2} + 852 \delta_{x,3}
                      + 630 \delta_{x,4} \right\} \ , \cr
C_{\Ga}^{(5)}(x) =& {1 \over 209952} \left\{3525 \delta_{x,1}
                      + 5870 \delta_{x,2} + 9007 \delta_{x,3}
                      + 12016 \delta_{x,4} + 6804 \delta_{x,5}
                  \right\} \ , \cr
C_{\Ga}^{(6)}(x) =& {1 \over 3779136} \left\{44744 \delta_{x,1}
                      + 77659 \delta_{x,2} + 111952 \delta_{x,3}
                      + 153196 \delta_{x,4} \right. \cr
&\qquad \qquad \left. + 162940 \delta_{x,5} + 74844 \delta_{x,6}
                  \right\}. \cr
}   \eqno({\rm \secH.2})$$
Using (\secH.2) one can calculate the correlation length $\xi_{\Ga}$ by
the procedure described in the previous section. We just mention
a few pairs $\lb\la,\xi_{\Ga}\rb$:
\sn
$\lb 0,0\rb$, $\lb 0.00005,0.094(2)\rb$,
$\lb 0.0005,0.121(3)\rb$, $\lb 0.005,0.167(6)\rb$, $\lb 0.05,0.27(2)\rb$,
$\lb 0.25,0.50(5)\rb$, $\lb 0.5,0.8(1)\rb$, $\lb 0.75,1.2(2)\rb$.
\sn
Fig.\ 2 shows a plot including more estimates for the correlation length.
At $\la = 0$ there is no correlation between different sites, i.e.\
the correlation length is zero. One observes that it increases drastically
for $\la > 0$. It is clearly different from zero even
for very small values of $\la$. We have also plotted the estimate (\secG.21)
in Fig.\ 2. This crude estimate fits the numerical results surprisingly well
for all values of $\la$ accessible to us. In particular, it nicely reproduces
the behaviour for small $\la$ as it is expected from our derivation of
the estimate.
In Fig.\ 2 we also plotted the properly normalized inverse mass
$m(\la)^{-1}$. The agreement is good for $\la > 0.3$. For $\la < 0.1$ one
observes a clear disagreement. Note that in this region
${C_{\Ga}(x) \over C_{\Ga}(x+1)} \approx e^{1 \over \xi_{\Ga}} < 10$
and one should therefore expect that at least in this region the finite
lattice spacing is important.
\medskip
It has been observed in $\q{\weA}$ that for $\phi = \vphi = 0$
the dispersion relation (\secD.6) agrees with a Klein-Gordon
dispersion relation up to order $\la^2$. Furthermore, it was shown
in $\q{\albCONF}$ that at the second order phase transition
$\phi = \vphi = 0$, $\la=1$ the dispersion relation is of
Klein-Gordon type with mass $m(1) = 0$. Using the abbreviation
$$K := 2 \bsin{{P \over 2}}
   \eqno({\rm \secH.3})$$
for the lattice analogue of the momentum
we can specialize (\secD.6) to $\phi = \vphi = 0$ and calculate
two further orders. This yields the dispersion relation
$$\eqalign{
\E(K) :=&
\Delta E_{1,0}(P, 0, 0) =
\Delta E_{2,0}(P, 0, 0) = \cr
= &{1 \over \sqrt{3}} {\Biggl\{}
         6 + 2 (K^2-2) \la
           - {\la^2 \over 3} (K^4 - 6 K^2 + 6)
           + {\la^3 \over 18} (2 K^6 - 14 K^4 + 19 K^2 + 18) \cr
& \qquad   - {\la^4 \over 324} (15 K^8 -152 K^6 +531 K^4 -738 K^2 + 358) \cr
& \qquad   + {\la^5 \over 2916} (63 K^{10} - 764 K^8 + 3214 K^6
                      - 5087 K^4 + 1121 K^2 + 2198)
          {\Biggr\}} + \O(\la^6) \cr
}    \eqno({\rm \secH.4})$$
for the two fundamental quasiparticles. The Klein-Gordon dispersion
relation predicts
$$\eqalign{
\E(K) =& \sqrt{ m(\la) + a(\la) K^2} \cr
      =& m(\la) + a(\la) {K^2 \over 2 m(\la)}
               - a(\la)^2 {K^4 \over 8 m(\la)^3}
               + a(\la)^3 {K^6 \over 16 m(\la)^5}
               - a(\la)^4 {5 K^8 \over 128 m(\la)^7} \cr
       &\qquad + a(\la)^5 {7 K^{10} \over 256 m(\la)^9}
               + \O(K^{12}) \cr
}    \eqno({\rm \secH.5})$$
where we have included a free normalization constant $a(\la)$ which
corresponds to the velocity of light. Rewriting
(\secH.4) in the form (\secH.5) leads to
$$\eqalign{
\E(K) = m(\la)
    &+ {\la \over 729} (-4435 \la^4 +3618 \la^3 -2754 \la^2 +1944 \la +5832)
       {K^2 \over 2 m(\la)} \cr
    &- {16 \la^2 \over 243} (-2221 \la^3 + 567 \la^2 + 324 \la + 972)
       {K^4 \over 8 m(\la)^3} \cr
    &+ {256 \la^3 \over 81} (-223 \la^2 + 144 \la + 162)
       {K^6 \over 16 m(\la)^5}
     - {4096 \la^4 \over 135} (134 \la + 135) {5 K^8 \over 128 m(\la)^7} \cr
    &+ 32768 \la^5 {7 K^{10} \over 256 m(\la)^9}
     + \O(\la^6). \cr
}    \eqno({\rm \secH.6})$$
In (\secH.6) the coefficients of $K^n$ with lowest order in $\la$
agree with the Klein-Gordon dispersion relation (\secH.5)
\footnote{${}^{6})$}{
According to $\q{\albCONF}$ one should have $\E(K) = 3 \abs{K}$ at
$\la = 1$. At $\la = 1$ the series (\secH.4) does not really
converge any more. Nevertheless, it seems that (\secH.4) is
compatible with $\E(K)^2 = 9 K^2$ at $\la =1$.
}.
However, using $m(\la)$ as in (\secH.1) and fitting $a(\la)$ from the
coefficient of $K^2$, all but these leading orders disagree
with (\secH.5).
Thus, although a Klein-Gordon dispersion relation is certainly a good
approximation to (\secH.4) it is unfortunately not the exact form.
\mn
It happens quite frequently in two-dimensional quantum field
theories that the dispersion relation is $\sin$-Gordon or $\sinh$-Gordon.
However, even these possibilities can be ruled out because the first
five orders of these dispersion relations agree with Klein-Gordon
($\sin(x^2) + \O(x^6) = x^2 = \sinh(x^2) + \O(x^6)$ ) but the deviation
from Klein-Gordon occurs already at order $K^4$. Therefore we
consider an even more general dispersion relation of the form
$$\eqalign{
\E(K) &= \sqrt{ m(\la) + {g\left(b(\la) a(\la) K^2 \right) \over b(\la)}}
     \ , \cr
g(x) &= x + x^2 + c_3 x^3 + c_4 x^4 + c_5 x^5 + \O(x^6)
}    \eqno({\rm \secH.7})$$
which contains the Klein-Gordon relation (\secH.5) for $b(\la) \to 0$.
In particular, (\secH.7) is a good approximation to
the Klein-Gordon dispersion relation for small $b(\la)$.
(\secH.7) would also include both the $\sin$-Gordon and
$\sinh$-Gordon relations for $c_2 = 0$ but since this has
already been ruled out $c_2$ has been
absorbed in $b(\la)$. Determining
from the first orders of the Taylor expansion of (\secH.7) with
respect to $K^2$ first $m(\la)$, then $a(\la)$, $b(\la)$ and $c_3$ it
turns out that $c_3$ depends on $\la$. Thus (\secH.7) can be ruled out
if the function $g$ is required to be universal for all $\la$.
\bn
\chapsubtitle{\secI.\ Convergence of single-particle excitations}
\mn
As far as the proof of the quasiparticle picture is concerned the main
open question is the convergence of the single-particle states, or
equivalently the existence of the limits $N \to \infty$ of the
corresponding eigenvalues of the Hamiltonian. We have argued in section
{\secD} that convergence of the perturbation expansions
is sufficient to guarantee the existence of the limits
$N \to \infty$. Therefore we will discuss the radius of convergence
for the perturbation expansion of the single-particle
excitations in this section.
\mn
For bounded operators --in particular finite dimensional ones--
one could use criteria involving operator norms similar to those
for v.\ Neumann series. Unfortunately, the potential for
$\Delta H_N^{(n)}$ as defined in (\secB.2) and (\secB.14)
is unbounded if $N$ is not fixed. Thus, we have to apply the
slightly more complicated Kato-Rellich theory of regular
perturbations. Reviews of this subject can be found e.g.\
in the monographs $\q{\reedsimon}\q{\katobook}$. The main
results we are going to use were originally published in
$\q{\rellich}\q{\kato}$. The theory of Kato and Rellich applies
in particular to operators of the form (\secC.1), i.e.\
$H(\la) = H_0 + \la V$.
\medskip
Suppose that the single-particle eigenvalues $\Delta E$ have a non-zero
distance from the scattering eigenvalues (the continuous
spectrum) at $\la = 0$. Then it is clear from the discussion
in the previous sections that these eigenvalues are
non-degenerate and isolated. In particular, the resolvent
$(\Delta H_N^{(n)}(\la) - z)^{-1}$ is bounded for $\abs{\Delta E - z} > 0$.
Restricting to the hermitean case,
this is sufficient to guarantee that the $\Delta H_N^{(n)}(\la)$
are an analytic family in the sense of Kato. In this case,
the Kato-Rellich theorem ($\q{\reedsimon}$ Theorem XII.8)
may be used to guarantee a non-zero
radius of convergence $r_0 > 0$ for the single-particle
eigenvalues of $\Delta H_N^{(n)}(\la)$.
\mn
In order to obtain explicit estimates of the radius of convergence
one needs the inequality
$$\norm{ V \state{a} } \le {\cal V} \norm{H_0 \state{a} }
                         + {\cal K} \norm{ \state{a} }
   \eqno{\rm (\secI.1)}$$
on ${\cal D}(H_0)$ which in our case is dense in the the complete
Hilbert space $\H$. Then, the isolated point eigenvalues of $H(\la)$ are
convergent at least for
$$\la < r_1 := {\cal V}^{-1}   \eqno{\rm (\secI.2)}$$
as long as these eigenvalues {\it do not come in contact with
continuous spectrum} $\q{\kato}$. On the one hand this criterion
is very simple, on the other hand one must estimate not
only the constant ${\cal V}$ but also examine the level
crossings between single-particle excitations and scattering
states. There is another estimate $r_2$ that guarantees the
separation of eigenvalues as
well but gives smaller radii of convergence. For self-adjoint
$H_0$ with isolated eigenvalue $E^{(0)}_0$ where the
nearest eigenvalue $E^{(0)}_1$ has distance
$\epsilon := \abs{E^{(0)}_1 - E^{(0)}_0}$
($\epsilon^{-1} = \norm{g(E^{(0)}_0)}$) the perturbation
expansion of $E_0(\la)$ is convergent for
$$\la < r_2 := {\epsilon \over 2
           \left( {\cal K} + {\cal V} (\abs{E^{(0)}_0} + \epsilon) \right) }
   \eqno{\rm (\secI.3)}$$
and there are no crossings with neighbouring levels. In order to
compare the estimates (\secI.2) and (\secI.3) let us assume
${\cal K} = 0$ and $\abs{E^{(0)}_0} = \epsilon$. For this
almost optimal case one has $r_1 = 4 r_2$ showing that the criterion
(\secI.3) is much more restrictive.
\medskip
Let us now apply these general results to the present case of
$\Zed_n$-spin quantum chains. For non-degenerate single-particle
eigenvalues the Kato-Rellich theorem can be applied to guarantee
a positive radius of convergence $r_0$. Then we know from section
{\secE} and appendix {\appB} that the spectrum of $\Delta H_N^{(n)}(\la)$
is a quasiparticle spectrum for $\la < r_0$. This fact can be used
to calculate the constant ${\cal V}$ and obtain explicit
estimates $r_1$ (where level crossings still have to be discussed)
or $r_2$. One can obtain the estimate (\secI.1) with ${\cal K}=0$
using Schwarz' inequality:
$${\cal V} := \sup_{\state{a} \in \H}
{\astate{a} \Delta V \state{a} \over \norm{\Delta H_{N,0}^{(n)} \state{a}}}.
   \eqno{\rm (\secI.4)}$$
In general, this supremum need not be finite but then it is very
difficult to ensure convergence at all. In our case,
the important observation is that due to the quasiparticle
picture we can evaluate (\secI.4) exclusively
from the single-particle excitations. To see this one performs
a first order expansion in $\la$ for any composite particle state,
compares coefficients and uses the quasiparticle property to
expand the expectation values of $\Delta H_{N,0}^{(n)}$ and $\Delta V$
in single-particle excitations. Thus, ${\cal V}$ can be calculated
as
$${\cal V} = \max_{Q,P}
     {{}_P\apstate{s^Q} \Delta V \pstate{s^Q}_P
          \over \norm{\Delta H_{N,0}^{(n)} \pstate{s^Q}_P } }.
   \eqno{\rm (\secI.5)}$$
\medskip
In order to implement this program explicitly we specialize to
the case of $\Zed_3$ with the parametrization (\secB.10).
At $\la = 0$ both single-particle eigenvalues are isolated
for $-{\pi \over 2} < \vphi < {\pi \over 2}$. This guarantees
a non-zero radius of convergence $r_0$.
\mn
The simplest case is the parity conserving case $\phi = \vphi =0$.
Here, the maxima are located at zero momentum $P=0$ and
both charge sectors are degenerate. Furthermore, we have
$\norm{\Delta H_{N,0}^{(3)} \pstate{s^Q}_0} = \epsilon = E_0^{(0)}$.
{}From (\secH.1) we can therefore read off ${\cal V} = {2 \over 3}$,
or in terms of radii of convergence
$$r_1 = {3 \over 2}, \qquad r_2 = {3 \over 8}, \qquad
{\rm for} \quad n=3, \ \phi = \vphi = 0 .
   \eqno{\rm (\secI.6)}$$
$r_2 = 0.375$ is certainly too small which
can easily be seen applying a naive ratio test to (\secH.1).
Extrapolating ${m^{(\nu)} \over m^{(\nu+1)}}$ to $\nu = \infty$
one obtains an estimate for the radius of absolute convergence
of about $1.3$. Thus, for $\phi = \vphi = 0$ the radius of convergence
is expected to be close to the boundary of the phase $\la = 1$ which
is also supported by the calculations in section {\secH}.
\mn
For general angles $0 \le \vphi < {\pi \over 2}$, the free part of
the Hamiltonian $\norm{\Delta H_{N,0}^{(3)} \pstate{s^Q}_0}$ is minimized
for $Q=1$ and the potential
${}_P\apstate{s^Q} \Delta V \pstate{s^Q}_P$ is maximal
for $P = {\phi \over 3}$. Thus, we read off from (\secD.6)
${\cal V} = \left( \sqrt{3} \bsin{{\pi - \vphi \over 3}} \right)^{-1}$.
Furthermore, one has $\epsilon = 8 \bsin{{\pi - \vphi \over 3}}
- 4 \bsin{{\pi + \vphi \over 3}}$ and
$E_0^{(0)} = 4 \bsin{{\pi - \vphi \over 3}}$.
This amounts to the following radii
$$r_1 = \sqrt{3} \bsin{{\pi - \vphi \over 3}} , \qquad
 r_2 = {\sqrt{3} \bsin{{\pi - \vphi \over 3}} \left(
 2 \bsin{{\pi - \vphi \over 3}}- \bsin{{\pi + \vphi \over 3}} \right)
\over
2  \left(
3 \bsin{{\pi - \vphi \over 3}}- \bsin{{\pi + \vphi \over 3}} \right)}
   \eqno{\rm (\secI.7)}$$
for $n=3$, $0 \le \vphi < {\pi \over 2}$.
For $\vphi \to {\pi \over 2}$ the situation is contrary to that at
$\phi = \vphi = 0$. The $Q=2$ particle state becomes degenerate with
two $Q=1$ scattering states at $\vphi = {\pi \over 2}$ such that
the radius of convergence must tend to zero for $\vphi \to {\pi \over 2}$.
Whereas $r_2$ has precisely this property, $r_1$ tends to $0.866\ldots$
which is certainly too large.
\medskip
Because for small $\vphi$ we would prefer the large radius of
convergence $r_1$ but at $\vphi \approx {\pi \over 2}$ this is much
too large and $r_2$ seems more appropriate we have to enhance the
estimate given by $r_1$ by a discussion of level crossings between
single-particle states and scattering states.
For $0 \le \vphi < {\pi \over 2}$
the first level crossing of this kind will take place between
the $Q=2$ single-particle excitation and a two $Q=1$ particles
scattering state.
\sn
It is very difficult
to determine those values of $\la$ explicitly and precisely
where they take place. Therefore, we will use the first order
approximation of the perturbation expansion.
We are looking for those values of $\la$ where a single point
$P$ exists such that $x := 2 \Delta E_{1,0}({P \over 2},\phi,\vphi) -
\Delta E_{2,0}(P,\phi,\vphi)$ vanishes. The fact that we are looking
for no real crossings but $x=0$ implies ${{\rm d} x \over {\rm d}P} = 0$.
Inserting (\secD.6) and (\secD.7) up to first order leads to the
condition
$$\bsin{{P \over 2} - {\phi \over 3}} = \bsin{P + {\phi \over 3}}
    \eqno({\rm \secI.8})$$
Eq.\ (\secI.8) has a solution
$$P = {2 \pi \over 3}    \eqno({\rm \secI.9})$$
that does not depend on $\phi$. Now we can solve the linear equation
$x\mid_{\la_0}=0$ for the value $\la_0$. One obtains
$$\la_0 = {\bcos{{\vphi \over 3}} - \sqrt{3} \bsin{{\vphi \over 3}}
            \over
           \bcos{{\phi \over 3}} + \sqrt{3} \bsin{{\phi \over 3}} } \, .
    \eqno({\rm \secI.10})$$
Fig.\ 3 shows a plot of the estimates (\secI.7) and (\secI.10) for
the self-dual case $\vphi = \phi$. Note that $r_1$ and $r_2$ are independent
of $\phi$. However, we have assumed that the Hamiltonian is hermitean and
therefore $\phi$ must be real.
For $\vphi$ close to ${\pi \over 2}$,
$\la_0$ is smaller than $r_2$ which is an apparent contradiction because
there are no level crossings for $\la < r_2$.
Remember, however, that $\la_0$ has been calculated approximately such that
this difference is not significant. At $\phi = \vphi = 0$ we find
$\la_0 \approx 1$. This is reassuring because perturbation
expansions should not be valid beyond the second order phase transition
at $\la = 1$. Although $\la_0$ was estimated by looking at non-zero
momenta there are also level crossings in the zero momentum sectors
at $\la = 1$. Thus, the radius of convergence is indeed smaller than
$r_1$. Still, our results agree in magnitude with the intuitive
expectations from the `ratio test'. The dots in Fig.\ 3 indicate the two
models whereof the spectra have been presented in Fig.\ 2 and Fig.\ 3 of
$\q{\lett}$. For the left dot one expects a converging perturbation expansion
whereas in the other case it should not converge (compare $\q{\lett}$).
Indeed, both estimates $r_2$ and $\la_0$ make this distinction.
\bn
For completeness we have also included an estimate for the boundary
of the massive high-temperature phase in Fig.\ 3. At this boundary,
levels of the $Q=1$ particle with generically non-zero momentum
cross with the ground state. Its explicit location has been
obtained estimating the minimum of the dispersion relation (\secD.6)
with $P = {\phi \over 3}$ and solving the second order approximation
$\Delta E_{1,0} ({\phi \over 3}, \phi, \phi) = 0$ for $\la$.
At $\phi = 0$ the agreement with the exact value $\la = 1$ is excellent.
For small non-zero angles the true value is smaller than 1 but our
approximation gives values that are slightly larger than 1.
Also at $\phi = {\pi \over 2}$ we observe a small deviation from
the exact value $\la = 0.901292\ldots$ $\q{\mccoyadv}$:
Our estimate yields $\la = 0.866\ldots$ (the agreement with $r_1$ is
a coincidence).
\sn
The level crossings transition $\la_0$ divides the massive high-temperature
phase of the $\Zed_3$-chiral Potts model into two parts
which we label I and II. In part I the derivation of the quasiparticle
picture as presented in section {\secE}
and appendix {\appB} is rigorous. Thus, in regime I the spectrum is
a quasiparticle spectrum with two fundamental particles existing
for all momenta. In $\q{\lett}$ we have presented numerical evidence that
regime II probably also exhibits a quasiparticle spectrum with two fundamental
particles where the $Q=2$ particle has the unusual property that it exists
only in a limited range of the momentum $P$. At $\vphi=\phi={\pi \over 2}$
this statement has been proven rigorously in $\q{\dkcoy}$. We expect
that the idea to approximate multi-particle states by putting single-particle
states of `small' chains with a sufficient separation on a longer chain and
to use the finite correlation length in order to ensure vanishing of boundary
terms (which we cannot show directly like in appendix {\appB} any more)
will apply also in regime II for general angles $\phi$, $\vphi$.
However, in contrast to section {\secE} we loose control over the fundamental
$Q=2$ excitation because the perturbation series does not converge any more
and there is no guarantee for the completeness of this construction.
At least it is plausible to still expect a quasiparticle spectrum in regime
II with two fundamental particles of which the $Q=2$ particle may have a
Brillouin zone that is smaller than the interval $\lbrack 0, 2 \pi \rbrack$.
\bn
\chapsubtitle{\secJ.\ Conclusion and outlook}
\mn
In this paper we have
presented an argument using perturbation theory proving that the massive
high-temperature phases of all $\Zed_n$-spin quantum chains exhibit
quasiparticle spectra with $n-1$ fundamental particles. Since the
argument relies on perturbation theory it applies rigorously only to
very high temperatures. Due to the perturbative nature of the details
we were not able to give it any predictive power for those case where some
of the fundamental particles cross with scattering states. For these cases
one needs different methods, e.g.\ Bethe ansatz techniques
$\q{\dkcoy}\q{\kedem}$ or numerical methods $\q{\lett}$.
Nevertheless, the basic idea of approximating a multi-particle state
by single-particle states sitting on subparts of the chain might be
applicable in the entire massive high-temperature phase. One could even
speculate that a similar argument can be applied to $\Zed_n$-spin models
in higher dimensions as well.
%%% Modification
% additional paragraph/sentence on finite-size effects has been added
\sn
A refined (but less rigorous) version of this argument
can be used to control the finite-size effects of $k$-particle states
showing in particular that the {\it energy} of the excitations does
not pick up any corrections at order ${1 \over N}$.
\sn
Furthermore, our derivation of the quasiparticle picture involving $n-1$
fundamental particles applies to the scaling region near the conformal point
$\la = 1$, $\phi = \vphi = 0$, the only non-rigorous part of the proof being
the radius of convergence.
This region ($\la < 1$, $\phi$, $\vphi$ small) corresponds to
perturbations of conformal field theories with the thermal operator
$\q{\zamPA-\mussardo}$ and a small additional perturbation of the type
presented in $\q{\cardy}$ that breaks parity.
\sn
Using duality $\q{\han}$ our results about the quasiparticle spectra
can be pulled over to the massive low-temperature phase
of $\Zed_n$-spin quantum chains.
\sn
Having derived such a quasiparticle picture the main open problem
is to find the corresponding massive field theory and to obtain
the associated scattering matrix.
\mn
We also studied the correlation functions using a perturbation expansion
for the ground state of the $\Zed_3$-model.
Although this approach is limited to short ranges, we were not
only able to estimate correlation lengths in the massive high-temperature
phase but it also turned out that the correlation functions have
oscillatory contributions. For very high temperatures the oscillation
length is proportional to the inverse of one of the chiral
angles $L \sim \phi^{-1}$. We further observed that the oscillation
length is closely related to the minimum of the dispersion relations
for general values of the parameters. The relation $L P_{\min}
= 2 \pi$ is valid on the lattice with a much better accuracy than
the well-known relation $\xi \sim m^{-1}$. For special values of the
parameters we were able to derive the relation $L P_{\min} = 2 \pi$
from a form factor decomposition but one should certainly understand it
better in the general case.
\bn
\chapsubtitle{Acknowledgements}
\mn
%%% Modification
% double mentioning of W.\ Eholzer corrected
I would like to thank W.\ Eholzer, F.\ E{\ss}ler, H.\ Hinrichsen,
R.\ H\"ubel, M.\ R\"osgen and K.\ Yildirim for many useful
discussions. I am particularly grateful to G.\ von Gehlen for many
comments and encouragement.
\vfill
\eject
\chapsubtitle{Appendix {\appA}: Perturbation expansions for two-particle
                                states}
\mn
In this appendix we will explicitly diagonalize the matrix (\secD.14).
Although this can be done for all four different cases simultaneously
it is more illustrative to treat first one case separately.
The simplest case is actually $N$ even and
${N P \over 2 \pi}$ odd. First, we perform the following
diagonal change of bases
$$B \pstate{t^{\pm}_j}_P := e^{-i \Ph (j-1)} \pstate{t^{\pm}_j}_P
. \eqno({\rm \appA.1})$$
If we also identify the $\pstate{t^{\pm}_j}_P$ with the standard
basis of $\Complex^{{N \over 2} -1}$
the ${N \over 2} -1$ times ${N \over 2} -1$ matrix $W$ satisfies
$$W = B \left(
\matrix{0 & 1 & 0 & \cdots & 0 \cr
        1 & 0 & 1 & \ddots & \vdots \cr
        0 & 1 & \ddots & \ddots & 0 \cr
   \vdots & \ddots & \ddots & 0 & 1 \cr
        0 & \cdots & 0 & 1 & 0 \cr
}\right) B^{-1}
. \eqno({\rm \appA.2})$$
The eigenvalues and eigenvectors of the matrix on the r.h.s.\ of
(\appA.2) are well-known in the literature (see e.g.\ $\q{\jones}$,
example 1.2.5).
We can use them to write down immediately the eigenvalues and eigenvectors
of the matrix $W$:
$$\eqalign{
\pstate{\tau^{\pm}_k}_P :&= B {2 \over \sqrt{N}}
\left( \matrix{\bsin{{2 k \pi \over N}} \cr
           \bsin{{2 k 2 \pi \over N}} \cr
           \vdots \cr
           \bsin{{2 k ({N \over 2} - 1) \pi \over N}} \cr
}\right), \cr
W \pstate{\tau^{\pm}_k}_P &= 2 \bcos{{2 k \pi \over N}}
           \pstate{\tau^{\pm}_k}_P
\ , \qquad 1 \le k \le {N \over 2} - 1. \cr
}  \eqno({\rm \appA.3})$$
Thus, we have calculated a first order expansion for the next excitations
in the charge sectors $2 Q = 2$, $2 Q = 1$ for $N \ge 4$:
$$\eqalign{
\Delta E_{2,k}(P,\phi,\vphi) &= 8 \bsin{{\pi - \vphi \over 3}}
        - \la {8 \over \sqrt{3}} \bcos{\Ph - \phit}
                                 \bcos{{2 k \pi \over N}}
        + \O(\la^2) \ , \cr
\Delta E_{1,k}(P,\phi,\vphi) &= \Delta E_{2,k}(P,-\phi,-\vphi) \ ,
\qquad  1 \le k \le {N \over 2} - 1 \ ,
\quad {\rm for} \ N \ {\rm even} , \ {N P \over 2 \pi} \ {\rm odd}. \cr
}  \eqno({\rm \appA.4})$$
The same argument can be applied to the remaining cases. The only
additional consideration is that we need a second change of bases $M$ in
$\Complex^{N-1}$. For $N$ odd it is defined by
$$
M := {1\over \sqrt{2}} \left(
\matrix{1 & 0 & \cdots & \cdots & 0 & 1 \cr
        0 & \ddots & \ddots & \Ddots & \Ddots & 0 \cr
   \vdots & \ddots &  1 & 1 & \Ddots & \vdots \cr
   \vdots & \Ddots & -1 & 1 & \ddots & \vdots \cr
        0 & \Ddots & \Ddots & \ddots & \ddots & 0 \cr
        -1 & 0 & \cdots & \cdots & 0 & 1 \cr
}\right) \, \qquad N-1 \times N-1, N \ {\rm odd}
  \eqno({\rm \appA.5a})$$
whereas for $N$ even it is defined as follows
$$
M := {1\over \sqrt{2}} \left(
\matrix{1 & 0 & \cdots & \cdots & \cdots & 0 & 1 \cr
        0 & \ddots & \ddots & & \Ddots & \Ddots & 0 \cr
   \vdots & \ddots &  1 & 0 & 1 & \Ddots & \vdots \cr
   \vdots &   & 0 & \sqrt{2} & 0 & & \vdots \cr
   \vdots & \Ddots & -1 & 0 & 1 & \ddots & \vdots \cr
        0 & \Ddots & \Ddots &  & \ddots & \ddots & 0 \cr
        -1 & 0 & \cdots & \cdots & \cdots & 0 & 1 \cr
}\right) \, \qquad N-1 \times N-1, N \ {\rm even}.
  \eqno({\rm \appA.5b})$$
With the definition (\appA.1) of the $N-1 \times N-1$
matrix $B$ we can write $W$ as
$$\pmatrix{W^{\rm even} & 0 \cr
           0 & W^{\rm odd} \cr} = B M \left(
\matrix{0 & 1 & 0 & \cdots & 0 \cr
        1 & 0 & 1 & \ddots & \vdots \cr
        0 & 1 & \ddots & \ddots & 0 \cr
   \vdots & \ddots & \ddots & 0 & 1 \cr
        0 & \cdots & 0 & 1 & 0 \cr
}\right) M^{-1} B^{-1}
 \eqno({\rm \appA.6})$$
where $W^{\rm even}$ is the $\lb {N \over 2} \rb \times
\lb {N \over 2} \rb$ matrix $W$ for even lattice momenta
($\delta_P^N = 1$) and $W^{\rm odd}$ is the
$\lb {N-1 \over 2} \rb \times \lb {N-1 \over 2} \rb$ matrix
$W$ for odd lattice momenta (with reversed order of basis
vectors).
\mn
Again, we can use the well-known results $\q{\jones}$ to write
down the eigenvalues and eigenvectors of the matrix $W$:
$$\eqalign{
\pstate{\tau^{\pm}_k}_P :&= B {2 \over \sqrt{N}}
\left( \matrix{\bsin{{(2 k - \delta_P^N) \pi \over N}} \cr
         \bsin{{(2 k - \delta_P^N) 2 \pi \over N}} \cr
         \vdots_{} \cr
         \bsin{{(2 k - \delta_P^N) (\lb{N \over 2}\rb - 1) \pi \over N}} \cr
         \bsin{{(2 k - \delta_P^N) \lb{N \over 2}\rb \pi \over N}}
\cases{{1 \over \sqrt{2}} & for $N$ even \cr
                        1 & for $N$ odd \cr
} \cr
}\right), \cr
W \pstate{\tau^{\pm}_k}_P &=
2 \bcos{{(2 k - \delta_P^N) \pi \over N}} \pstate{\tau^{\pm}_k}_P
\ , \qquad 1 \le k \le \left\lb{N \over 2}\right\rb. \cr
}  \eqno({\rm \appA.7})$$
Inserting definitions one immediately obtains the final results
(\secD.18) and (\secD.19) from (\appA.7).
\medskip
Let us now turn to the second order. For simplicity we restrict
to the $Q=2$-sector and odd $N$. We abbreviate the resolvent by $g$
and its values by:
$$R_{3,1} := - 4 \sqrt{3} \bcos{{\vphi \over 3}} \, , \qquad
R_{1,2} := - 4 \sqrt{3} \bcos{{\pi - \vphi \over 3}} \, , \qquad
R_{0,1} := 4 \sqrt{3} \bcos{{\pi + \vphi \over 3}} \, .
  \eqno({\rm \appA.8})$$
Then we obtain for the matrix elements between the states
$\pstate{t^{+}_k}_P$ of the second order expression:
$$\eqalign{
B^{-1} \, q \, r(V) \, g \, r(V) \, q \, B =
{8 \over 3}\Biggl\{&\left\lb
{\bcos{P-{2 \phi \over 3}} + 1 \over R_{0,1}}
 -{1 \over R_{1,2}} + {1 \over R_{3,1}}\right\rb
\left(
\matrix{1 & 0 & \cdots & 0 \cr
        0 & 0 & \cdots & 0 \cr
   \vdots & \vdots & \ddots & \vdots \cr
        0 & 0 & \cdots & 0 \cr
}\right) \cr
&+{\bcos{{P \over 2} + {2 \phi \over 3}} \over R_{1,2}}
\left(
\matrix{0 & 1 & 0 & \cdots & 0 \cr
        1 & 0 & 1 & \ddots & \vdots \cr
        0 & 1 & \ddots & \ddots & 0 \cr
   \vdots & \ddots & \ddots & 0 & 1 \cr
        0 & \cdots & 0 & 1 & 0 \cr
}\right) \cr
&+{\bcos{P - {2 \phi \over 3}} \over R_{3,1}}
\left(
\matrix{0 & 0 & 1 & \cdots & 0 \cr
        0 & 0 & 0 & \ddots & \vdots \cr
        1 & 0 & \ddots & \ddots & 1 \cr
   \vdots & \ddots & \ddots & 0 & 1 \cr
        0 & \cdots & 1 & 1 & 0 \cr
}\right) \cr
&+\left\lb{2 \over R_{1,2}} + {N-4 \over R_{3,1}} \right\rb \id
\Biggr\} . \cr
}  \eqno({\rm \appA.9})$$
For a chain of length $N$ ($N$ odd), the matrix (\appA.9) is of
size $\lb{N \over 2}\rb \times \lb{N \over 2}\rb$. It is straightforward
to take the matrix elements of this matrix between the first order
eigenstates (\appA.7). We omit the explicit form of this second
order expression because it would be almost as long as (\appA.9).
\vfill
\eject
\chapsubtitle{Appendix {\appB}: Proof of quasiparticle picture}
\mn
In this appendix we present a modified version of the proof in section
{\secE} that the spectrum of the Hamiltonian (\secB.2) can be explained
in terms of $n-1$ fundamental quasiparticles.
\sn
The main steps of the proof will be as presented in section {\secE}.
However, instead of considering general multi-particle states we will
go directly back to the single-particle excitations. The corresponding
perturbative eigenstates are given by (\secD.2). One also has to be
careful where it is permitted to deal directly with $\Delta H_N^{(n)}$
or where one should rather consider $H_N^{(n)}$ first.
\sn
In this appendix we will concentrate on the vanishing of boundary terms.
Not all arguments presented in section {\secE} will be spelled out in
detail again. In particular, the limiting procedures are taken for
granted. This explicit presentation has been shifted to this appendix
because the explicit formulae are a bit nasty although the ideas are
quite simple.
\medskip
In order to be able to discuss $r$-particle states we first write
down the generalization of (\secE.3) to any partition of $N$ in
$r$ arbitrary integers $N_j > 1$ ($N=\sum_{j=1}^r N_j$):
$$\eqalign{
H_{N}^{(n)} &= H_{N_1}^{(n)} \otimes \id + \ldots +
               \id \otimes H_{N_j}^{(n)} \otimes \id + \ldots +
               \id \otimes H_{N_r}^{(n)}
                   + \sum_{j=1}^r \O(H_{N_j}^{(n)}) \ , \cr
T_{N}  &= \left( \id + \O(T_{N_r}) \right) \ldots
                \left( \id + \O(T_{N_1}) \right)
                T_{N_1} \otimes \ldots \otimes T_{N_r}. \cr
}    \eqno({\rm \appB.1})$$
Let $\nu_j := \sum_{i=1}^{j-1} N_i$ ($\nu_0 := 0$). Then, the boundary
operators $\O(H_{N_j}^{(n)})$ and $\O(T_{N_j})$ are given by
$$\eqalign{
\O(H_{N_j}^{(n)}) &= \la \sum_{k=1}^{n-1} \a_k \Ga_{\nu_j + N_j}^k \left\{
                     \Ga_{\nu_j + 1}^{n-k} - \Ga_{\nu_j + N_j + 1}^{n-k}
                     \right\} \ , \cr
\left( \id + \O(T_{N_j}) \right)
&\state{i_1 i_2 \ldots i_{\nu_j + N_j} i_{\nu_j + N_j+1} \ldots
       i_{\nu_{j+1} + N_{j+1}} i_{\nu_{j+1} + N_{j+1} + 1} \ldots i_N} \cr
&=\state{i_1 i_2 \ldots i_{\nu_{j+1} + N_{j+1}} i_{\nu_j + N_j+1} \ldots
       i_{\nu_j + N_j} i_{\nu_{j+1} + N_{j+1} + 1} \ldots i_N}. \cr
}    \eqno({\rm \appB.2})$$
\indent
It will be useful to verify first that additivity of energy and momentum
holds for $\la = 0$. To this end we show that a composite particle state
$$\pstate{N;Q_{\rm tot}}_{P_{\rm tot}} :=
\pstate{s^{Q_1}}_{P_1} \otimes \ldots \otimes \pstate{s^{Q_r}}_{P_r}
    \eqno({\rm \appB.3})$$
has approximately energy $E := \sum_{j=1}^r E_{Q_j}$, momentum
$P_{\rm tot} := \sum_{j=1}^r P_j$ and total charge
$Q_{\rm tot} := \sum_{j=1}^r Q_j \mod n$. Recall that we have
defined all $\pstate{s^{Q_j}}_{P_j}$ in the complete Hilbert space
$\H$ but it is useful to think of them as elements of $\H_{N_j}$,
i.e.\ $\pstate{s^{Q_j}}_{P_j} \in \H_{N_j}$.
Applying (\appB.1) to these states one obtains for the energy:
$$\eqalign{
\left(H_{N}^{(n)} - E \right) & \pstate{N;Q_{\rm tot}}_{P_{\rm tot}}
= \sum_{j=1}^r \O(H_{N_j}^{(n)}) \pstate{N;Q_{\rm tot}}_{P_{\rm tot}} \cr
& + \sum_{j=1}^r \pstate{s^{Q_1}}_{P_1} \otimes \ldots \otimes
   \left(H_{N_j}^{(n)} - E_{Q_j} \right)
    \pstate{s^{Q_j}}_{P_j} \otimes \ldots \otimes \pstate{s^{Q_r}}_{P_r} \cr
&= 0 \cr
}    \eqno({\rm \appB.4})$$
where we have used that $\O(H_{N_j}^{(n)})$ vanishes at $\la = 0$
(compare (\appB.2) ). Thus, we verified that additivity of energy
is exact at $\la = 0$.
\mn
Using (\appB.1) and (\appB.2) one obtains for the momentum
$$\eqalign{
T_{N} \pstate{N;Q_{\rm tot}}_{P_{\rm tot}} =&
\prod_{j=1}^r \left\{ \id + \O(T_{N_j}) \right\}
\left(T_{N_1} \pstate{s^{Q_1}}_{P_1}\right) \otimes \ldots \otimes
\left(T_{N_r} \pstate{s^{Q_r}}_{P_r}\right) \cr
=& \prod_{j=1}^r \left\{\id + \O(T_{N_j}) \right\}
e^{i P_{\rm tot}} \pstate{N;Q_{\rm tot}}_{P_{\rm tot}} \cr
=& e^{i P_{\rm tot}} \pstate{N;Q_{\rm tot}}_{P_{\rm tot}}
+e^{i P_{\rm tot}} \pstate{s^{Q_1}}_{P_1}
  \lotimes \ldots \lotimes \pstate{s^{Q_r}}_{P_r} \cr
&\phantom{ e^{i P_{\rm tot}} \pstate{N;Q_{\rm tot}}_{P_{\rm tot}} }
-e^{i P_{\rm tot}} \pstate{s^{Q_1}}_{P_1}
  \otimes \ldots \otimes \pstate{s^{Q_r}}_{P_r}
}    \eqno({\rm \appB.5})$$
where `$\lotimes$' denotes the modifications that occur when shifting
in the entire tensor product instead of acting in its individual parts.
Locally, these modifications look as follows:
$$\eqalign{
\pstate{s^{Q_j}}_{P_j} \lotimes & \pstate{s^{Q_{j+1}}}_{P_{j+1}}
- \pstate{s^{Q_j}}_{P_j} \otimes \pstate{s^{Q_{j+1}}}_{P_{j+1}} \cr
=&{e^{-i P_{j+1}} \over \sqrt{N_{j+1}} }
\sum_{k=1}^{N_j} {e^{-i k P_j} \over \sqrt{N_j}}
\state{0 \ldots \underbrace{Q_j}_{{\rm position} \ k-1} \ldots 0 Q_{j+1}}
\otimes \state{0 \ldots 0} \cr
&- {e^{-i P_j} \over \sqrt{N_j} } \state{0 \ldots 0 Q_j}
\otimes \pstate{s^{Q_{j+1}}}_{P_{j+1}} \cr
\to & \quad 0. \cr
}    \eqno({\rm \appB.6})$$
In (\appB.6) we have used the explicit form (\secB.9)
of the single-particle states $\pstate{s^{Q_j}}_{P_j}$. The
vanishing of the boundary terms for $N_j \to \infty$ is ensured
by the normalization factors $N_j^{-{1 \over 2}}$. Thus, we
have also shown that $\pstate{N;Q_{\rm tot}}_{P_{\rm tot}}$
approximates an eigenstate of the translation operator
to total momentum $P_{\rm tot}$ for $\la = 0$.
\medskip
For $\la > 0$ we have to consider single-particle eigenstates
${\pstate{N_j;Q_j}}_{P_j}$ that are derived by
perturbation series from the states $\pstate{s^{Q_j}}_{P_j}$.
These states have the form
$${\pstate{N_j;Q_j}}_{P_j} =
\pstate{s^{Q_j}}_{P_j} + \sum_{\nu > 0} \la^{\nu}
\sum_{{i^{(j)}_1 + \ldots + i^{(j)}_{N_j} = Q_j \mod n} \atop
     {\# \{ i^{(j)}_k \ne 0\} \le 2 \nu + 1} }
\kappa_{i^{(j)}_1, \ldots ,i^{(j)}_{N_j}}^{(\nu)}
\pstate{i^{(j)}_1 \ldots i^{(j)}_{N_j}}_{P_j}.
    \eqno({\rm \appB.7})$$
It is important to note that the explicit form of the Hamiltonian (\secB.2)
implies that at most $2 \nu + 1$ spins $i_k$ are different
from zero in the $\nu$th order of the perturbation expansion.
In passing we mention that we do not need the explicit form of the
$\kappa_{i^{(j)}_1, \ldots ,i^{(j)}_{N_j}}^{(\nu)}$ and therefore the
argument also applies to the more general Hamiltonian (\secB.1) without
modification.
\sn
We should stress that the states (\appB.7) are in general not convergent
for $N_j \to \infty$ although the corresponding eigenvalues of
$\Delta H_{N_j}^{(n)}$ and $T_{N_j}$ converge (compare e.g.\
(\secG.13) ). However, after re-normalizing $\pstate{N_j;Q_j}_{P_j}$
to norm $1$ we could again think of it as lying in $\H$ for all $N_j$
because the $\nu$th order is independent of $N_j$ for $\nu < N_j$.
\sn
An $r$-particle state now is approximated by
$$\pstate{N;Q_{\rm tot}}_{P_{\rm tot}} :=
\pstate{N_1;Q_1}_{P_1} \otimes \ldots \otimes \pstate{N_r;Q_r}_{P_r}.
    \eqno({\rm \appB.8})$$
Note that (\appB.8) cannot be directly related to (\appB.3) by
a perturbation expansion.
\mn
First, we consider the translation operator. Thus, we have
to generalize (\appB.5) to the states (\appB.7), (\appB.8):
$$\eqalign{
T_{N} {\pstate{N;Q_{\rm tot}}}_{P_{\rm tot}} =&
\prod_{j=1}^r \left\{ \id + \O(T_{N_j}) \right\}
\left(T_{N_1} {\pstate{N_1;Q_1}}_{P_1}\right)
    \otimes \ldots \otimes
\left(T_{N_r} {\pstate{N_r;Q_r}}_{P_r}\right) \cr
=& \prod_{j=1}^r \left\{\id + \O(T_{N_j}) \right\}
e^{i P_{\rm tot}} {\pstate{N;Q_{\rm tot}}}_{P_{\rm tot}} \cr
=& e^{i P_{\rm tot}} {\pstate{N;Q_{\rm tot}}}_{P_{\rm tot}}
+e^{i P_{\rm tot}} {\pstate{N_1;Q_1}}_{P_1}
  \lotimes \ldots \lotimes {\pstate{N_r;Q_r}}_{P_r} \cr
&\phantom{ e^{i P_{\rm tot}}
   {\pstate{N;Q_{\rm tot}}}_{P_{\rm tot}} }
 -e^{i P_{\rm tot}} {\pstate{N_1;Q_1}}_{P_1}
  \otimes \ldots \otimes {\pstate{N_r;Q_r}}_{P_r}.
}    \eqno({\rm \appB.9})$$
The modifications introduced by $\lotimes$ in (\appB.9) are more
complicated than (\appB.6). Locally, they look as follows:
$$\eqalign{
{\pstate{N_j;Q_j}}_{P_j}& \lotimes
           {\pstate{N_{j+1};Q_{j+1}}}_{P_{j+1}}
- {\pstate{N_j;Q_j}}_{P_j} \otimes
           {\pstate{N_{j+1};Q_{j+1}}}_{P_{j+1}} \cr
=&{1 \over \sqrt{N_j} \sqrt{N_{j+1}} }
\sum_{\nu,\mu \ge 0} \la^{\nu+\mu} \Biggl\{
\sum^{\quad *}_{{{i^{(j+1)}_1 + \ldots + i^{(j+1)}_{N_{j+1}}
              = {Q_{j+1}} \mod n} \atop
     {\# \{ i^{(j+1)}_k \ne 0\} \le 2 \mu + 1}} \atop
             i^{(j+1)}_{m_{j+1}} \ne i^{(j)}_{k_j}}
\sum_{{i^{(j)}_1 + \ldots + i^{(j)}_{N_j} = {Q_j} \mod n} \atop
     {\# \{ i^{(j)}_k \ne 0\} \le 2 \nu + 1}} \cr
&e^{-i k_j P_j}
\kappa_{i^{(j)}_1, \ldots ,i^{(j)}_{N_j}}^{(\nu)}
\state{i^{(j)}_{k_j+1} \ldots i^{(j)}_{N_j} i^{(j)}_{1}
                     \ldots i^{(j)}_{k_j-1}i^{(j+1)}_{m_{j+1}}} \otimes \cr
&e^{-i m_{j+1} P_{j+1} }
\kappa_{i^{(j+1)}_1, \ldots ,i^{(j+1)}_{N_{j+1}}}^{(\mu)}
\state{i^{(j+1)}_{m_{j+1}+1} \ldots}
\Biggr\} \cr
&-{1 \over \sqrt{N_j}} \sum_{\nu \ge 0} \la^{\nu} \Biggl\{
\sum^{\quad **}_{{{i^{(j)}_1 + \ldots + i^{(j)}_{N_j} = {Q_j} \mod n} \atop
     {\# \{ i^{(j)}_k \ne 0\} \le 2 \nu + 1}} \atop
             i^{(j)}_{m_j} \ne 0} \cr
& e^{-i m_j P_j}
\kappa_{i^{(j)}_1, \ldots ,i^{(j)}_{N_j}}^{(\nu)}
\state{i^{(j)}_{m_j+1} \ldots i^{(j)}_{N_j} i^{(j)}_{m_j}}
\Biggr\} \otimes {\pstate{N_{j+1};Q_{j+1}}}_{P_{j+1}} \cr
\to & \quad 0. \cr
}    \eqno({\rm \appB.10})$$
The explicit form of (\appB.10) may be slightly confusing. Note, however,
that the vanishing of the boundary terms is guaranteed by the same argument
as in (\appB.6). It is sufficient to guarantee
that the coefficients of $\la^{\nu}$ in the boundary terms become small
with respect to the coefficients of the eigenstates. The index set
of the sums * and ** has precisely this property. The sum * has at most
$2 \mu + 1$ non-zero terms whereas the complete momentum decomposition
has $N_{j+1}$ terms. Thus, the complete sum * is suppressed by the
factor $N_{j+1}^{-{1 \over 2}}$. Also the sum ** has $2 \nu + 1$ terms
(or less) compared to $N_j$ for the complete fourier transform such
that it is suppressed by the normalization factor $N_j^{-{1 \over 2}}$.
This shows that ${\pstate{N;Q_{\rm tot}}}_{P_{\rm tot}}$
indeed approximates an eigenstate of the translation operator with
total momentum $P_{\rm tot}$.
\sn
Translating these statements into weak language we conclude the following:
The scalar product of (\appB.10) with an arbitrary momentum eigenstate
tends to zero if we rescale both states such that they lie on the unit
sphere. On the other hand, for any $N$ a normalized true eigenstate
of the translation operator $T_N$
exists such that the scalar product of this state with
${\pstate{N;Q_{\rm tot}}}_{P_{\rm tot}}$ tends to
its norm for $N \to \infty$.
\mn
Now we have to consider the generalization of (\appB.4) to $\la > 0$:
$$\eqalign{
\left(H_{N}^{(n)} - E \right)
&    {\pstate{N;Q_{\rm tot}}}_{P_{\rm tot}}
= \sum_{j=1}^r \O(H_{N_j}^{(n)})
     {\pstate{N;Q_{\rm tot}}}_{P_{\rm tot}} \cr
& + \sum_{j=1}^r {\pstate{N_1;Q_1}}_{P_1} \otimes \ldots \otimes
   \left(H_{N_j}^{(n)} - E_{Q_j} \right)
     {\pstate{N_j;Q_j}}_{P_j} \otimes \ldots
     \otimes {\pstate{N_r;Q_r}}_{P_r} \cr
=&\sum_{j=1}^r \O(H_{N_j}^{(n)})
     {\pstate{N;Q_{\rm tot}}}_{P_{\rm tot}}. \cr
}    \eqno({\rm \appB.11})$$
The boundary terms explicitly read as follows:
$$\eqalign{
\O(H_{N_j}^{(n)}) & {\pstate{N;Q_{\rm tot}}}_{P_{\rm tot}} =
{\pstate{N_1;Q_1}}_{P_1} \otimes \ldots \cr
&\sum_{k=1}^{n-1} \a_k \sum_{\nu \ge 0} \la^{\nu+1}
\sum_{{i^{(j)}_1 + \ldots + i^{(j)}_{N_j} = Q_j \mod n} \atop
     {\# \{ i^{(j)}_k \ne 0\} \le 2 \nu + 1} }
\kappa_{i^{(j)}_1, \ldots ,i^{(j)}_{N_j}}^{(\nu)}
{1 \over \sqrt{N_j}} \Biggl\{ \sum_{m=0}^{N_j -1 } \cr
&e^{-i m P_j}
\state{(i^{(j)}_m - k \mod n) i^{(j)}_{m+1}
         \ldots i^{(j)}_{m-2} (i^{(j)}_{m-1} + k \mod n) }
\otimes {\pstate{N_{j+1};Q_{j+1}}}_{P_{j+1}} \cr
& - e^{-i m P_j}
\state{i^{(j)}_m \ldots i^{(j)}_{m-2} (i^{(j)}_{m-1} + k \mod n) }
\otimes \left( \Ga_1^{n-k}
{\pstate{N_{j+1};Q_{j+1}}}_{P_{j+1}} \right) \Biggr\} \cr
&\quad \ldots \otimes {\pstate{N_r;Q_r}}_{P_r}. \cr
}    \eqno({\rm \appB.12})$$
The crucial point in (\appB.12) is that the states
$\state{i^{(j)}_m \ldots i^{(j)}_{m-1}}$ do not combine to momentum
eigenstates any more. More precisely, at fixed order $\nu$ of the
perturbation expansion one obtains at most $2 \nu + 3$ terms of
a complete momentum eigenstate. Therefore, these states are suppressed
for $N_j$ sufficiently large by the normalization factor
$N_j^{-{1 \over 2}}$. In other words: If we project (\appB.12) at
fixed order in $\la$ onto any momentum eigenstate and correct by the
norm of ${\pstate{N;Q_{\rm tot}}}_{P_{\rm tot}}$ the result tends
to zero. Note that we may not draw direct conclusions for the
limit of $H_N^{(n)}$ because the single-particle energies $E_{Q_j}$
do not converge. But for $\Delta H_N^{(n)}$ the single-particle energies
$\Delta E_{Q_j,0}$ converge and from (\appB.11), (\appB.12) we may
conclude that the states (\appB.8) behave precisely like $r$-particle
states in the limit $N \to \infty$.
\medskip
We have shown so far that the quasiparticle excitations describe
a subset of the spectrum of $\Delta H_N^{(n)}$ in the weak limit.
To complete the proof we have to argue that this is already the
complete spectrum. This is guaranteed by the fact that for any finite
$M$ the complete Hilbert space $\H_M$ can be mapped onto a subspace
of $\H_N$ ($N$ sufficiently large) that is spanned precisely
by the states (\appB.8). One natural choice is the mapping
$$\pstate{Q_1 \ldots Q_M}_{P_{\rm tot}} \mapsto
{\pstate{N_1;Q_1}}_{P_1} \otimes
\ldots \otimes {\pstate{N_M;Q_M}}_{P_M} \in \H_N.
    \eqno({\rm \appB.13})$$
This completes the proof.
\bigskip
Let us conclude with a summary of what we have assumed and what we were
able to prove.
Of course, the explicit form of the Hamiltonian (\secB.2) played an
important r\^ole. We needed three facts:
\sn
\item{1)} For some values of the parameters ($\la = 0$) the quasiparticle
          spectrum is trivially guaranteed.
\item{2)} In the vicinity of this point ($\la > 0$) only nearest neighbours
          interact.
\item{3)} The Hamiltonian (\secB.2) possesses a $\Zed_n$-symmetry.
\sn
It might seem that the third property was convenient mainly for notational
reasons because it straightforwardly encoded property 1). However, we
also needed the explicit form of the Hamiltonian (\secB.2) in order to
ensure the absence of further selection rules (at least at $\la = 0$).
Thus, although we did not rely heavily on property 3), we doubt that our
proof of the quasiparticle spectrum can easily be generalized to models
having more complicated selection rules.
\sn
We further required that
\sn
\item{4)} The perturbation expansions for the {\it single}-particle states
          converge.
\sn
Note that we did not assume the Hamiltonian $H_N^{(n)}$ to be hermitean nor
did we require it to be diagonalizable -- only the existence of the
single-particle eigenvalues is needed.
\sn
Already in section {\secD} we inferred from property 4) (and 2) ) that the
limits $N \to \infty$ of the single-particle eigenvalues of
$\Delta H_N^{(n)}$ exist. The proof presented in this appendix shows
that under these assumptions
\sn
\item{a)} The weak limits of the operators $T_N$ and
          $\Delta H_N^{(n)}$ exist,
\item{b)} The weak limits can be `diagonalized', i.e.\ the
          projection-valued measure of (\secB.16) does indeed exist, and
\item{c)} In this limit their spectrum can be expressed in terms
          of quasiparticle excitations. In particular, the spectrum of
          the weak limit of $\Delta H_N^{(n)}$ is explicitly known if
          the dispersion relations of the single-particle excitations
          can be calculated.
\vfill
\eject
\chapsubtitle{Appendix {\appC}: Symmetries of the Hamiltonian and the
                                oscillation length}
\mn
In this appendix we first discuss the behaviour of the Hamiltonian
%%% Modification
% credit changed
(\secB.2) under parity for special values of the parameters. One finds
symmetries that were observed numerically in $\q{\yildirim}$
for the integrable submanifold and can be derived e.g.\ along the lines
of appendix B of $\q{\mccoyadv}$. The resulting identities will subsequently
be used in order to derive the values of the oscillation length
given in (\secG.7) from the form factor expansion (\secG.5).
\mn
\leftline{{\bf Symmetries of the Hamiltonian:}} \noindent
%%%%%%%%%%%%%%%%%%%%%%%%%%%%%%%%%%%%%%%%%%%%%%%%
%%% Some major modifications in this section
%%%%%%%%%%%%%%%%%%%%%%%%%%%%%%%%%%%%%%%%%%%%%%%%
% a) the original proof was erranous
% b) it can be simplified considerably
Denote the projection of the Hamiltonian $H_N^{(n)}$ in eq.\ (\secB.2)
onto the spaces $\H_N^{P,Q}$ in eq.\ (\secB.15) by `$H_N^{(n)}(P,Q)$'.
Furthermore, introduce a parity operator $\Parity$ by the following
action on the states (\secB.5):
$$r(\Parity) \state{i_1 \ldots i_j \ldots i_N}
            = \state{i_1 i_N i_{N-1} \ldots i_j \ldots i_2} \, .
  \eqno{(\rm \appC.1)}$$
Note that $\Parity \si_{1+x} \Parity = \si_{1-x}$,
$\Parity \Ga_{1+x} \Parity = \Ga_{1-x}$.
Then one has the following identities (see also $\q{\yildirim}$):
$$\eqalign{
\a_k = \a_{n-k}
   \quad & \Rightarrow \quad \Parity H_N^{(n)}(P,Q) \Parity
                      = H_N^{(n)}(-P,Q) \, ,\cr
\ab_k^{*} = \ab_{n-k} \hbox{ and } \a_k \in \Real
   \quad & \Rightarrow \quad \Parity H_N^{(n)}(P,Q) \Parity =
                  \left(H_N^{(n)}(-P,Q)\right)^{+} \, , \cr
\ab_k^{*} = \ab_{n-k} \hbox{ and } \a_k^{*} = e^{-{2 \pi i z k}} \a_k
   \quad & \Rightarrow \quad \Parity H_N^{(n)}(P_{{\rm m},Q} + P,Q) \Parity =
                  \left(H_N^{(n)}(P_{{\rm m},Q}-P,Q)\right)^{+} \, \cr
}  \eqno{(\rm \appC.2)}$$
where the symmetry of the last line holds for
%%% Modification
% unnecessary assumption deleted
% any $Q$ that is invertible
%%%% Modification
%% unnecessary assumption deleted
%in $\Zed_n$ %, chains of length $N$ that are divisible by $n$
%and
those $P_{{\rm m},Q}$ satisfying
%%% Modification
% Additional assumption necessary
$P_{{\rm m},Q} Q^{-1} + \pi z \equiv 0$ mod $\pi$ as well as
$e^{i 2 P_{{\rm m},Q}}$ being an $n$th root of unity.
%%%% Modification
%% footnote added
%\footnote{${}^{7})$}{Close inspection of the argument presented below
%shows that these assumptions can be weakened. E.g.\ for the $\Zed_4$
%chain the $e^{i 2 P_{{\rm m},Q}}$ are in fact 2nd roots of unity which
%can be exploited in order to cover also the case $Q=2$ for $n=4$.
%}.
Note that with the parametrization (\secB.10) the cases covered
by (\appC.2) are precisely those covered by (\secG.7) with $z = {2 \over n}$.
In this case, $P_{{\rm m},Q} = \pi (1 - {2 Q \over n})$ is
a solution to $P_{{\rm m},Q} Q^{-1} + {2 \pi \over n} = 0$ lying in the
interval $\lbrack -\pi,\pi \rbrack$ -- the other solution is shifted
by $\pi$. The solution $P_{{\rm m},Q} = \pi (1 - {2 Q \over n})$
corresponds to the minimum in the dispersion relation of the single-particle
state in this charge sector (see $\q{\weA}\q{\yildirim}$).
%%% Modification
% unnecessary assumption deleted
%The relation $P_{{\rm m},Q} = \pi (1 - {2 Q \over n})$ seems to
%be valid for general $Q$ but we need the above constraints
%in the argument presented below.
\mn
The first two lines of (\appC.2) follow immediately by looking at
$\Parity H_N^{(n)} \Parity$, keeping in mind that the translation operator
defined in (\secB.8) satisfies $\Parity T_N \Parity = T_N^{-1} = T_N^{+}$.
The derivation of the third line of (\appC.2) is more complicated.
For $Q$ invertible in $\Zed_n$ it can be shown choosing a suitable basis
(see appendix B of $\q{\thesis}$).
For $z = {2 \over n}$ and $N \equiv 0 \mod n$ one can follow the lines of
eqs.\ (B.12) -- (B.16) in appendix B of $\q{\mccoyadv}$ to elegantly prove
the third line of (\appC.2).
\mn
In the case $z = {2 \over n}$ and $N \equiv 0 \mod n$ we introduce an
operator $U$ following $\q{\mccoyadv}$ by
$$U := \Parity \left( \prod\nolimits_{x=1}^N \si_x^{-2x} \right) \, .
  \eqno{(\rm \appC.3)}$$
Now, observing that
$$U \si_{1+x} U^{-1} = \si_{1-x} \, , \qquad
U \Ga_{1+x} U^{-1} = \om^{2 x} \Ga_{1-x}
  \eqno{(\rm \appC.4)}$$
one concludes that $U \, H_N^{(n)} \, U^{-1} = \left(H_N^{(n)}\right)^{+}$
for $\a_k^{*} = \om^{-2k} \a_k$. Finally one verifies that
$$T_N U \pstate{i_1 \ldots i_N}_{P}
 = \Parity T_N^{-1} \left( \prod\nolimits_{x=1}^N \si_x^{-2x} \right)
   \pstate{i_1 \ldots i_N}_{P}
 = U \hat{Q}^{-2} e^{-i P}
   \pstate{i_1 \ldots i_N}_{P}
  \eqno{(\rm \appC.5)}$$
where $\hat{Q}$ is the charge operator given by eq.\ (\secB.7).
Eq.\ (\appC.5) implies that the operator $U$ maps a state of charge $Q$
and momentum $P$ to a state of charge $Q$ and momentum
$-{4 \pi Q \over n} - P$. After putting things together one obtains
the desired result.
\mn
\leftline{{\bf Oscillation length from symmetries of the Hamiltonian:}}
\noindent
Assume that the Hamiltonian $H(P,Q)$ projected onto momentum and charge
eigenspaces with eigenvalues $P$ and $Q$ has one of the following symmetries:
$$\Parity H(P_{{\rm m}, Q}+P,Q) \Parity = H(P_{{\rm m}, Q}-P,Q)
\qquad \hbox{or} \qquad
\Parity H(P_{{\rm m}, Q}+P,Q) \Parity = \left(H(P_{{\rm m}, Q}-P,Q)\right)^{+}
  \eqno({\rm \appC.6})$$
with some $P_{{\rm m}, Q}$ depending on the charge sector $Q$. Assume
furthermore that $P_{\vac} = 0$ and that $\Xi_1 \vac$ has charge $Q$.
Then the oscillation length $L$ of the correlation function $C_{\Xi}(x)$
satisfies
$$L P_{{\rm m}, Q} = 2 \pi \, .
    \eqno({\rm \appC.7})$$
Note that this is true for more general Hamiltonians $H(P,Q)$, but it
covers in particular the case (\appC.2) for the $\Zed_n$-chiral Potts
model.
\mn
For a proof of (\appC.7) we start from the form factor expansion (\secG.5)
which in the present case becomes
$$C_{\Xi}(x)= \sum_{r} \int_0^{2 \pi} {\rm d}P \,
        e^{i P x } \, { \abs{\astate{P,Q;r} \Xi_1 \vac}^2
                 \over \normvac}
    \eqno({\rm \appC.8})$$
where we have only written the quantum numbers $P$ and $Q$ explicitly
and comprised the other ones in the label `$r$'. First we observe
that $\Parity \Xi_1 \Parity = \Xi_1$. If the Hamiltonian satisfies
$\Parity H(P_{{\rm m}, Q}+P,Q) \Parity = H(P_{{\rm m}, Q}-P,Q)$, then
eigenstates of momentum $P_{{\rm m}, Q}+P$ are mapped under parity to
eigenstates of momentum $P_{{\rm m}, Q}-P$. This means that
$\astate{(P_{{\rm m}, Q}+P),Q;r} \Xi_1 \vac =
\astate{(P_{{\rm m}, Q}-P),Q;r} \Xi_1 \vac$. If the symmetry involves
the adjoint of the Hamiltonian one finds
$\astate{(P_{{\rm m}, Q}+P),Q;r} \Xi_1 \vac =
\astate{(P_{{\rm m}, Q}-P),Q;r} \Xi_1 \vac^{*}$. Thus, the following
identity is valid in both cases:
$$\abs{ \astate{(P_{{\rm m}, Q}+P),Q;r} \Xi_1 \vac}^2 =
\abs{\astate{(P_{{\rm m}, Q}-P),Q;r} \Xi_1 \vac}^2 \, .
    \eqno({\rm \appC.9})$$
Now we return to the form factor expansion (\appC.8):
$$\eqalign{
C_{\Xi}(x) &= \sum_{r} \left\{
        \int\limits_{P_{{\rm m}, Q}}^{P_{{\rm m}, Q} + \pi} {\rm d}P \,
        e^{i P x } \, { \abs{\astate{P,Q;r} \Xi_1 \vac}^2 \over \normvac}
     +  \int\limits_{P_{{\rm m}, Q} - \pi}^{P_{{\rm m}, Q}} {\rm d}P \,
        e^{i P x } \, { \abs{\astate{P,Q;r} \Xi_1 \vac}^2 \over \normvac}
     \right\} \cr
&= \sum_{r} \int\limits_{0}^{\pi} {\rm d}P \, \left\{
        e^{i (P_{{\rm m}, Q}+P) x } \,
        { \abs{\astate{(P_{{\rm m}, Q}+P),Q;r} \Xi_1 \vac}^2 \over \normvac}
   \right. \cr & \qquad \qquad \qquad \left.
     +  e^{i (P_{{\rm m}, Q}-P) x } \,
        { \abs{\astate{(P_{{\rm m}, Q}-P),Q;r} \Xi_1 \vac}^2 \over \normvac}
     \right\} \cr
&= e^{i P_{{\rm m}, Q} x} \sum_{r}
        \int\limits_{0}^{\pi} {\rm d}P \, 2 \cos(P x) \,
        { \abs{\astate{(P_{{\rm m}, Q}+P),Q;r} \Xi_1 \vac}^2 \over \normvac}
         \cr
}    \eqno({\rm \appC.10})$$
\ifnum\journal=\jstat
\else
 \eject
 \noindent
\fi
where the last equality follows from (\appC.9).
This shows that $C_{\Xi}(x)$ is of the form
$$C_{\Xi}(x) = e^{{2 \pi i x \over L}} f(x)
    \eqno({\rm \appC.11})$$
with $L$ satisfying (\appC.7) and $f(x)$ is given by the remaining
integral in (\appC.10) which is clearly real.
\mn
Note that if the Hamiltonian has several different $P_{{\rm m},Q}$ such
that (\appC.6) holds (which applies to (\appC.2)) one obtains different
expressions for $C_{\Xi}(x)$ involving different $L$ and $f(x)$. The suitable
one among them can be singled out by demanding e.g.\ $f(x) > 0$ for all $x$.
Our explicit computations indicate that this requirement (which means that
the oscillations are exclusively encoded in the phase factor) indeed leads
to the oscillations lengths presented in (\secG.7).
\vfill
\eject
\chapsubtitle{References}
\mn
\bibitem{\ostlund} S.\ Ostlund, {\it Incommensurate and Commensurate Phases
                 in Asymmetric Clock Models},
                 Phys.\ Rev.\ {\bf B24} (1981) p.\ 398
\bibitem{\rietalA} M.\ Marcu, A.\ Regev, V.\ Rittenberg,
                 {\it The Global Symmetries of Spin Systems Defined on
                 Abelian Groups. I},
                 J.\ Math.\ Phys.\ {\bf 22} (1981) p.\ 2740
\bibitem{\rietalB} P.\ Centen, M.\ Marcu, V.\ Rittenberg,
                 {\it Non-Universality in $\Zed_3$ Symmetric Spin Systems},
                 Nucl.\ Phys.\ {\bf B205} (1982) p.\ 585
\bibitem{\hkn}   S.\ Howes, L.P.\ Kadanoff, M.\ denNijs,
                 {\it Quantum Model for Commensurate-In\-commen\-surate
                 Transitions},
                 Nucl.\ Phys.\ {\bf B215} (1983) p.\ 169
\bibitem{\gehri} G.\ von Gehlen, V.\ Rittenberg, {\it $\Zed_n$-Symmetric
                 Quantum Chains with an Infinite Set
                 of Conserved Charges and $\Zed_n$ Zero Modes},
                 Nucl.\ Phys.\ {\bf B257} (1985) p.\ 351
\bibitem{\dogra} L.\ Dolan, M.\ Grady, {\it Conserved Charges from
                 Self-Duality},
                 Phys.\ Rev.\ {\bf D25} (1982) p.\ 1587
%%% Modification
% additional reference
\bibitem{\perk} J.H.H.\ Perk, {\it Star-Triangle Equations, Quantum Lax Pairs,
                 and Higher Genus Curves}, Proc.\ Symp.\ Pure Math.\ {\bf 49}
                 (1989) p.\ 341
\bibitem{\onsager} L.\ Onsager, {\it Crystal Statistics.\ I.\ A
                 Two-Dimensional Model with an Order-Disorder Transition},
                 Physical Review {\bf 65} (1944) p.\ 117
\bibitem{\yang} H.\ Au-Yang, B.M.\ McCoy, J.H.H.\ Perk, Sh.\ Tang, M.L.\ Yan,
                 {\it Commuting Transfer Matrices in the Chiral Potts Models:
                 Solutions of Star-Triangle Equations with Genus $>1$},
                 Phys.\ Lett.\ {\bf 123A} (1987) p.\ 219
\bibitem{\baxterA} R.J.\ Baxter, J.H.H.\ Perk, H.\ Au-Yang,
                 {\it New Solutions of the Star-Triangle Relations for
                 the Chiral Potts Model},
                 Phys.\ Lett.\ {\bf 128A} (1988) p.\ 138
\bibitem{\baxterB} R.J.\ Baxter, {\it The Superintegrable Chiral Potts Model},
                 Phys.\ Lett.\ {\bf 133A} (1988) p.\ 185
\bibitem{\albertiniA} G.\ Albertini, B.M.\ McCoy, J.H.H.\ Perk,
                 {\it Commensurate-Incommensurate Transition in the Ground
                 State of the Superintegrable Chiral Potts Model},
                 Phys.\ Lett.\ {\bf 135A} (1989) p.\ 159
\bibitem{\albertiniB} G.\ Albertini, B.M.\ McCoy, J.H.H.\ Perk,
                 {\it Level Crossing Transitions and the Massless
                 Phases of the Superintegrable Chiral Potts Chain},
                 Phys.\ Lett.\ {\bf 139A} (1989) p.\ 204
\bibitem{\mccoyadv} G.\ Albertini, B.M.\ McCoy, J.H.H.\ Perk,
                 {\it Eigenvalue Spectrum of the
                 Superintegrable Chiral Potts Model},
                 Adv.\ Studies in Pure Math.\ {\bf 19} (1989) p.\ 1
\bibitem{\perkadv} H.\ Au-Yang, J.H.H.\ Perk, {\it Onsager's Star-Triangle
                 Equation: Master Key to Integrability},
                 Adv.\ Studies in Pure Math.\ {\bf 19} (1989) p.\ 57
\bibitem{\scm}   B.M.\ McCoy, {\it The Chiral Potts Model: from Physics to
                 Mathematics and back},
                 Special functions ICM 90, Satellite Conf.\ Proc., ed.\
                 M.\ Kashiwara and T.\ Miwa,
                 Springer (1991) p.\ 245
\bibitem{\roanA} S.-S.\ Roan, {\it A Characterization of ``Rapidity''
                 Curve in the Chiral Potts Model},
                 Commun.\ Math.\ Phys.\ {\bf 145} (1992) p.\ 605
\bibitem{\daviesA} B.\ Davies, {\it Onsager's Algebra and Superintegrability},
                 Jour.\ Phys.\ A: Math.\ Gen.\ {\bf 23} (1990) p.\ 2245
\bibitem{\daviesB} B.\ Davies, {\it Onsager's Algebra and the Dolan-Grady
                 Condition in the non-Self-Dual Case},
                 J.\ Math.\ Phys\ {\bf 32} (1991) p.\ 2945
\bibitem{\roanB} S.-S.\ Roan, {\it Onsager's Algebra, Loop Algebra and
                 Chiral Potts Model},
                 preprint Max-Planck-Institut f\"ur Mathematik
                 Bonn MPI/91-70
\bibitem{\ahn}   C.\ Ahn, K.\ Shigemoto, {\it Onsager Algebra and
                 Integrable Lattice Models},
                 Mod.\ Phys.\ Lett.\ {\bf A6} (1991) p.\ 3509
\bibitem{\fatzamA} V.A.\ Fateev, A.B.\ Zamolodchikov,
                 {\it Nonlocal (Parafermion) Currents in Two-Dimen\-sional
                 Conformal Quantum Field Theory and Self-Dual Critical Points
                 in $\Zed_N$-Sym\-metric Statistical Systems},
                 Sov.\ Phys.\ JETP {\bf 62} (1985) p.\ 215
\bibitem{\alcaraz} F.C.\ Alcaraz, A.L.\ Santos, {\it Conservation Laws for
                 $\Zed(N)$ Symmetric Quantum
                 Spin Models and Their Exact Ground State Energies},
                 Nucl.\ Phys.\ {\bf B275} (1986) p.\ 436
\bibitem{\fateev} V.A.\ Fateev, A.B.\ Zamolodchikov,
                 {\it Conformal Quantum Field Theory Models in Two
                 Dimensions Having $\Zed_3$ Symmetry},
                 Nucl.\ Phys.\ {\bf B280} (1987) p.\ 644
\bibitem{\lykyanov} V.A.\ Fateev, S.L.\ Lukyanov,
                 {\it The Models of Two-Dimensional Conformal Quantum
                 Field Theory with $\Zed_n$ Symmetry},
                 Int.\ Jour.\ of Mod.\ Phys.\ {\bf A3} (1988) p.\ 507
\bibitem{\cardy} J.L.\ Cardy, {\it Critical Exponents of the Chiral
                 Potts Model from Conformal Field Theory},
                 Nucl.\ Phys.\ {\bf B389} (1993) p.\ 577
\bibitem{\zamPA} A.B.\ Zamolodchikov,
                 {\it Higher-Order Integrals of Motion in Two-Dimensional
                 Models of the Field Theory with Broken Conformal Symmetry},
                 JETP Lett.\ 46 (1987) p.\ 160
\bibitem{\zamPB} A.B.\ Zamolodchikov, {\it Integrals of Motion in Scaling
                 3-State Potts Model Field Theory},
                 Int.\ Jour.\ of Mod.\ Phys.\ {\bf A3} (1988) p.\ 743
\bibitem{\zamPC} A.B.\ Zamolodchikov,
                 {\it Integrable Field Theory from Conformal Field Theory},
                 Adv.\ Studies in Pure Math.\ {\bf 19} (1989) p.\ 641
\bibitem{\fatzamB} V.A.\ Fateev, A.B.\ Zamolodchikov,
                 {\it Integrable Perturbations of $\Zed_N$ Parafermion
                 Models and the $O(3)$ Sigma Model},
                 Phys.\ Lett.\ {\bf B271} (1991) p.\ 91
\bibitem{\mussardo} G.\ Mussardo, {\it Off-Critical Statistical Models:
                 Factorized Scattering Theories and Bootstrap Program},
                 Phys.\ Rep.\ {\bf 218} (1992) p.\ 215
\bibitem{\weA}   G.\ von Gehlen, A.\ Honecker, {\it Multi-Particle Structure
                 in the $\Zed_n$-Chiral Potts Models},
                 Jour.\ Phys.\ A: Math.\ Gen.\ {\bf 26} (1993) p.\ 1275
\bibitem{\gehlenph} G.\ von Gehlen,
                 {\it Phase Diagram and Two-Particle Structure
                 of the $\Zed_3$-Chiral Potts Model}, Proceedings of
                 {\it International Symposium on Advanced Topics of
                 Quantum Physics}, ed.\ J.Q.\ Liang, M.L.\ Wang, S.N.\ Qiao,
                 D.C.\ Su, Science Press Beijing (1992) p.\ 248
%%% Modification
% reference has appeared
\bibitem{\lett} G.\ von Gehlen, A.\ Honecker, {\it Excitation Spectrum and
                 Correlation Functions of the $\Zed_3$-Chiral Potts Quantum
                 Spin Chain}, % preprint BONN-TH-94-20
                 Nucl.\ Phys.\ {\bf B435} (1995) p.\ 505
\bibitem{\dkcoy} S.\ Dasmahapatra, R.\ Kedem, B.M.\ McCoy,
                 {\it Spectrum and Completeness of the 3 State
                 Superintegrable Chiral Potts Model},
                 Nucl.\ Phys.\ {\bf B396} (1993) p.\ 506
%%% Modification
% reference has appeared
\bibitem{\kedem} R.\ Kedem, B.M.\ McCoy, {\it Quasi-Particles in the
                 Chiral Potts Model},
%                preprint ITPSB 94-013, hep-th/9405089
                 Int.\ Jour.\ of Mod.\ Phys.\ {\bf B8} (1994) p.\ 3601
\bibitem{\han}   N.S.\ Han, A.\ Honecker,
                 {\it Low-Temperature Expansions and Correlation
                 Functions of the $\Zed_3$-Chiral Potts Model},
                 Jour.\ Phys.\ A: Math.\ Gen.\ {\bf 27} (1994) p.\ 9
\bibitem{\chrihen} Ph.\ Christe, M.\ Henkel,
                 {\it Introduction to Conformal
                 Invariance and its Applications to Critical Phenomena},
                 Lecture Notes in Physics m16, Springer-Verlag (1993)
\bibitem{\zam}   A.B.\ Zamolodchikov,
                 {\it Infinite Additional Symmetries in Two-Dimensional
                 Conformal Quantum Field Theory},
                 Theor.\ Math.\ Phys.\ 65 (1986) p.\ 1205
\bibitem{\baym}  G.\ Baym, {\it Lectures on Quantum Mechanics},
                 Benjamin/Cummings Publishing (1969), 3rd printing (1974),
                 chapter 11
\bibitem{\tang}  G.\ Albertini, B.M.\ McCoy, J.H.H.\ Perk, S.\ Tang,
                 {\it Excitation Spectrum and Order Parameter for the
                 Integrable $N$-State Chiral Potts Model},
                 Nucl.\ Phys.\ {\bf B314} (1989) p.\ 741
\bibitem{\hela}  M.\ Henkel, J.\ Lacki, {\it Integrable Chiral $\Zed_n$
                 Quantum Chains and a New Class of Trigonometric Sums},
                 Phys.\ Lett.\ {\bf 138A} (1989) p.\ 105
%%% Modification
% additional reference
\bibitem{\camp} W.J.\ Camp, {\it Decay of Order in Classical Many-Body
                Systems.\ II.\ Ising Model at High Temperatures},
                Phys.\ Rev.\ {\bf B6} (1972) p.\ 960
%%% Modification
% additional reference
\bibitem{\thesis} A.\ Honecker,
                 {\it Quantum Spin Models and Extended Conformal Algebras},
                 Ph.D.\ thesis BONN-IR-95-12 (1995), hep-th/9503104
\bibitem{\jones} F.M.\ Goodman, P.\ de la Harpe, V.F.R.\ Jones,
                 {\it Coxeter Graphs and Towers of Algebras},
                 Springer-Verlag (1989)
%%% Modification
% more precise reference
\bibitem{\yildirim} K.\ Yildirim, {\it Parit\"atssymmetrie im Chiralen
                 $\Zed_n$-Symmetrischen Integrablen Potts-Modell},
                 Diplomarbeit BONN-IB-95-02 (1995)
\bibitem{\cardyB} J.L.\ Cardy, {\it Effect of Boundary Conditions on
                 the Operator Content of Two-Dimen\-sional
                 Conformally Invariant Theories},
                 Nucl.\ Phys.\ {\bf B275} (1986) p.\ 200
\bibitem{\schuetz} G.\ von Gehlen, V.\ Rittenberg, G.\ Sch\"utz,
                 {\it Operator Content of $n$-State Quantum Chains in the
                 $c=1$ Region},
                 J.\ Phys.\ A: Math.\ Gen.\ {\bf 21} (1988) p.\ 2805
\bibitem{\automos} A.\ Honecker, {\it Automorphisms of ${\cal W}$-Algebras
                 and Extended Rational Conformal Field Theories},
                 Nucl.\ Phys.\ {\bf B400} (1993) p.\ 574
\bibitem{\kogut} J.B.\ Kogut, {\it An Introduction to Lattice Gauge
                 Theory and Spin Systems},
                 Reviews of Mod.\ Phys.\ {\bf 51} (1979) p.\ 659
%%% Modification
% misprint in title corrected
\bibitem{\krallm} T.W.\ Krallman, {\it Phasendiagramm des Chiralen
                 Pottsmodells}, Diplomarbeit BONN-IR-91-11 (1991)
\bibitem{\albcoy} G.\ Albertini, B.M.\ McCoy,
                 {\it Correlation Functions of the Chiral Potts Chain
                 from Conformal Field Theory and Finite-Size Corrections},
                 Nucl.\ Phys.\ {\bf B350} (1990) p.\ 745
\bibitem{\lehmann} H.\ Lehmann, {\it \"Uber Eigenschaften von
                 Ausbreitungsfunktionen und Renormierungs\-kon\-stanten
                 Quantisierter Felder}, Il Nuovo Cimento
                 {\bf 11} (1954) p.\ 342
\bibitem{\gehkal} G.\ von Gehlen, L.\ Kaldenbach,
                 {\it Off-Criticality Behaviour of the
                 $\Zed_5$-Quantum Spin Chain},
                 in preparation
\bibitem{\albCONF} G.\ Albertini, S.\ Dasmahapatra, B.M.\ McCoy, {\it Spectrum
                 Doubling and the Extended Brillouin Zone in the Excitations
                 of the Three States Potts Spin Chain}, Phys.\ Lett.\
                 {\bf A170} (1992) p.\ 397
\bibitem{\reedsimon}  M.\ Reed, B.\ Simon, {\it Methods of Modern Mathematical
                 Physics IV: Analysis of Operators},
                 Academic Press (1978), chapter XII
\bibitem{\katobook} T.\ Kato, {\it Perturbation Theory for Linear Operators},
                 Springer-Verlag (1976), second edition
\bibitem{\rellich} F.\ Rellich,
                 {\it St\"orungstheorie der Spektralzerlegung.\ IV.},
                 Math.\ Ann.\ {\bf 117} (1940) p.\ 356
\bibitem{\kato} T.\ Kato, {\it On the Convergence of the Perturbation
                 Method. I.}, Progr.\ Theor.\ Phys.\ {\bf 4} (1949) p.\ 514
\vfill
\eject
\chapsubtitle{Figure captions}
\mn
\parindent=1.3cm
\def\figitem#1{\item{\hbox to 1.1cm{#1 \hfill}}}
\figitem{Fig.\ 1:}
Correlation function $C_{\Ga}(x)$ stretched by $e^{x \over \xi_{\Ga}}$
in comparison to the fits (\secG.18a) at $\phi = \vphi = {\pi \over 2}$,
$\la = {1 \over 2}$. The `error bars' are given by
%%% Modification
% wrong sign in exponent changed
$a e^{x - 6 \over \xi_{\Ga}}$ which conveys an idea how much the
values have actually been stretched. The oscillatory contribution
to $C_{\Ga}(x)$ is clearly visible.
\figitem{Fig.\ 2:}
Correlation length for $C_{\Ga}(x)$ in the massive high-temperature phase
on the parity conserving line $\phi = \vphi = 0$. The points indicate
estimates obtained from a perturbative evaluation of $C_{\Ga}(x)$. The lines
indicate the approximation (\secG.21) for $\xi_\Ga$
and the properly normalized inverse mass gap $m(\la)^{-1}$.
\figitem{Fig.\ 3:}
Radii of convergence and boundary of the massive high-temperature phase
for the hermitean $\Zed_3$-chain. $r_1$ is an estimate ensuring
convergence if {\it no level crossings between point and continuous
spectrum occur}. The estimate $r_2$ also ensures the absence of
level crossings. The perturbation series are definitely
convergent for $\la < r_2$ although the true radius of convergence is
larger. It extends until the value $\la_0$ where the first
level crossings between fundamental quasiparticles and scattering
states occur. $\la_0$ has been approximated using a first order
perturbation expansion which is surprisingly accurate.
\figitem{}
The boundary of the massive high-temperature phase close to
$\la = 1$ has been approximated using a second order perturbation
expansion.
\figitem{}
Note that $r_1$ and $r_2$ are independent of $\phi$ up to the order
calculated whereas for $\la_0$ we put $\vphi = \phi$.
\vfill
\end